\documentclass[fleqn,11pt]{article}

\usepackage{graphicx}
\usepackage{float}
\usepackage{standalone}
\usepackage{mathrsfs}
\usepackage{wrapfig}
\usepackage{float}
\usepackage{stmaryrd}
\usepackage{amsmath}
\usepackage{subfigure}
\usepackage{pgf,tikz}
\usetikzlibrary{arrows}
\usetikzlibrary{babel}
\usepackage{wrapfig}
\usetikzlibrary{calc,patterns,angles,quotes}
\usepackage{vmargin}
\usepackage{longtable}
\usepackage{subfigure}

\usepackage{subcaption}
\usepackage{algpseudocode}
\usepackage{array}
\usepackage{multirow}
\usepackage{multicol}
\usepackage{longtable}
\usepackage{booktabs}

\usepackage{titlesec}

\usepackage{pdfpages}
\usepackage{pdflscape}
\usepackage{xfrac}
\usepackage{authblk}
\usepackage[utf8]{inputenc}
\usepackage{enumitem} 
\usepackage{natbib}
\usepackage{extarrows}
\usepackage[ruled,linesnumbered]{algorithm2e}
\usepackage{multirow}
\usepackage{pdflscape}
\usepackage[toc,page]{appendix}

\usepackage{amssymb}
\usepackage{amsmath}
\usepackage{amsthm}
\usepackage{mathabx}
\usepackage{empheq}
\usepackage{mathtools}

\usepackage[hang, flushmargin]{footmisc}
\usepackage[colorlinks = true,
            linkcolor = blue,
            urlcolor  = blue,
            citecolor = blue,
            anchorcolor = blue]{hyperref}

\numberwithin{equation}{section}

\usepackage{mathrsfs}

\usepackage{array}
\newcolumntype{P}[1]{>{\centering\arraybackslash}p{#1}}

\def\dotminus{\mathbin{\ooalign{\hss\raise1ex\hbox{.}\hss\cr
  \mathsurround=0pt$-$}}}

\begin{document}

\title{Learning collective variables\\ that respect permutational symmetry}
\author[1]{Jiaxin Yuan\thanks{jyuan98@umd.edu}}
\author[1]{Shashank Sule\thanks{ssule25@umd.edu}}
\author[2]{Yeuk Yin Lam\thanks{lam00185@umn.edu}}
\author[1]{Maria Cameron\thanks{mariakc@umd.edu}}
{\scriptsize
\affil[1]{Department of Mathematics, University of Maryland, College Park, MD 20742, USA}
\affil[2]{School of Mathematics, University of Minnesota, Twin Cities, MN 55455, USA}
}

\maketitle

\abstract{
In addition to translational and rotational symmetries, clusters of identical interacting particles possess permutational symmetry. Coarse-grained models for such systems are instrumental in identifying metastable states, providing an effective description of their dynamics, and estimating transition rates. We propose a numerical framework for learning collective variables that respect translational, rotational, and permutational symmetries, and for estimating transition rates and residence times. It combines a sort-based featurization, residence manifold learning in the feature space, and learning collective variables with autoencoders whose loss function utilizes the orthogonality relationship (Legoll and Lelievre, 2010). The committor of the resulting reduced model is used as the reaction coordinate in the forward flux sampling and to design a control for sampling the transition path process. We offer two case studies, the Lennard-Jones-7 in 2D and the Lennard-Jones-8 in 3D. The transition rates and residence times computed with the aid of the reduced models agree with those obtained via brute-force methods.}


\tableofcontents

\section{Introduction}
\label{sec:intro}
Molecular dynamics (MD) simulations offer a platform for investigating the properties and behaviors of molecular systems at an atomic level. 
Typically, MD simulations aim to explore the configurational space, identify metastable states, and estimate the transition rates between them. The two main challenges in achieving these goals are high dimensionality and a wide range of timescales. The dimension of the configurational space is $3N$, where $N$, the number of atoms, is often large. The shortest timescale of interatomic bond vibrations that determines the timestep of MD simulations is of the order of a femtosecond, while conformational changes of interest happen on much larger timescales. 
%For example, the expected time of the transition from C7eq to C7ax in alanine dipeptide is of the order of a microsecond~\cite{vani2022computing,Evans2022}.

Collective variables (CVs), differentiable functions of atomic coordinates, facilitate the study of complex molecular systems. They are typically used for 
\begin{itemize}
    \item biasing MD simulations,
    \item defining the metastable states, and
    \item visualizing the results.
\end{itemize} 
In small organic molecules, such as butane and alanine dipeptide, the dihedral angles formed by triplets of atoms along the molecule's backbone serve as natural collective variables. The MD force fields often admit significant changes in dihedral angles while the lengths of covalent bonds and interbond angles change relatively little. However, for large and complex molecules or atomic clusters, finding suitable CVs is much more challenging. In 2005, Ma and Dinner~\cite{MaDinner2005} proposed using a neural network to learn the committor, facilitating the study of isomerization in the alanine dipeptide. Since then, machine learning methods have shown great promise for identifying CVs and exploring the conformational space of biomolecules -- see Refs.~\cite{Gkeka2020review,Noe2021review,Ferguson2020review} for a comprehensive overview and references therein. Over the past decade, researchers have successfully employed autoencoders~\cite{Kramer1992AutoassociativeNN} to discover CVs in molecular dynamics (MD) simulations. Time-lagged autoencoders~\cite{Noe2018autoencoders} are trained to predict future configurations, $\mathbf{x}_\tau \rightarrow \mathbf{x}_{\tau+1}$, and identify low-dimensional embeddings that capture the slow dynamics of diffusion processes. The RAVE algorithm~\cite{Tiwary2018RAVE}, a widely adopted method, utilizes a variational autoencoder to learn the latent distribution and interprets CVs in terms of pre-selected order parameters. Ferguson's group~\cite{Ferguson2018autoencoders_on-the-fly,Ferguson2018autoencoders_innovations} developed neural network architectures within the autoencoder framework to accommodate periodicity and eliminate translational and rotational invariance in CV discovery. In recent work by Belkacemi et al.~\cite{belkacemi2023autoencoders}, CVs learned by autoencoders~\cite{Kramer1992AutoassociativeNN} allowed the authors to identify six metastable states in HSP90 and map them to a 2D space.  

In this work, we address the problem of learning CVs for a class of systems that lack obvious physically motivated reaction coordinates: systems of interacting particles with permutational symmetry. Examples are the Lennard-Jones clusters modeling heavy rare gas (Ar, Xe, Cr) clusters at low temperatures~\cite{Wales_2004_book}, and sticky spheres such as micron-size colloidal particles interacting via the depletion force~\cite{HolmesCerfon2017Sticky}.

The goal of this work is to 
\begin{itemize}
    \item develop an algorithm for learning CVs respecting the permutational symmetry that allows us to $(i)$ define the metastable states and $(ii)$ accurately estimate the transition rate between them, and
    \item apply it to Lennard-Jones clusters and compare the utility of the learned CVs with the standard ones, the second and third central moments of the coordination numbers~\cite{Tribell02010LJ7,Tiwary2019CNum}.
\end{itemize} 
Our original goal was to learn CVs such that the transition rates in the resulting coarse-grained model match those in the original model. However, we have realized that this requirement is incompatible with the invariance of the coarse-grained model under permutational symmetry. Therefore, we have split this project into two. In the first study, Sule et al.~\cite{Sule2025butane} explore the task of preserving rate under rotational and translational symmetry through a case study of a normal-butane molecule. Only a thoughtfully chosen feature map and a carefully designed CV result in an accurate rate, while other seemingly reasonable options yield exaggerated rate estimates. 
This work is devoted to learning CVs respecting rotational, translational, and \emph{permutational} symmetry and using the committor of the resulting coarse-grained model in forward flux sampling~\cite{Allen_2009} and stochastic control~\cite{YUAN_2024_optimalcontrol} to find the transition rates and residence times.

We adjust a framework for identifying collective variables (CVs) proposed in the parallel work~\cite{Sule2025butane} for systems with permutational symmetry. The framework consists of five steps.
\begin{enumerate}
    \item \emph{A feature map that achieves invariance under rotational, translational, and permutational symmetry.} We map the atomic coordinates into a vector invariant under rotations and translations, e.g., a vector of pairwise distances squared or a vector of coordination numbers~\cite{Tribell02010LJ7,Tiwary2019CNum}, and then sort this vector. 
    
    \item \emph{Learning the residence manifold}. This step is done by means of diffusion maps~\cite{CoifmanLafon2006,Coifman_2008,banisch2017_target_measure_diffusionMap} and diffusion nets~\cite{2015diffusionnets}.
    
    \item \emph{Learning the CVs whose level sets are orthogonal to the residence manifold.} To do it, we first design a confining potential $V_1$ whose zero-level set is the residence manifold. Then, we learn CVs whose gradients are orthogonal to the gradient of $V_1$ at the residence manifold via the use of autoencoders following Refs.~\cite{Noe2018autoencoders,Ferguson2018autoencoders_on-the-fly,Ferguson2018autoencoders_innovations,Tiwary2018RAVE,Pande2018autoencoders,Bolhuis2021autoencodes,Belkacemi_2021}. This orthogonality requirement is motivated by the orthogonality condition obtained by Legoll and Lelievre~\cite{Legoll_2010} for a scalar CV and then extended to vector CVs by Duong et al.~\cite{Sharma_2018}. This condition reduces the coarse-grained error measured by the relative entropy of the probability density of the CV following the original process and the probability density of the coarse-grained process in this CV. 
    
    \item \emph{Computing the free energy and the diffusion tensor for the reduced dynamics in the learned CVs.} This step is standard. The calculation of the diffusion matrix follows the recipe proposed in Maragliano et al.~\cite{Maragliano}. The calculation of the free energy consists of flattening the energy landscape by means of well-tempered metadynamics~\cite{Branduardi2012MetadynamicsWA,Dama2014WelltemperedMC,Valsson2016metad}, binning the stochastic trajectory in the flattened landscape, and then restoring the free energy. 
    
    \item \emph{Computing the committor for the low-dimensional coarse-grained model}. This committor is used as the reaction coordinate for the forward flux sampling~\cite{Allen_2005} and to design a controller for sampling transition trajectories~\cite{YUAN_2024_optimalcontrol}. 
    \end{enumerate}

Using this framework, we have conducted two case studies: the standard benchmark problem, Lennard-Jones-7 (LJ7) in 2D, and a more complex and less studied example, Lennard-Jones-8 (LJ8) in 3D. For both cases, we demonstrated the power of the proposed algorithm for learning CVs and its robustness with respect to implementation details and quality of representations at intermediate steps. The rate estimates obtained with the aid of the learned CVs are consistent with the brute force rates. For comparison, we also used the standard set of CVs, the second and third central moments of the coordination numbers, $(\mu_2,\mu_3)$~\cite{Tribell02010LJ7,EVANS2023,YUAN_2024_optimalcontrol}, for LJ7, and $(\mu_2,\mu_3)$ and LDA-based~\cite{DHS2001} CVs for LJ8. The rates obtained with $(\mu_2,\mu_3)$ are comparable in both cases with the brute force, while the LDA-based CVs yielded notably larger errors in rate estimates.

The rest of the paper is organized as follows. Section~\ref{sec:LJreview} reviews Lennard-Jones clusters. Section \ref{sec:background} discusses theory and computational techniques that serve as building blocks of the proposed framework. Section \ref{sec:method} introduces the proposed algorithm for learning CVs. Sections \ref{sec:LJ7results} and \ref{sec:LJ8results} are devoted to the case studies of LJ7 in 2D and LJ8 in 3D, respectively. The performance and features of the proposed algorithm and the results of the two case studies are discussed in Section \ref{sec:discussion}. The conclusions are drawn in Section \ref{sec:conclusion}. Additional technical information is distributed between the Appendices and Supplementary Information (SI). The codes developed for this work are available on GitHub~\cite{margotyjx,mar1akc}:\\
 \href{https://github.com/margotyjx/OrthogonalityCVLearning/tree/main/LrCV_permsym}{https://github.com/margotyjx/OrthogonalityCVLearning/tree/main/LrCV\_permsym} 
and \\\href{https://github.com/mar1akc/LJ7in2D_LJ8in3D_learningCVs}{https://github.com/mar1akc/LJ7in2D\_LJ8in3D\_learningCVs}.

%%%%%%%%%%%%%%%%%%%%%%%%%%%%%%%%%%%%%%

\section{Lennard-Jones clusters}
\label{sec:LJreview}
Lennard-Jones clusters of up to $\sim$100 particles are prototypical interacting particle systems, small enough that the number of particles matters. The Lennard-Jones potential 
\begin{equation}
    \label{eq:LJpot}
    V = 4\sum_{i<j}\left(r^{-12}_{i,j} - r^{-6}_{i,j}\right),
\end{equation}
where $r_{i,j}$ denotes the distance between atoms $i$ and $j$, is inspired by the dipole-dipole interaction and is used to model nonbonded interactions between atoms of ionic solids, for example, in zeolites~\cite{ZeoliteLJ2011}. Lennard-Jones clusters featured in seminal works of Wales' group--see Ref.~\cite{Wales_2004_book} and references therein--are benchmarks for global optimization~\cite{Wales1997hopping} and establishing thermodynamic~\cite{BerryWales1989} and dynamical~\cite{Dellago1998LJ,WALES2002DPS,String2002,picciani2011,EVANS2023} properties in finite systems. 

Wales and collaborators have developed a methodology~\cite{WALES2002DPS} and software~\cite{Wales_software} for mapping energy landscapes onto stochastic networks whose vertices and edges represent, respectively, local energy minima and saddles that separate those minima. The transition rates between adjacent minima are found using Kramers' formula~\cite{kramersRate,Berezh2004} adjusted to account for the point group order~\cite{WALES2002DPS}.

\begin{figure}[htbp]
    \centering
\includegraphics[width = 0.9\textwidth]{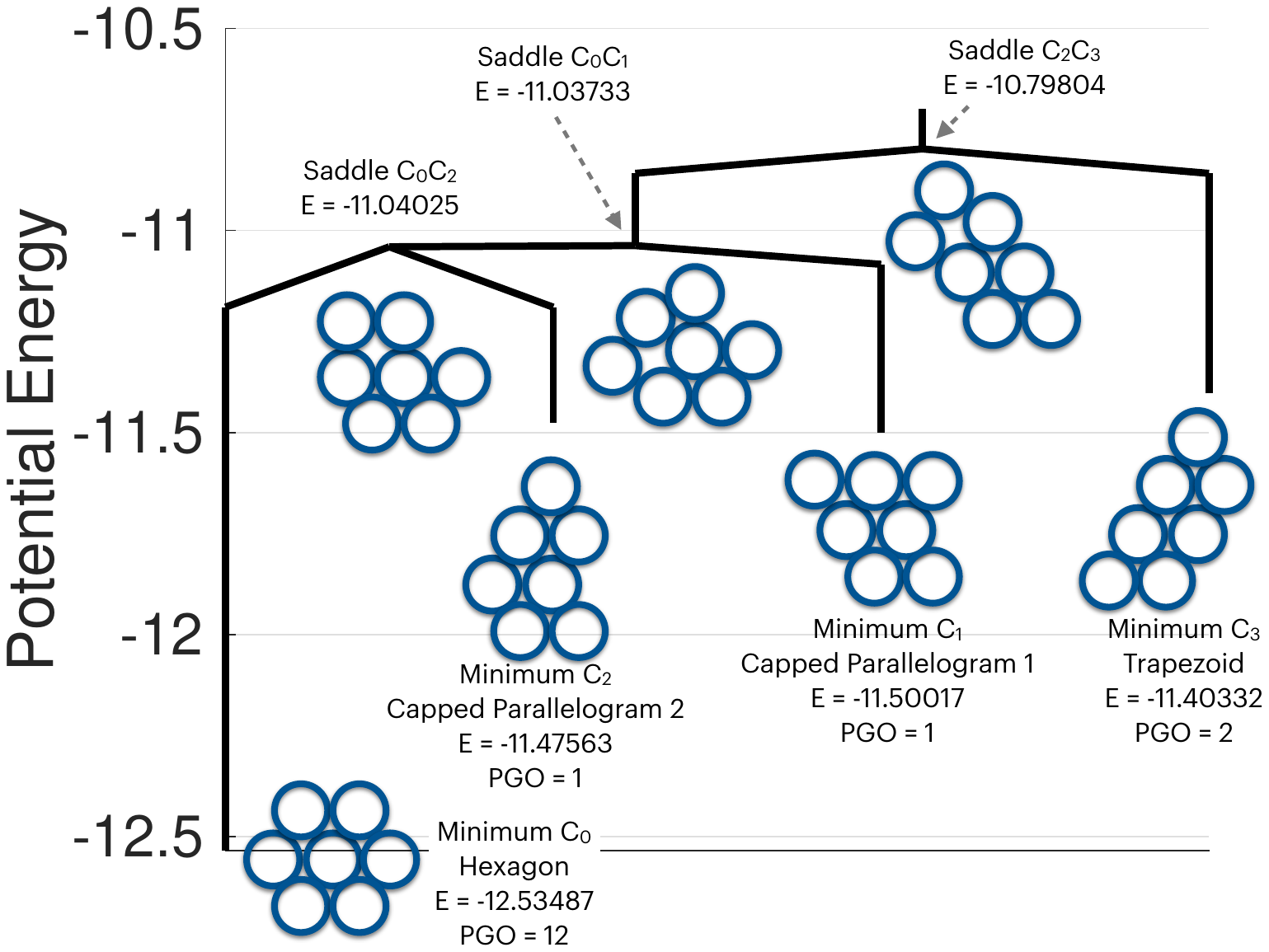}% Here is how to import EPS art
\caption{The disconnectivity graph for the Lennard-Jones-7 cluster in 2D and descriptions of its local potential energy minima and saddles. $E$ is the energy value, and PGO is the point group order.}
\label{fig:LJ7digraph}
\end{figure}

\begin{figure}[htbp]
    \centering
\includegraphics[width =0.9\textwidth]{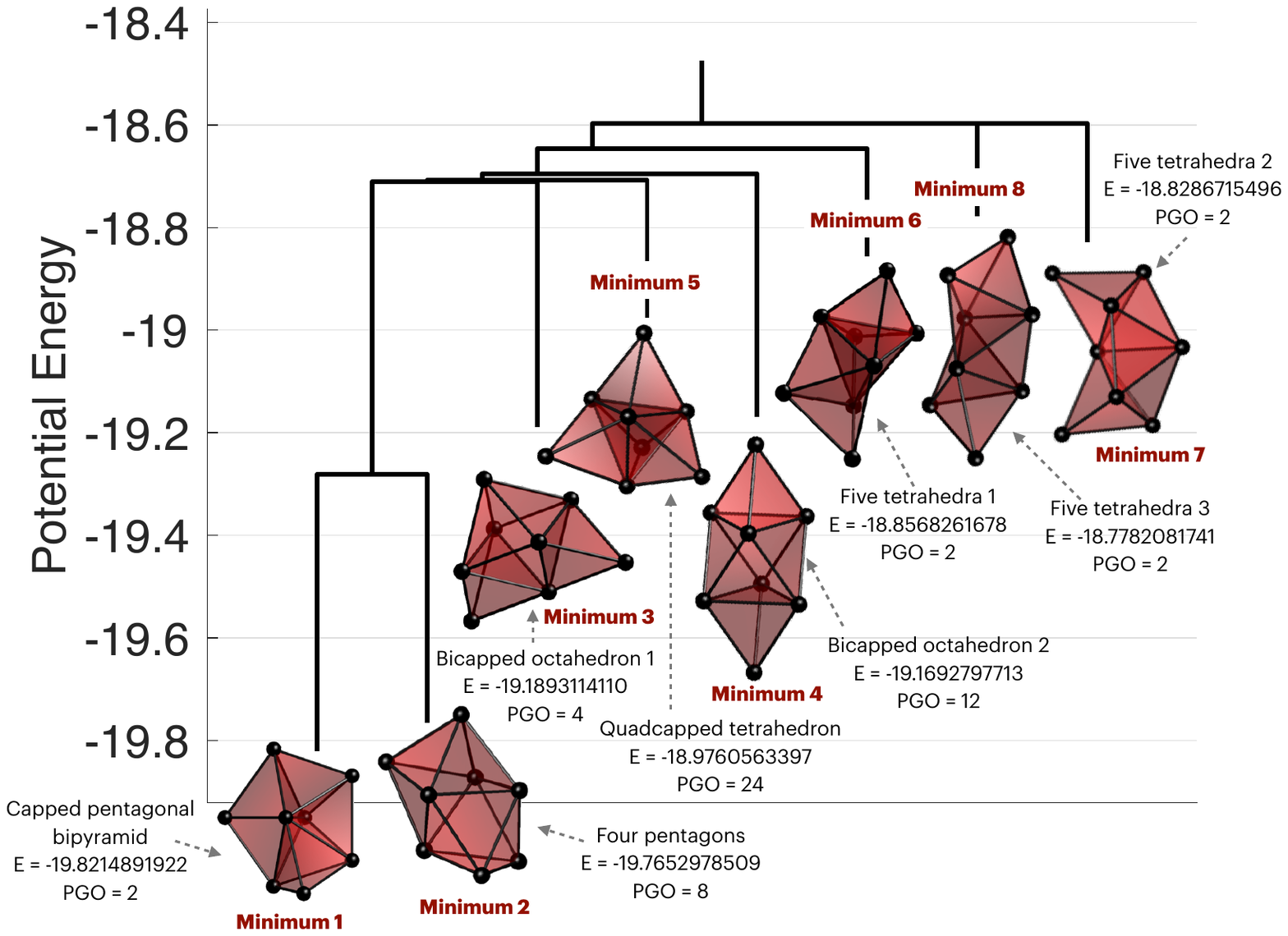}% Here is how to import EPS art
\caption{The disconnectivity graph for the Lennard-Jones-8 cluster in 3D and descriptions of its local potential energy minima. $E$ is the energy value, and PGO is the point group order.}
\label{fig:LJ8digraph}
\end{figure}

Wales's group visualizes these networks using disconnectivity graphs~\cite{Wales_2004_book}, i.e., trees whose leaves correspond to local potential energy minima, with their $y$-coordinates representing their energy values, and joined at $y$-coordinates equal to the lowest saddle separating them in the energy landscape. Disconnectivity graphs for Lennard-Jones-7 (LJ7) in 2D~\cite{WALES2002DPS} and Lennard-Jones-8 (LJ8) in 3D~\cite{Forman_Cameron_2017} are shown in Figs. \ref{fig:LJ7digraph} and \ref{fig:LJ8digraph} respectively. 

Mapping of Lennard-Jones clusters onto stochastic networks greatly facilitates the study of the dynamics of the cluster~\cite{Cameron2014FlowsIC,Cameron2014,CameronGan2016}. 
However, the computation of such a network is a formidable task, and the representation of clusters' dynamics via Markov jump processes introduces errors in transition rates. These errors are inherited from the rates between neighboring minima estimated via Kramers-Langer's formula~\cite{Kramers1940,Berezh2004} that may be inaccurate when the barriers are broad and flat~\cite{Roux2022}.

\begin{figure}
    \centering
\includegraphics[width = 0.9\textwidth]{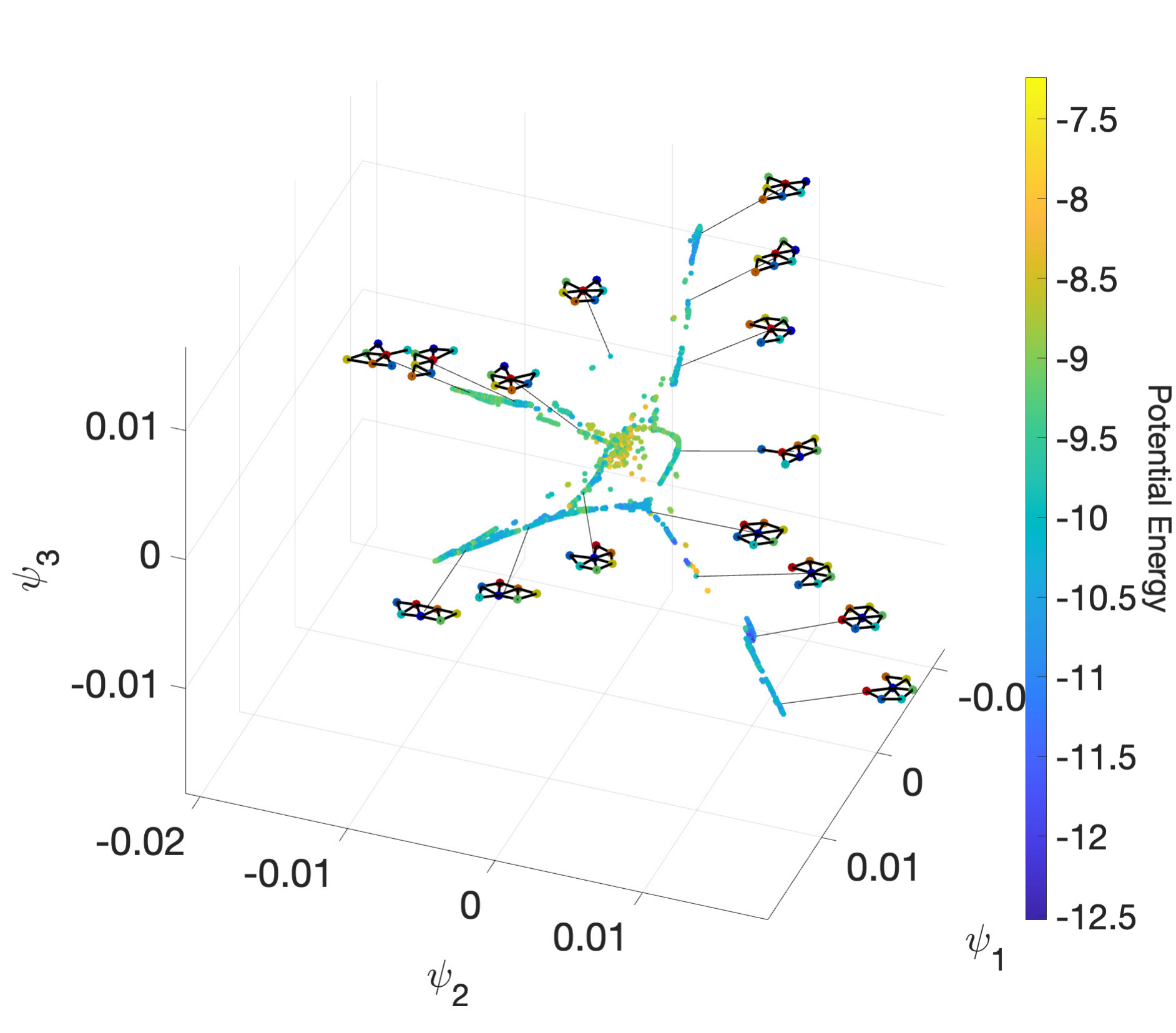}% Here is how to import EPS art
\caption{\label{fig:LJ7dmap} The diffusion map of a set of 30467 representative configurations of LJ7 in 2D into the span of three dominant nonconstant eigenvectors of the Markov matrix. The dataset is collected by binning a long trajectory in $(\mu_2,\mu_3)$ (the second and third central moments of the coordination numbers) in the forcefield modified by well-tempered metadynamics~\cite{Dama2014WelltemperedMC}. The configurations along the trajectory were aligned by solving the orthogonal Procrustes problem at each time step. The normalization parameter was set to $\alpha =1$ to eliminate the effect of the sampling density. The bandwidth parameter $\epsilon = 4$, the sparsification of the kernel matrix was achieved by using 256 nearest neighbors.}
\end{figure}

 Coifman et al.~\cite{Coifman_2008} mapped a large collection of representative configurations of the 14-dimensional Lennard-Jones-7 (LJ7) in 2D into a three-dimensional space spanned by three dominant eigenvectors of the diffusion map. We generated a similar figure -- see Fig. \ref{fig:LJ7dmap}. 

\begin{figure}[htbp]
\includegraphics[width = 0.6\textwidth]{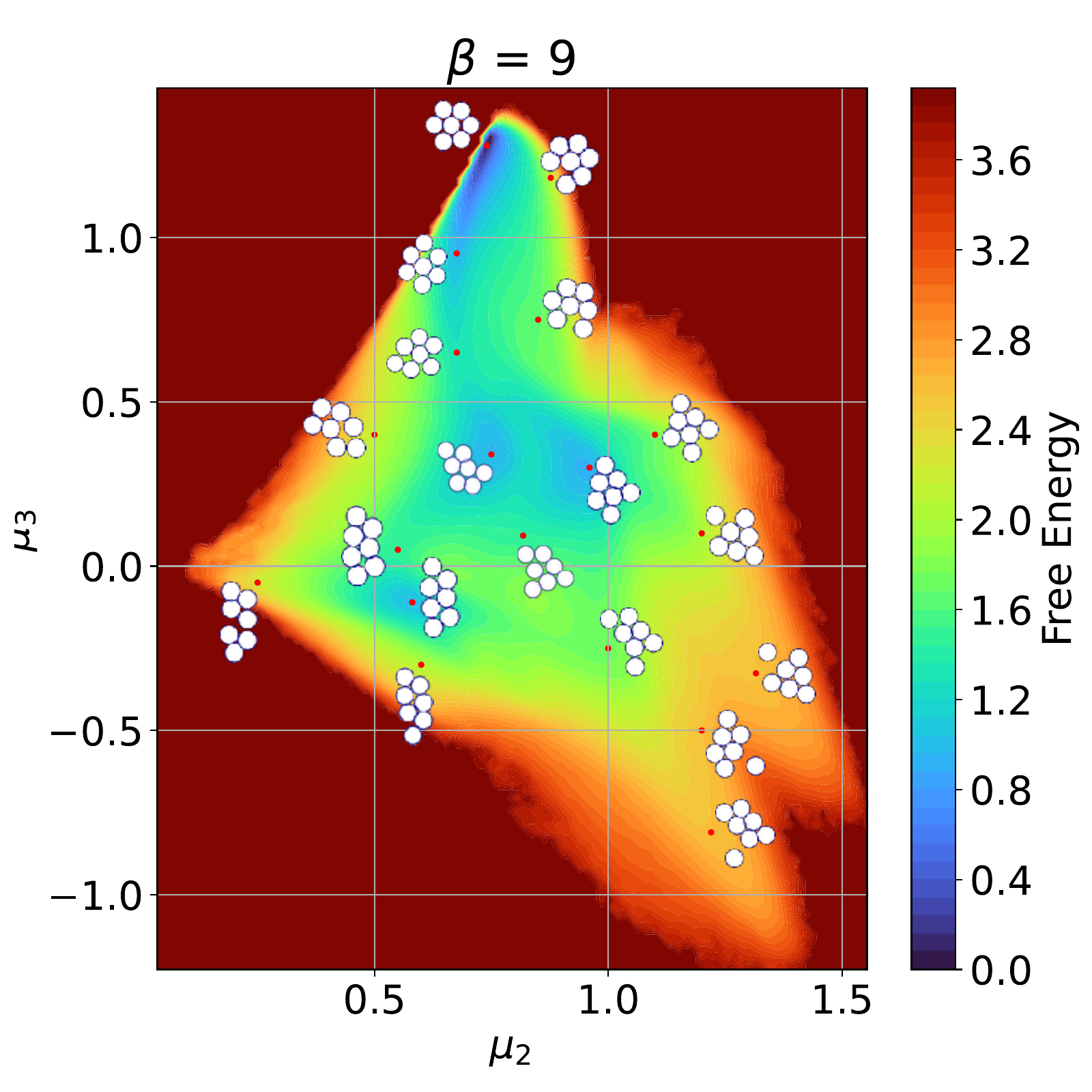}% Here is how to import EPS art
\caption{The free energy landscape of LJ7 in 2D in the second and third central moments  $(\mu_2,\mu_3)$ of the coordination numbers \eqref{eq:CNum}.}
\label{fig:LJ7confs} \end{figure}

In contrast with small organic molecules with a backbone consisting of covalent-bonded heavy atoms, such as butane and alanine dipeptide, where the natural CVs are the dihedral angles along the backbone, Lennard-Jones clusters do not have obvious physically motivated CVs. Nonetheless, two pairs of physically motivated CVs were crafted.
Picciani et al.~\cite{picciani2011} used bond-oriented order parameters, $Q_4$ and $Q_6$, as CVs for direct transition current sampling to study the transition in LJ38 between the icosahedral and face-centered cubic funnels. Tribello et al.~\cite{Tribell02010LJ7} proposed to use the second and third central moments, $\mu_2$ and $\mu_3$, of the numbers of the nearest neighbors, termed \emph{coordination numbers}, as CVs in the metadynamics algorithm to explore the energy landscape of LJ7 in 2D and calculated the free energy landscape. This idea was further developed in Tsai et al.~\cite{Tiwary2019CNum}: the number of nearest neighbors of atom $i$, $c_i$, was approximated by a continuous function, and $(\mu_2,\mu_3)$ were used to study nucleations. Later works~\cite{EVANS2023,YUAN_2024_optimalcontrol} used the second and third central moments,$(\mu_2,\mu_3)$,  
of the coordination numbers defined as
\begin{equation}
    \label{eq:CNum}
    c_i = \sum_{j\neq i} \frac{1-\left(\frac{r_{i,j}}{1.5}\right)^8}{1-\left(\frac{r_{i,j}}{1.5}\right)^{16}},
\end{equation}
to quantify the transition between the hexagon and the trapezoid in LJ7 in 2D. The free energy landscape of LJ7 in 2D in $(\mu_2,\mu_3)$ at $\beta = 9$ is displayed in Fig.~\ref{fig:LJ7confs}.

%%%%%%%%%%%%%%%%%%%%%%%%%%%%%%%%%%%%%%%%%%

\section{Background: theory and methods}
\label{sec:background}
In this section, we review the necessary background to motivate and explain the proposed framework for learning CVs and estimating rates.

\subsection{Effective dynamics in collective variables}
The effective dynamics of a single collective variable with underlying overdamped Langevin dynamics were analyzed and quantified by Legoll and Lelievre~\cite{Legoll_2010} and then extended to a vector of CVs with underlying overdamped Langevin or Langevin dynamics by Duong et al.~\cite{Sharma_2018}.

Let a system be governed by the  \textit{overdamped Langevin equation} in $\Omega_x\subset\mathbb{R}^{n}$,
\begin{equation}
\label{eq:overdampedLangevin}
dX_t = - \nabla V(X_t)dt + \sqrt{2\beta^{-1}}dW_t,
\end{equation} 
where the potential energy function $V(x)$ is smooth almost everywhere, has a finite number of isolated minima, and $\beta^{-1} = k_B T$ is the temperature in units of Boltzmann's constant. 
The invariant density is given by
\begin{align}
\label{eq: gibbs}
    \mu_X(x) = Z^{-1}e^{-\beta V(x)},\quad Z = \int_{\mathbb{R}^{n}}e^{-\beta V(x)}dx.
\end{align}
Let $\xi:\Omega_X\rightarrow\Omega_Y$ map $X_t$ to a $d$-dimensional CV space. The governing SDE for $\xi$ can be readily written out using Ito's formula; however, this SDE is not closed as it contains coefficient functions that depend on $X_t$. Legoll and Lelievre~\cite{Legoll_2010} and Duong et al.~\cite{Sharma_2018} proposed the following closure for the effective dynamics of $Y_t = \xi(X_t)$:
\begin{equation}\label{eq:dynamics_Lelievre}
    dY_t = b(Y_t)dt + \sqrt{2\beta^{-1}A(Y_t)}dW_t
\end{equation}
where 
\begin{align}\label{eq:b_CVs}
    & b(y) = \mathbb{E}_\mu \left[(D\xi \nabla U - \beta^{-1}\Delta \xi)(X_t)|\xi(X_t) = y \right],\\
    \label{eq:A_CVs}
    & A(y) = \mathbb{E}_\mu\left[|D\xi D\xi^\top|(X_t)|\xi(X_t) = y \right],
\end{align}
and $D\xi$ is the Jacobian matrix of $\xi$. 
Under the assumption that $D\xi D\xi^\top\succeq \delta I_d$ where $\delta > 0$ is a constant and $I_d$ is a $d\times d$ identity matrix, SDE~\eqref{eq:dynamics_Lelievre} is equivalent to~\cite{Sule2025butane} 
\begin{align}    
    dY_t &= [-M(Y_t)\nabla F(Y_t) + \beta^{-1}\nabla \cdot M(Y_t)]dt + \sqrt{2\beta^{-1}}M(Y_t)^{\frac{1}{2}}dW_t, \label{eq:dynamics_Maragliano}
\end{align}
obtained by Maragliano et al.~\cite{Maragliano} by an argument based on the variational property of the committor.
In \eqref{eq:dynamics_Maragliano},
$F(Y_t)$ and $M(Y_t)$ are the free energy and the diffusion tensor defined as 
\begin{equation}\label{eq:FE}
    F(y) = -\beta^{-1}\ln \left(\int_{\Omega_x} Z^{-1} e^{-\beta U(x)}\prod_{l = 1}^d \delta(\xi_l(x) - y_l)dx\right),
\end{equation}
\begin{align}
    M(y) &= e^{-\beta F(y)}\int_{\Omega_x}D\xi(x)D\xi^\top(x)Z^{-1} e^{-\beta V(x)}\prod_{l = 1}^d \delta(\xi_l(x) - y_l)dx.\label{eq:DMatrix}
\end{align}
The delta-function in the definitions of the free energy (\ref{eq:FE}) and diffusion matrix (\ref{eq:DMatrix}) is interpreted according to the co-area formula \cite{co-area}. I.e., for an integrable function $f: \mathbb{R}^n \rightarrow \mathbb{R}$,
\begin{align}
  \int_{\Omega_x} f(x) \delta(\xi_l(x) - y_l)dx &= \int_{\Omega_{y_l}}\left(\int_{\Sigma(y'_l)}f|\nabla \xi_l|^{-1}\delta(y'_l - y_l)d\sigma \right)dy'_l  \notag\\
&= \int_{\Sigma(y_l)}f\|\nabla \xi_l\|^{-1}d\sigma,\label{eq:coarea}
\end{align}
where $\Sigma(y_l) = \{x \in \mathbb{R}^n, y_l \in \mathbb{R}: \xi_l(x) = y_l\}$ is the level set of the function $\xi_l$, $d\sigma$ denotes the surface element of $\Sigma(y_l)$, and
it is assumed that $\|\nabla \xi_l\|^{-1} \neq 0$.

The invariant density for SDE \eqref{eq:dynamics_Maragliano} is 
\begin{equation}
    \label{eq:muz}
    \mu_Y(y)=Z^{-1}_Fe^{-\beta F(y)},\quad Z_F: = \int_{\Omega_Y}e^{-\beta F(y)}.
\end{equation}

\subsection{Coarse-graining error in the transition rate}
How accurately does the effective dynamics governed by \eqref{eq:dynamics_Lelievre} represent the dynamics \eqref{eq:overdampedLangevin}? Let metastable sets $A_y$ and $B_y$ be chosen in the CV space and then lifted to the original space, i.e.
\begin{align*}
    A_x &= \{x\in\mathbb{R}^n~|~\xi(x)\in A_y\},\\
    B_x &= \{x\in\mathbb{R}^n~|~\xi(x)\in B_y\}.
\end{align*}
An important quantity of interest is the transition rate between the sets $A$ and $B$, 
\begin{equation}
    \label{eq:nuAB}
    \nu_{AB}: = \lim_{T\rightarrow\infty}\frac{N_{AB}(T)}{T},
\end{equation}
where $T$ is the elapsed time and $N_{AB}(T)$ is the number of transitions from $A$ to $B$ observed during the time $T$.
The transition rate between $A$ and $B$ can be expressed via the committor function $q$ whose value at any point of space is the probability that the process starting at this point will first hit $A$ rather than $B$~\cite{EVE2006,EVE2010}. The committor is the solution of the following boundary value problem for the backward Kolmogorov equation:
\begin{equation}
    \label{eq:commBVP}
    \begin{cases}
        \mathcal{L}q = 0,&x\in\Omega\backslash (A\cup B)\\
        ~~q = 0,&x\in\partial A\\
        ~~q = 1,& x \in \partial B
    \end{cases},
\end{equation}
where $\mathcal{L}$ is the generator of the SDE governing the stochastic process.
The transition rates from $A_x$ to $B_x$ and from $A_y$ to $B_y$ are equal, respectively, to
\begin{equation}
    \label{eq:nuABx}
    \nu_{A_xB_x} = \beta^{-1}\int_{\Omega^x_{A_xB_x}}\|\nabla q(x)\|^2\mu_X(x) dx
\end{equation}
and 
\begin{equation}
    \label{eq:nuABy}
    \tilde{\nu}_{A_yB_y} = \beta^{-1}\int_{\Omega^y_{A_y B_y}}\nabla \tilde{q}(y)^\top M(y) \nabla \tilde{q}(y)\mu_Y(y) dy,
\end{equation}
where $q(x)$ and $\tilde{q}(y)$ are the committors in the original space and the CV space respectively,  $\Omega_{AB}^x:=\Omega_x\backslash (A_x\cup B_x)$, and $\Omega_{AB}^y:=\Omega_y\backslash (A_y\cup B_y)$. Zhang et al.~\cite{ZhangHartmannSchutte_2016} have proven that the transition rate in the CV space, $\tilde{\nu}_{A_yB_y}$, is always greater or equal to the rate between the corresponding sets in the original space, $\nu_{A_xB_x}$, and the relationship between these rates in the case of the governing SDE being the overdamped Langevin \eqref{eq:overdampedLangevin} is
\begin{equation}
    \label{eq:rateZHS}
    \tilde{\nu}_{A_yB_y} = \nu_{A_xB_x} +\beta^{-1}\int_{\Omega^x_{A_xB_x}}\left\|\nabla(q(x) - \tilde{q}(\xi(x))\right\|^2\mu_X(x)dx.
\end{equation}
Moreover, if the CV is the committor, i.e., if $\xi(x)\equiv q(x)$, then~\cite{ZhangHartmannSchutte_2016} $\tilde{\nu}_{A_yB_y} = \nu_{A_xB_x}$. This means that the committor is an optimal CV for the preservation of the transition rate in the reduced model. This fact motivates our choice of the committor of the reduced model as the reaction coordinate for the forward flux sampling. 

\subsection{Coarse-graining error measured by the relative entropy and the orthogonality condition}
Legoll and Lelievre~\cite{Legoll_2010} and Duong et al.~\cite{Sharma_2018} measured the error between $Y_t$, governed by \eqref{eq:dynamics_Lelievre}, and $\xi(X_t)$, where $X_t$ is governed by \eqref{eq:overdampedLangevin}, by the relative entropy of the probability densities $\phi(t,\cdot)$ of $Y_t$ and $\psi(t,\cdot)$ of $\xi(X_t)$: 
\begin{align} 
    \mathcal{H}(t) &:=H\left(\psi^{\xi}(t, \cdot)|\phi(t, \cdot)\right) = \int_{\Omega_y}\psi^{\xi}(t, y)\log\left(\frac{\psi^{\xi}(t, y)}{\phi(t, y)}\right)dy.\label{eq:relative_entropy1}
\end{align}
Under several technical assumptions, the difference $\mathcal{H}(t) - \mathcal{H}(0)$ is bounded from above by~\cite{Legoll_2010,Sharma_2018} 
\begin{equation}
    \mathcal{H}(t) - \mathcal{H}(0) \le C_1 + C_2\left[H\left(\psi(0,\cdot)|\mu_X\right) - H\left(\psi(t,\cdot)|\mu_X\right)\right],
\end{equation}
where $\psi(t,\cdot)$ is the probability density of $X_t$ and the expressions for $C_1$ and $C_2$ involve several technical constants that are hard to estimate in practice. The constant $C_1$ is zero if the Jacobian matrix of $\xi$, $D\xi$, is constant along its level sets. The constant $C_2$ is small if the following \emph{orthogonality condition} holds. Suppose that the potential function $V(x)$ in \eqref{eq:overdampedLangevin} is decomposed as
\begin{equation}
\label{eq:Vdecomp}
    V(x) = V_0(x) + \epsilon^{-1}V_1(x),
\end{equation}
where $V_1(x) \ge 0$, the zero level set of $V_1$ has co-dimension one, and $\epsilon$ is a small parameter. In this case, the invariant density of \eqref{eq:overdampedLangevin} is concentrated near the set where $V_1 = 0$, which we call the \emph{residence manifold}.
Then $C_2$ is $O(\epsilon)$ if~\cite{Legoll_2010,Sharma_2018}
\begin{equation}
    \label{eq:OC}
    D\xi(x)\nabla V_1(x) = 0,
\end{equation}
i.e., if \emph{the level sets of the CV are orthogonal to the residence manifold}.

Our method for learning CVs is based on the orthogonality condition \eqref{eq:OC}. We first learn the residence manifold in the feature space that respects the system's symmetries using \emph{diffusion maps} and \emph{diffusion nets}, and then learn CVs whose level sets are normal to the residence manifold using \emph{autoencoders}.

\subsection{Diffusion maps and diffusion nets}
\subsubsection{Diffusion maps}
The diffusion map algorithm~\cite{CoifmanLafon2006,Coifman_2008} is a widely used method for manifold learning. Let $\{x_i\}_{i = 1}^N \subset \mathbb{R}^n$ be a dataset consisting of independent samples drawn from a probability density $\rho(x)$ residing on a low-dimensional manifold $\mathcal{M}$. The algorithm begins by computing the pairwise similarities between data points using a kernel function. Commonly used is the Gaussian kernel
\begin{equation}
\label{eq:Gkernel}
    k_\epsilon(x_i, x_j) = \exp\left( -\frac{\|x_i - x_j\|^2}{2\epsilon} \right),
\end{equation}
where the \textit{kernel bandwidth} $\epsilon > 0$ is a hyperparameter. The $N\times N$ similarity matrix $K_{\epsilon}$ has entries $K_\epsilon(i,j) = k_\epsilon(x_i, x_j)$. A remarkable feature of diffusion maps is that the user does not need to know the sampling density. Instead, the sampling density at the data points is estimated as the vector of row-wise averages:
\begin{equation}
\label{eq:rho_eps}
    \rho^N_\epsilon(i) = \frac{1}{N} \sum_{j = 1}^N K_\epsilon(i, j).
\end{equation}
Indeed, as $\epsilon\rightarrow 0$ and $N\rightarrow\infty$, we have:
\begin{align*}
    \rho(x_i) &= \int_{\mathbb{R}^n}\rho(y)\delta(x_i-y)dy\\
    & = \lim_{\epsilon\rightarrow0}(2\pi\epsilon)^{-n/2}\int_{\mathbb{R}^n}\rho(y)k_{\epsilon}(x_i,y)dy \\
    & = \lim_{\epsilon\rightarrow 0}\frac{1}{(2\pi\epsilon)^{-n/2}}\lim_{N\rightarrow\infty}\frac{1}{N} \sum_{j = 1}^N K_\epsilon(i, j).
\end{align*}
Since the normalization factor of $(2\pi\epsilon)^{-n/2}$ would cancel out in the construction of the Markov matrix below, it is immediately omitted in \eqref{eq:rho_eps}.

To eliminate the influence of the sampling density $\rho$ on the Markov matrix, the kernel matrix is right-normalized as
\begin{equation}
\label{eq:K1}
    \tilde{K}_{\epsilon}: = K_\epsilon D_\epsilon^{-1},
\end{equation}
where $D_\epsilon: = {\sf diag}\{\rho^N_\epsilon(x_1), \ldots, \rho^N_\epsilon(x_N)\}$ is a diagonal matrix. Then the generator matrix constructed by the diffusion map approximates the Laplace-Beltrami operator. 
Alternatively, to construct an approximation to the generator of the overdamped Langevin dynamics with invariant density $\mu_X$ using enhanced sampling data, one needs to right-multiply $K_{\epsilon}$ by  $D_{\epsilon}^{-1}M^{1/2}$ where $M^{1/2} = {\sf diag}\{\mu_X(x_1), \ldots, \mu_X(x_N)\}$~~\cite{banisch2017_target_measure_diffusionMap}:
\begin{equation}
\label{eq:Kmu}
    \tilde{K}_{\epsilon}: = K_\epsilon D_{\epsilon}^{-1} M^{1/2}.
\end{equation}
One can use either of these options, \eqref{eq:K1} and \eqref{eq:Kmu}, in the proposed algorithm for learning CVs.

The transition matrix of a Markov chain defined on the dataset $\{x_i\}_{i = 1}^N$, the \emph{Markov Matrix},  is obtained by left-normalizing the kernel matrix to make its row sums equal to one:
\begin{equation}
\label{eq:Pmatrix}
    P_{\epsilon} = \tilde{D}_{\epsilon}^{-1}\tilde{K}_{\epsilon},
\end{equation}
where $\tilde{D}_{\epsilon}$ is the diagonal matrix with the row sums of $\tilde{K}_{\epsilon}$ along its diagonal. 

% The discrete generator on the dataset is constructed as:
% \begin{equation}
%     L_{\epsilon} = \frac{1}{\epsilon}\left(I - P_{\epsilon}\right).
% \end{equation}

Let $\lambda_0 >\lambda_1\ge \ldots$ be the eigenvalues of $P_{\epsilon}$ and $\psi_0$, $\psi_1$, $\ldots$, be the corresponding  right eigenvectors. The largest eigenvalue $\lambda_0$ is always equal to one, and the corresponding eigenvector $\psi_0$ has all entries equal. The other eigenvalues are strictly less than one. The eigenvectors $[\psi_1,\ldots,\psi_d]$, $d\le N-1$, are used to map the data into $\mathbb{R}^d$:
\begin{equation}
    \label{eq:dmap}
    z_i = \Psi(x_i): = [\psi_1(i),\ldots,\psi_d(i)],
\end{equation}
where $\psi_j(i)$ is the $i$th component of $\psi_j$, $j=1,\ldots,d$. We first try $d = 3$ hoping to obtain a two-dimensional manifold spanned by $\{z_i\}_{i=1}^N\subset\mathbb{R}^3$. This two-dimensional manifold is the residence manifold. The diffusion net algorithm~\cite{2015diffusionnets} is used to globally extend the diffusion map beyond the point cloud $\{x_i\}_{i=1}^N$.

\subsubsection{Diffusion net}
Although diffusion maps enable manifold learning for high-dimensional systems, they cannot directly accommodate out-of-sample data points. The diffusion net proposed by Mishne et al.~\cite{2015diffusionnets} is a machine learning-based method designed to extend diffusion maps to new data. The complete architecture consists of an encoder and a decoder. The encoder maps high-dimensional input data to a low-dimensional representation, while the decoder reconstructs the original data from this low-dimensional space.

Since our goal is to obtain analytically computable derivatives of the diffusion map on the dataset $\{x_i\}_{i=1}^N$, it suffices to train only the encoder neural network with a simpler loss function than the one proposed in Ref.~\cite{2015diffusionnets} Our loss is 
\begin{equation}
\label{eq:DNet_loss}
    \mathcal{L}_{\sf DNet}(\theta) = \frac{1}{2N} \sum_{i = 1}^N \|\Psi_{\sf DNet}(x_i;\theta) - \Psi(x_i)\|^2,
\end{equation}
where $\Psi_{\sf DNet}(x; \theta)$ is the encoder neural network and $\theta$ is the vector of its parameters.

\subsection{Forward flux sampling and stochastic control}
Our use of the committor in the reduced model is twofold:
\begin{itemize}
    \item as the reaction coordinate for the forward flux sampling~\cite{Allen_2005, Allen_2009}, and
    \item to design a stochastic control for sampling transition trajectories~\cite {YUAN_2024_optimalcontrol}.
\end{itemize}

\subsubsection{Forward Flux Sampling}
\label{sec:FFS}
Forward flux sampling (FFS)~\cite{Allen_2005, Allen_2009} is a splitting method for estimating escape rate $k_A$ and generating ensembles of transition paths from a region $A$ to a region $B$. The escape rate $k_A$ is different from the rate $\nu_{AB}$ defined by \eqref{eq:nuAB}. It is defined by
\begin{equation}
    \label{eq:kA}
    k_A = \lim_{T\rightarrow\infty}\frac{N_{AB}}{T_A},
\end{equation}
where $T_A$ is the total time within the interval $[0,T]$ during which a trajectory $X_t$, $0\le t\le T$, last hit $A$ rather than $B$.
Given a reaction coordinate $\lambda(x)$, $x\in\Omega$,
the phase space $\Omega$ is stratified using level sets of $\lambda$.
The region $A$ is defined as $\lambda < \lambda_A = \lambda_0$, and region $B$ as $\lambda > \lambda_B = \lambda_m$. The escape rate from $A$, $k_{A}$, is expressed as:
\begin{equation}
    \label{eqn:FFS_rate}
    k_{A} = \Phi_{A,0}\prod_{i=0}^{m-1} \mathbb{P}(\lambda_{i+1} \mid \lambda_i),
\end{equation}
where $\Phi_{A,0}$ is exit flux from $A$ and $P(\lambda_{i+1}\mid\lambda_i)$ is the probability that the trajectory starting at the interface $\lambda_i$ will next hit the interface $\lambda_{i+1}$ rather than $\lambda_0 = \partial A$. 

The implementation of FFS involves sampling both the initial flux $\Phi_{A,0}$ and the crossing probabilities between successive interfaces. The flux $\Phi_{A,0}$ is computed by initiating an unbiased simulation from region $A$ and monitoring how often trajectories cross the first interface $\lambda_0$. Each time a crossing of $\lambda_0$ from $A$ occurs, the system configuration is saved to be used to initiate trajectories for estimating the conditional probability $\mathbb{P}(\lambda_{1} \mid \lambda_0)$. Furthermore, if the trajectory enters the region $B$, i.e., crosses the interface $\lambda_m$, the unbiased simulation is interrupted and restarted at $A$.

The probabilities $\mathbb{P}(\lambda_{i+1} \mid \lambda_i)$ are estimated as the ratio of the number of trajectories initiated at $\lambda_i$ and hitting next $\lambda_{i+1}$ rather than $\lambda_0$ to the total number of trajectories initiated at $\lambda_i$. Whenever a trajectory initiated at $\lambda_i$ crosses $\lambda_{i+1}$, the configuration at $\lambda_{i+1}$ is saved to be used to initiate trajectories at $\lambda_{i+1}$.
We refer the reader to Ref.~\cite{Kratzer2013_FFS} for more details.

Although FFS is a widely used method for computing transition rates in rare-event systems via unbiased sampling, it has been shown~\cite{DicksonMakarov_2009} that its accuracy can suffer from a poor choice of reaction coordinate. This can lead to improper definitions of regions $A$ and $B$, and thus inaccurate rate estimates.

\subsubsection{Stochastic optimal control}
\label{sec:stochastic_control}
Let a process be governed by the overdamped Langevin dynamics~\eqref{eq:overdampedLangevin} and $A,B\subset\Omega$ be chosen disjoint metastable sets. Lu and Nolen~\cite{Lu2015} introduced the \emph{transition path process}, i.e., the process restricted to the trajectories of the original process that start at $\partial A$ and end at $\partial B$ without returning to $A$ in-between, is governed by
\begin{equation}
\label{eq:controlled}
    dX_t = \left[- \nabla V(X_t) + 2\beta^{-1} \nabla \log q(X_t)\right] dt + \sqrt{2\beta^{-1}} dW_t,
\end{equation}
where $q(x)$ is the committor.
Gao et al.~\cite{Gao_2023} established a connection of \eqref{eq:controlled} with stochastic optimal control, proving that the trajectories of \eqref{eq:controlled} minimize a certain cost functional. This means that $2\beta^{-1}\nabla\log q(x)$ is the optimal control, the minimal modification to the drift in the $L_2$ sense that ensures all trajectories go from $A$ to $B$ without returning to $A$.

Zhang et al.~\cite{Zhang_Marzouk_2022} showed that for a problem with a finite time horizon, even a rough approximation to the optimal control makes a very effective controller. So is true for sampling the transition path process~\cite{YUAN_2024_optimalcontrol}.

%%%%%%%%%%%%%%%%%%%%%%%%%%%%%%%
%%%%%%%%%%%%%%%%%%%%%%%%%%%%%%%

\section{Proposed methodology}
\label{sec:method}
In this section, we detail the proposed framework for identifying CVs and estimating transition rates for systems with permutational symmetry. 
% Our objective is to find CVs that minimize the coarse-graining error as much as possible.
% Concretely, we begin by transforming the data into a symmetric feature space through sorting. The residence manifold is then identified using the target measure diffusion map. The CVs can either be extracted from the eigenfunctions of $L_{\epsilon, \mu}$ or learned using a machine learning framework, such as an autoencoder, which enforces the orthogonality condition~\eqref{eq: criterion for CV}.

\subsection{Featurization}
Designing an effective feature map for atomic coordinates data is crucial for the success of learning good CVs. Look at Figure \ref{fig:LJ7dmap} obtained by applying the diffusion map algorithm to LJ7 in 2D data. The data were generated by running a long trajectory in the energy landscape flattened by a metadynamics potential~\cite{Dama2014WelltemperedMC} using the second and third central moments of the coordination numbers~\eqref{eq:CNum} as CVs. The atomic coordinates were aligned after every time step by solving the Procrustes problem. A 129$\times$129 grid was introduced in the $(\mu_2,\mu_3)$ space. At each time step, a grid cell visited by the long trajectory was identified, and up to four vectors of atomic coordinates were saved from each cell visited. Observe that some potential energy minima in Fig.~\ref{fig:LJ7dmap}, in particular, minimum $C_2$, have multiple images corresponding to different permutations of the particles. Moreover, the shape of the point cloud resulting from a diffusion map into the span of the three dominant eigenvectors significantly changes when more input data is added or when different input data generated by the same algorithm is used.  The reason for such inconsistency is the failure to account for the symmetries present in LJ7, in particular, the permutational symmetry. Hence, it is necessary to map the input data into a feature vector that is invariant under the system's symmetries first. 

What is a good feature map? First, we mapped the input data consisting of vectors of atomic coordinates, $[x_1,\ldots,x_7,y_1,\ldots,y_7]\in\mathbb{R}^{14}$ in the case of LJ7, into a vector of pairwise distances squared, $d^2\in\mathbb{R}^{21}$,
\begin{equation}
    \label{eq:d2def}
    d^2:=[r_{2,1}^2,r_{3,1}^2,\ldots,r_{7,1}^2,r_{3,2}^2,\ldots,r_{7,6}^2],
\end{equation}
for LJ7. This map accounts for translations, rotations, and reflections. Then we mapped the $d^2$-data using a basis of symmetric polynomials following Ref.~\cite{VilarSymmetry2022}. Unfortunately, this feature map, followed by the diffusion map, did not lead to an interpretable manifold in 3D. 

The second attempt resulted in success. We sorted each vector of pairwise distances squared in increasing order and obtained the ${\sf sort}[d^2]$-data. This map is invariant under the permutational symmetry. It is continuous but only piecewise smooth, which violates the regularity assumption (Ref.~\cite{Sharma_2018}, Assumption 2.3).  The resulting 2D CV yields a nice free energy landscape, and the committor of the reduced system serves as a reaction coordinate and stochastic control, producing accurate estimates for transition rates and residence times -- see Section ~\ref{sec:LJ7results}. We note here that such sorting-based featurizations are becoming popular in the wider machine learning literature due to their attractive stability properties \cite{balan2023g, morris2024position} and their computational efficiency \cite{dym2024low}. To our knowledge, however, such sorting-based embeddings have not been previously used as a featurization tool for interacting particle systems. 

However, the ${\sf sort}[d^2]$ feature map was not as successful for LJ8 in 3D. The free energy computed for the resulting 2D CV had a unique minimum and was therefore not suitable for studying transitions between the neighborhoods of different potential energy minima.

Then we designed a feature map by sorting the vector of coordination numbers~\eqref{eq:CNum}, ${\sf sort}[c]\in\mathbb{R}^{N_a}$ where $N_a$ is the number of atoms. This feature map has several advantages. First, the dimension of the feature vector is equal to the number of particles, which is essential for large systems. Second, the coordination numbers are focused on the nearest neighborhoods of atoms, which amplifies the distinction between different atomic packings. The ${\sf sort}[c]$ feature map was successful for both LJ7 in 2D and LJ8 in 3D. In both cases, it led to free energy landscapes with separated minima of interest and accurate estimates for transition rates and residence times -- see Sections \ref{sec:LJ7results} and \ref{sec:LJ8results}. 

To summarize, we prioritize the ${\sf sort}[c]$ feature map for interacting particle systems in 2D and 3D.

\subsection{Learning the residence manifold}
We use the diffusion map or the target measure diffusion map to learn a residence manifold $\mathcal{M}$ in the feature space, ${\sf sort}[d^2]$ or ${\sf sort}[c]$. The Markov matrix $P_{\epsilon}$ ~\eqref{eq:Pmatrix} is dense by construction, though it has some very small values due to the exponentially decaying kernel.
There are two common ways to sparsify it: setting the kernel to zero if the distance between its arguments is less than a certain threshold, e.g., $3\sqrt{\epsilon}$, or using the ``$k$-nearest-neighbors (kNN)" method. We chose the second way as it is simpler and its performance quality is good enough for our purposes. The kNN method keeps only $k$  largest values in each row of the similarity matrix $K_{\epsilon}$ with entries~\eqref{eq:Gkernel} and zeroes out the rest, and then symmetrizes $K_{\epsilon}$ by replacing it with the half-sum of $K_{\epsilon}$ and $K_{\epsilon}^\top$. 

We aim to find 2D CVs. Hence, we aim to obtain a 2D manifold embedded in a 3D space. Therefore, we compute the top four eigenvalues $1=\lambda_0>\lambda_1\ge\lambda_2\ge\lambda_3$ of the Markov matrix $P_{\epsilon}$ ~\eqref{eq:Pmatrix} and the corresponding eigenvectors, discard the eigenvector $\phi_0$ as it has equal entries, and retain the eigenvectors 
\begin{equation}
\label{eq:Psi123}
    \Psi: = [\psi_1,\psi_2,\psi_3].
\end{equation}

Let $\Phi$ be the feature space. The images of the data points $\{x_i\}_{i=1}^N$  under the feature map will be denoted by $\{\phi_i\}_{i=1}^N$. The images of these points in the feature space under the diffusion map are
\begin{equation}
\label{eq:Psii}
    \Psi(\phi_i): = [\psi_1(i),\psi_2(i),\psi_3(i)],\quad 1\le i\le N.
\end{equation}

Our next goal is to learn a non-negative function $V_1:\mathbb{R}^3\rightarrow\mathbb{R}$ in \eqref{eq:Vdecomp}, whose zero-level set defines the residence manifold $\mathcal{M}$. Since $V_1$ can be an arbitrary smooth function whose zero-level set is $\mathcal{M}$, we design it to be a smoothed indicator function of $\mathcal{M}$. Thus, we represent $V_1$ as a neural network $V_1(\cdot;\theta)$, where $\theta$ is the vector of its parameters, with the outer nonlinear layer being the {\tt sigmoid} function which makes the range of $V_1(\cdot;\theta)$ be $(0,1)$.

Since $\Psi(\phi_i)\in\mathcal{M}$, we have $V_1(\Psi(\phi_i)) = 0$ for $1\le i\le N$. Hence, to learn a nontrivial $V_1$, we generate training points beyond $\mathcal{M}$ by sampling points within a bounding box of $\mathcal{M}$ from the uniform distribution and retaining only those points whose distance to $\mathcal{M}$ exceeds a threshold $r$. Let $\{p_j\}_{j=1}^m\in\mathbb{R}^3$ be the resulting set of points not lying on $\mathcal{M}$. 

To train the function $\hat{M}$, we minimize the loss:
\begin{equation}
\label{eq:V1loss}
    \mathcal{L}_{\sf{manifold}}(\theta) = \frac{a_1}{N} \sum_{i=1}^N V_1(\Psi(\phi_i);\theta) + \frac{a_2}{m} \sum_{j=1}^m \frac{1}{V_1(p_j;\theta)}.
\end{equation}
The scalars $a_1$ and $a_2$ balance the two terms in the loss function. Since we prioritize enforcing the condition $V_1= 0$ on $\mathcal{M}$, we use $a_1 = 1$ and $a_2 = 0.002$.

The function $\Psi(\cdot)$ in \eqref{eq:Psii} is undefined beyond the point cloud $\{\phi_i\}_{i=1}^N$. On the other hand, to learn CVs with desired properties, we need to impose the orthogonality condition \eqref{eq:OC}. Hence $V_1(\Psi(\phi))$ must be differentiable. Therefore, we train a diffusion net $\Psi_{\sf DNet}$ with the loss \eqref{eq:DNet_loss} to extend the diffusion map $\Psi$ beyond the point cloud $\{\phi_i\}_{i=1}^N$ and hence enable the differentiation of $V_1$ with respect to $\phi$.

% {\color{red} MARGOT: Is this description correct?}

\subsection{Learning collective variables}
We use the autoencoder (AE) framework to learn CVs. The CV $\xi_{\theta}(\phi)$
is parametrized as the encoder, and $\theta$ represents the vector of trainable parameters. 

We propose the following modifications to the original AE framework to preserve the dynamics in the feature space $\Phi$ invariant under rotations, translations, reflections, and permutations:

\begin{enumerate}
    \item Instead of using the atomic coordinates dataset $\{x_i\}_{i=1}^N$, we input the data in the feature space, $\{\phi_i\}_{i=1}^N$, invariant under symmetries of the system, in particular, the permutational symmetry. The feature space $\Phi$ is the image of the feature map $\phi(x) = {\sf sort}[d^2](x)$ or ${\sf sort}[c](x)$.
    
    \item Since the dimensionality of the feature space $\Phi$ is significantly higher than the CV dimension, $d=2$, we adjust the reconstruction loss to recover  $\Psi(\phi_i)\in\mathbb{R}^3$ rather than $\phi_i$ after decoding, i.e., the decoder with the vector of trainable parameters $\Theta$ performs the map ${\sf{Decoder}}_{\Theta}:\mathbb{R}^2\rightarrow\mathbb{R}^3$.  Hence, the reconstruction $\hat{R}$ is defined as 
    \begin{equation}
        \label{eq:reconstruction}
        \hat{R} = {\sf{Decoder}}_\Theta \circ \xi_{\theta}~:~\mathbb{R}^{{\sf dim}(\Phi)}\rightarrow\mathbb{R}^2\rightarrow\mathbb{R}^3.
    \end{equation} 
    
    % \item For differentiability and out-of-sample generalization, we train a diffusion net $\Psi_{\sf DNet}$ with the loss \eqref{eq:DNet_loss} that maps the feature space $\Phi$ to the residence manifold in $\mathcal{M}\subset\mathbb{R}^{3}$.
    
    \item In addition to the reconstruction loss from the standard AE framework, we enforce the orthogonality condition~\eqref{eq:OC}.
    
    \item Finally, we impose an independence regularization on the CVs to prevent them from becoming linearly dependent.
\end{enumerate}

The resulting loss function is:
\begin{align}
    \mathcal{L}_{\sf AE}(\theta,\Theta) &= \underbrace{\|{\sf Decoder}_{\Theta}\left[\xi(\Psi_{\sf DNet}(\phi))\right] - \Psi_{\sf DNet}\|^2}_{\text{\sf \tiny reconstruction loss}} \notag \\
    &+ \underbrace{\|D_{\phi}\xi_{\theta}(\phi) \nabla_{\phi} V_1(\Psi_{\sf DNet}(\phi))\|^2}_{\text{\sf \tiny orthogonality constraint}} \label{eq:CV_loss} \\
    &+ \underbrace{\|D_{\phi} \xi_{\theta}(\phi) \, D_{\phi}\xi_{\theta}(\phi)^\top - I\|^2}_{\text{\sf \tiny linear independence}}, \notag
\end{align}
where the gradient $\nabla_{\phi}$ and the Jacobian operator $D_{\phi}$ are taken in the feature space $\Phi$.

% {\color{red} MARGOT: is this correct?}
 
\subsection{Estimating transition rates and sampling transition trajectories}
\label{sec:rates&control}
Once CV $\xi(\phi)$, $\xi:\Phi\rightarrow\Omega_y\subset\mathbb{R}^2$, is learned, we compute the free energy $F(y)$ and the diffusion matrix $M(y)$, $y\in\Omega_y$ using standard methods described in Appendix \ref{app:FE&DM}. Then we define metastable sets $A$ and $B$ in the CV space $\Omega_y$ and solve the committor problem \eqref{eq:commBVP}: 
\begin{align}
\label{eq:LgenCV}
   \begin{cases} L\tilde{q}(y) = \nabla\cdot\left(e^{-\beta F(y)}M(y)\nabla \tilde{q}(y)\right) = 0,\\
     \tilde{q}(\partial A) = 0, \quad  \tilde{q}(\partial B) = 1.\end{cases}
\end{align}
The committor problem \eqref{eq:LgenCV} can be solved in a variety of ways. We find it easiest to solve it using the finite element method (FEM) as detailed in Appendix F1 of Ref.~\cite{YUAN_2024_optimalcontrol} and then approximate the FEM committor by a neural network to have a globally defined and easy-to-evaluate function.

The found committor  $\tilde{q}$ is lifted into the space of atomic coordinates as
\begin{equation}
    q(x):=\tilde{q}(\xi(\phi(x))),
\end{equation}
where $\phi(x) = {\sf sort}[d^2](x)$ or ${\sf sort}[c](x)$, $x\in\mathbb{R}^n$ is the vector of atomic coordinates, $[d^2]$ is the vector of interatomic distances squared, and $[c]$ is the vector of coordination numbers~\eqref{eq:CNum}.
Then $q(x)$ and $1-q(x)$ are used as the reaction coordinates in forward flux sampling (FFS) described in Section~\ref{sec:FFS}
to find the escape rates $k_A$ and $k_B$ from $A$ and $B$ respectively. The escape rates $k_A$ and $k_B$ relate to the rate $\nu_{AB}$ via~\cite{EVE2010}
\begin{equation}
    \label{eq:rates_relationship}
    k_A = \frac{\nu_{AB}}{\rho_A},\quad k_A = \frac{\nu_{AB}}{\rho_B},
\end{equation}
where $\rho_A$ and $\rho_B$ are the probabilities that, at a randomly picked time $t$, an infinite trajectory last hit $A$ rather than $B$ and the other way around, respectively.
Since 
\begin{equation}
    \label{eq:rhosum}
    \rho_A+\rho_B = 1,
\end{equation}
$k_A$ and $k_B$ allow us to find $\rho_A$, $\rho_B$, and $\nu_{AB}$ as follows:
\begin{align}
    \nu_{AB} &= \frac{1}{\frac{1}{k_A}+\frac{1}{k_B}},\label{eq:nABviakAkB}\\
    \rho_A &= \frac{1}{1 + k_A/k_B},\quad 
    \rho_B = \frac{1}{1 + k_B/k_A}.  \label{eq:rhoArhoB}
\end{align}

While the transition trajectory can be found by FFS by splicing the pieces of trajectories between consecutive level sets of the reaction coordinate, it is more convenient to generate them using stochastic control. We use 
\begin{equation}
    \label{eq:stoch_control}
    2\beta^{-1}\nabla\log q(x) = 2\beta^{-1}\nabla\log \tilde{q}(\xi(\phi(x)))
\end{equation} 
as the approximate optimal control--see Section \ref{sec:stochastic_control}.  The gradient in \eqref{eq:stoch_control} is taken with respect to atomic coordinates $x$. Since this is not exactly the optimal control, we discard the trajectories that return to the set $A$. The generated trajectories from $A$ to $B$ are used for estimating the probability density of reaction trajectories projected into the CV space by binning them in the CV space.
% The expected crossover time allows us to estimate the probability $\rho_{AB}$ that an infinite trajectory is reactive at a random time $t$~\cite{Lu2015}:
% \begin{equation}
%     \rho_{AB} = \nu_{AB}\mathbb{E}[\tau_{AB}].
% \end{equation}
% {\color{red} I don't think we actually included this quantity because of the inconsistency with the brute force rate. Should we include here?}

\subsection{Summary}
The proposed framework for learning CVs and estimating transition rates is summarized in Algorithm~\ref{alg:learnCVs}.

\begin{algorithm}[t]
        \KwIn{$\{x_i\}_{i=1}^N$, where $N$ is the number of data points,  $x_i\in\mathbb{R}^n$ are the vectors of atomic coordinates, $n = 2N_a$ or $n = 3N_a$, where $N_a$ is the number of atoms and 2 or 3 is the number of degrees of freed of each atom\;   }
        \KwOut{$\bullet$ CVs $\xi(\phi(x))\in\mathbb{R}^2$, where $\phi(x)$ is the feature map, $\phi(x) = {\sf sort}[d^2](x)$ or ${\sf sort}[c](x)$, $[d^2]\in\mathbb{R}^{N_a(N_a-1)/2}$ is the vector of pairwise distances squared, and $[c]\in\mathbb{R}^{N_a}$ is the vector of coordination numbers~\eqref{eq:CNum}; \\
    \phantom{\textbf{Output: }}$\bullet$ Rates $k_A$, $k_B$~(Eq.~\eqref{eq:kA}), $\nu_{AB}$~(Eq.~\eqref{eq:nuAB})\; \\
    % \phantom{\textbf{Output: }}$\bullet$ The expected crossover time $\mathbb{E}[\tau_{AB}]$; \\
    \phantom{\textbf{Output: }}$\bullet$  The probabilities $\rho_A$ and $\rho_B$\;
        }
        
       Map the dataset $\{x_i\}_{i=1}^N$ to a feature space: $\phi_i = \phi(x_i)$, $\phi(x) = {\sf sort}[d^2](x)$ or ${\sf sort}[c](x)$\;

       Use the diffusion map algorithm on the input dataset in the feature space $\{\phi_i\}_{i=1}^N$ and obtain a point cloud $\{\Psi(\phi_i)\}_{i=1}^{N}\subset\mathbb{R}^3$\;

       Learn the function $V_1:\mathbb{R}^3\rightarrow\mathbb{R}$ whose zero-level set is the residence manifold with the loss \eqref{eq:V1loss}.

        Train the diffusion net with the loss~\eqref{eq:DNet_loss} 
        to obtain a differentiable map $\Psi_{\sf DNet}(\phi)$ that approximates $\{\Psi(\phi_i)\}_{i=1}^N$\;

        Train an autoencoder network with the loss \eqref{eq:CV_loss} to learn CVs $\xi(\phi(x))$\;

        Compute the free energy $F(y)$ and the diffusion matrix $M(y)$ in the CV space\;
        
        Define the sets $A$ and $B$ in the CV space and solve the committor problem \eqref{eq:LgenCV} to find the committor $\tilde{q}(y)$\;

     Lift the committor to the space of atomic coordinates, $q(x) = \tilde{q}(\xi(\phi(x)))$, and use it as the reaction coordinate in the forward flux sampling algorithm to find $k_A$, $k_B$, $\nu_{AB}$, $\rho_A$, $\rho_B$\;
        
        Define the stochastic control according to \eqref{eq:stoch_control}. Generate transition trajectories and estimate the probability density of transition trajectories in the CV space.
        
		\caption{Learning CVs and estimating rates}
		\label{alg:learnCVs}
    \end{algorithm}

Algorithm~\ref{alg:learnCVs} requires training for a total of four neural networks: the diffusion net $\Psi$, the confining potential $V_1$, the autoencoder to learn the CV $\xi$, and the committor network $q$. All neural networks have a feedforward architecture. We use the following activation functions:
\begin{enumerate}
    \item the exponential linear unit (ELU), defined as
\begin{equation}
    \label{ELUdef0}
    \sf{ELU}(x) = 
        \begin{cases}
            x & \text{if } x > 0\\
%            \alpha (\exp(x) - 1) & \text{if } x \leq 0.
            \exp(x) - 1 & \text{if } x \leq 0       \end{cases},
\end{equation}
is used for the diffusion net and the autoencoder;
\item the hyperbolic tangent, {\sf tanh}, is chosen for the confining potential $V_1$;
\item {\sf tanh} or the rectified linear unit, {\sf ReLU}, is used for approximating the committor.
\end{enumerate}
The number of hidden layers ranges from 2 to 3, and the number of neurons per layer varies from 10 to 45, depending on the complexity of the function being approximated. The Adam optimizer~\cite{Adam_2017} is used for all training. The number of epochs is between 1000 and 2000, and the learning rate is from $10^{-3}$ to $5\cdot10^{-3}$. The neural architectures and training details used for each case are reported in Table S1 in the Supplementary Information (SI).

\section{Test case 1: Lennard-Jones-7 in 2D}
\label{sec:LJ7results}
The first test case for Algorithm \ref{alg:learnCVs} is LJ7 in 2D. The potential energy is defined by \eqref{eq:LJpot}. Its minima and the lowest saddles separating them are depicted in Fig. \ref{fig:LJ7digraph} as a disconnectivity graph. The deepest minimum is the hexagon denoted by $C_0$, and the shallowest and the most disconnected from the hexagon in the configurational space is the trapezoid, $C_3$. Additionally, there are two intermediate minima, capped parallelograms $C_1$ and $C_2$.  A popular test problem is the estimation of the transition rate between the hexagon and the trapezoid~\cite{Dellago1998LJ,WALES2002DPS,YUAN_2024_optimalcontrol}. We use Algorithm \ref{alg:learnCVs} with the feature maps ${\sf sort}[d^2]$ and ${\sf sort}[c]$ to learn CVs. In addition, we use the standard set of CVs, $(\mu_2,\mu_3)$, which are the second and third central moments of the coordination numbers \eqref{eq:CNum}, for comparison.

In the remainder of this section, we provide an overview of our workflow for this test problem. Additional implementation details can be found in the SI.

\subsection{Generating the input data}
\label{sec:LJ7data}
We consider the dynamics of LJ7 governed by the overdamped Langevin SDE \eqref{eq:overdampedLangevin}. To prevent the atoms from evaporating, we added a restraining spring force with spring constant $\kappa_r = 100$ that kicks in if an atom is at a distance greater than 2 from the center of mass of the cluster. The time integrator used is the Metropolis-Adjusted Langevin Algorithm (MALA)~\cite{Roberts1996ExponentialCO} with a time step of $5\cdot10^{-5}$ reduced units. Data for learning CVs were obtained at $\beta = 5$ using the well-tempered metadynamics algorithm~\cite{Dama2014WelltemperedMC} with $(\mu_2,\mu_3)$, the second and third central moments of the coordination numbers~\eqref{eq:CNum}, as biasing CVs. Then a long trajectory was integrated in the resulting biased potential and one configuration in each cell of the $129\times129$ mesh in the $(\mu_2,\mu_3)$ space visited by the long trajectory was saved. The resulting dataset consists of 16,641 data points. Further details are provided in Appendix~\ref{app:makedata}.

\subsection{Machine learned CVs in sorted pairwise distances squared}
The rotationally, translationally, and permutationally invariant feature map ${\sf sort}[d^2]$ transforms the vector of atomic coordinates of LJ7, $x = [x_1,\ldots,x_7,y_1,\ldots,y_7]\in\mathbb{R}^{14}$, to a vector $\phi(x)\in\mathbb{R}^{21}$ of pairwise distances squared. The residence manifold is then identified by applying the target measure diffusion map to the dataset $\{\phi_i\}_{i=1}^n$. The similarity matrix $K_{\epsilon}$ is sparsified using the $k$-nearest neighbors approximation with $k = 256$. The kernel bandwidth was set to $\epsilon = 4.0$.

The values of $k$ and $\epsilon$ are selected to ensure that the resulting manifold is smooth and lies in a three-dimensional space. Given the dataset size, $k$ is chosen to be less than one-fourth of the total number of data points. The kernel bandwidth $\epsilon$ should be proportional to the input dimension, which is 21 in this case. If $\epsilon$ is too small, the resulting manifold becomes fragmented or discontinuous.  
The diffusion map yields a point cloud $\{\Psi(\phi_i)\}_{i=1}^n\subset \mathbb{R}^3$ that spans an approximately two-dimensional residence manifold.

\begin{figure}[htbp]
    \centering
\includegraphics[width = 0.6\textwidth]{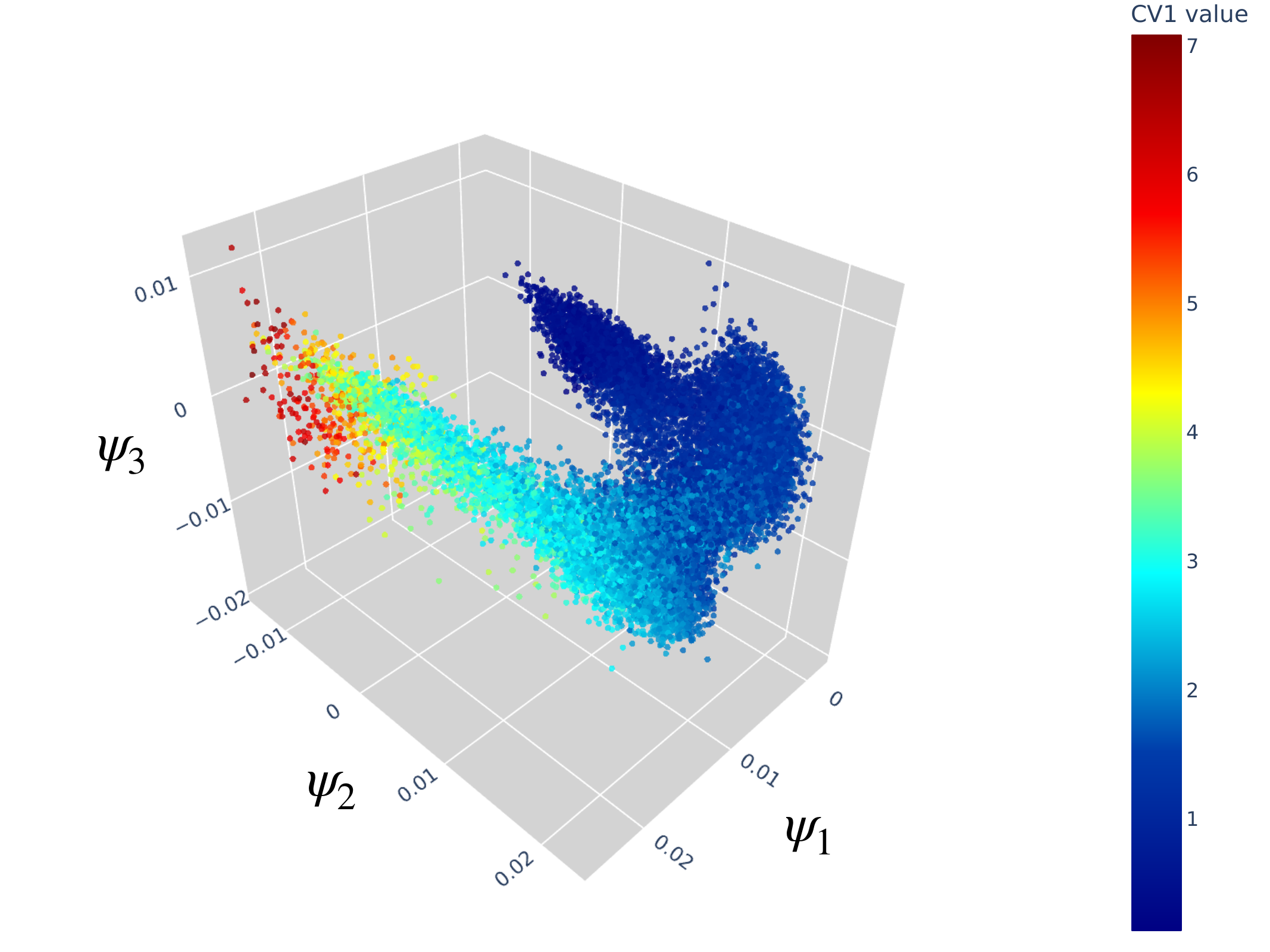}% Here is how to import EPS art
\caption{ The learned residence manifold in the space spanned by the three dominant eigenvectors of the target measure diffusion map for LJ7 in 2D with the ${\sf sort}[d^2]$ feature map. The manifold is colored according to the first CV learned by Algorithm~\ref{alg:learnCVs}.}
\label{fig:LJ7_manifold_sortd2}
\end{figure}

To ensure differentiability of the residence manifold, we approximate the mapping $\Psi:\mathbb{R}^{21}\rightarrow\mathbb{R}^3$ with a diffusion net. We use a two-layer feed-forward neural network with 45 and 25 neurons in the hidden layers, respectively. The resulting learned manifold embedded in $\mathbb{R}^3$ is shown in Fig.~\ref{fig:LJ7_manifold_sortd2}.

Next, we learn the confining potential $V_1$ whose zero-level set is the manifold shown in Fig. \ref{fig:LJ7_manifold_sortd2}. A point cloud surrounding the manifold is generated by randomly sampling points within the bounding box of the learned manifold, which is taken to be $[-0.02,0.02]^3$, and retaining only those points that are at least a distance of 0.005 away from the learned manifold. The function $V_1$ is then parameterized by a neural network with two hidden layers containing 45 and 30 neurons, respectively. 

Subsequently, the CVs are identified using an autoencoder trained to minimize the loss function~\eqref{eq:CV_loss}. The autoencoder consists of an encoder and decoder, each with two hidden layers containing 30 neurons.

\begin{figure}[htbp]
    \centering
    \includegraphics[width=0.6\textwidth]{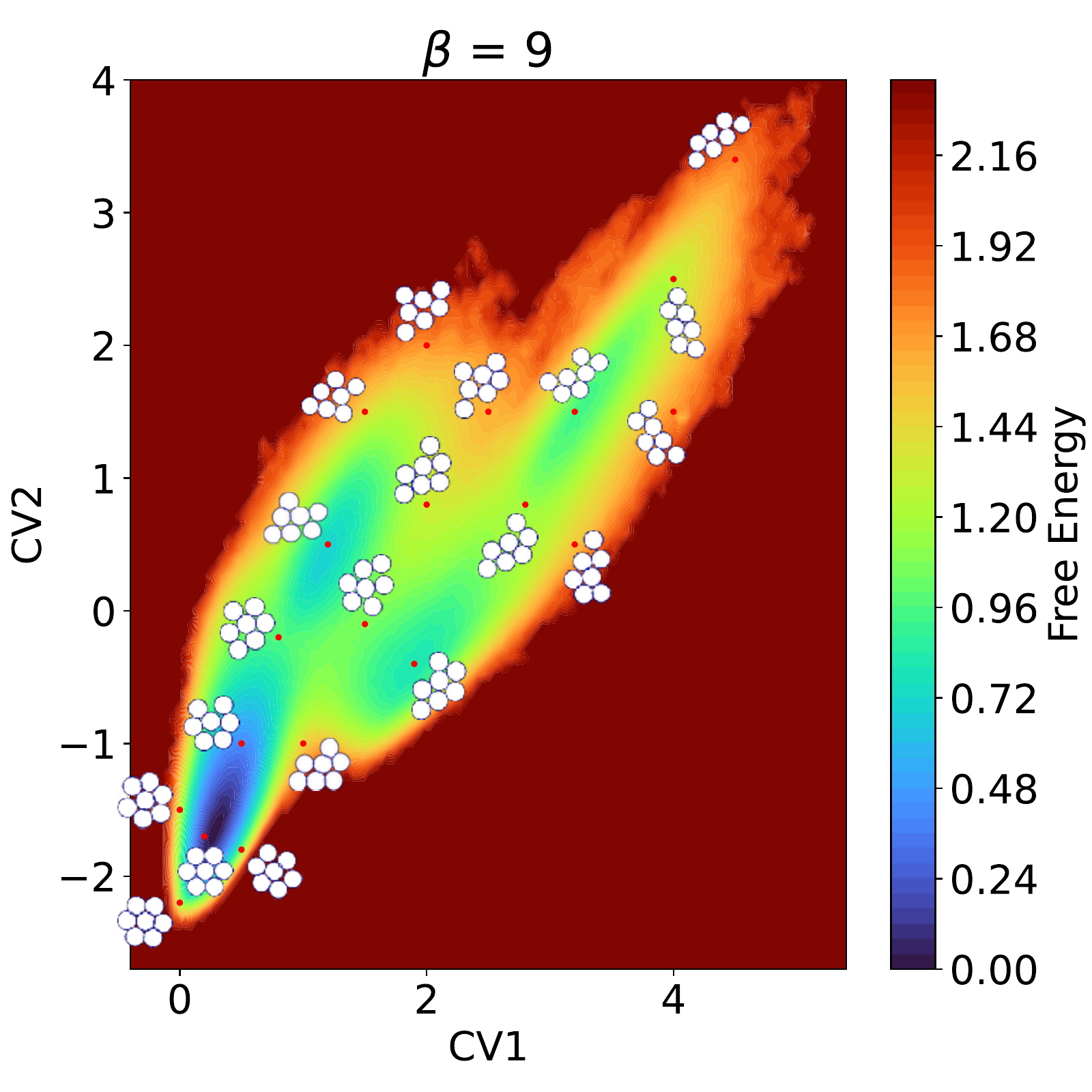}
    \caption{The free energy with respect to the learned CVs with the ${\sf sort}[d^2]$ feature map at $\beta = 9$ and a collection of representative atomic configurations.}
    \label{fig:LJ7_FE_confs_sortd2}
\end{figure}

\begin{figure}[htbp]
    \centering
    \includegraphics[width=0.6\textwidth]{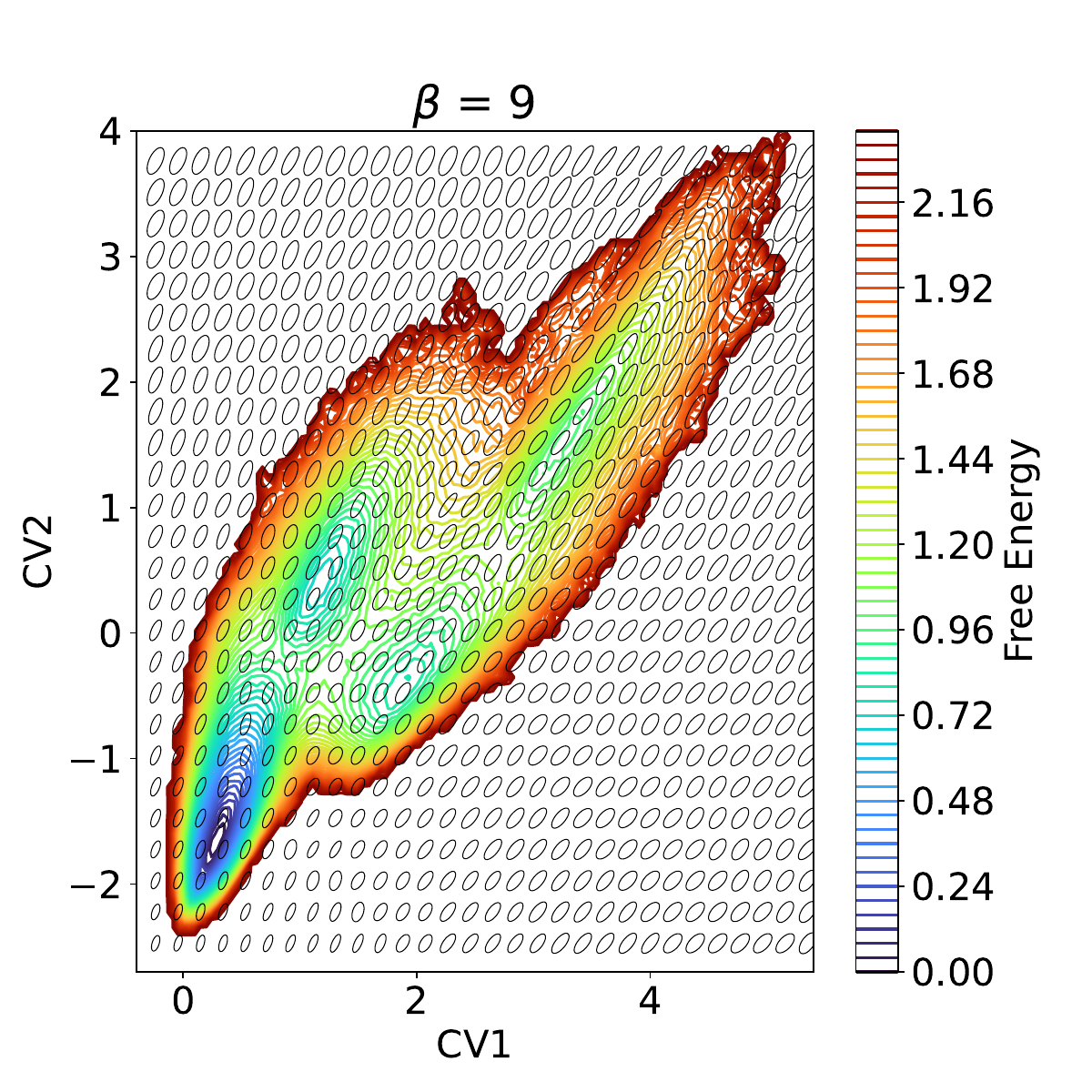}
    \caption{The diffusion matrix  with respect to the learned CVs with the ${\sf sort}[d^2]$ feature map at $\beta = 9$.}
    \label{fig:LJ7_DM_sortd2}
\end{figure}

An additional piece of information we have leveraged for this test case is the minimum energy path (MEP) from $C_0$ to $C_3$. The MEP represents the most probable transition path between the two given local minima of the potential energy surface and can be obtained using the string method~\cite{String_weinan}. The arc length along the MEP provides a natural candidate for a collective variable.

Since the arc length is defined only along the MEP, we incorporate this information as a boundary condition. Specifically, for data points on the MEP, we impose a constraint such that the first component of the CV output equals the arc length from the starting point of the MEP. We modify the loss~\eqref{eq:CV_loss} as:
\begin{equation}
\label{eq:lossMEP}
    \mathcal{L}_{\sf{MEP}} = \mathcal{L}_{\sf AE} + \sum_{y_i \in \text{MEP}} \left\| \xi({\tt sort}[d^2](y_i)) - S(y_i) \right\|^2,
\end{equation}
where $\{y_0, y_1, \ldots\}$ are configurations along the MEP indexed by time, and $S(y_i)$ is the arc length from $y_0$ to $y_i$.

In this system, the relatively small number of local minima allows for straightforward computation of the MEP. However, in systems with a larger number of local minima, finding the right MEP becomes significantly more challenging. Therefore, we do not include this strategy as part of the standard protocol; instead, we treat it as supplementary information when available.

We calculated the free energy and the diffusion matrix with respect to the learned CV with the feature map ${\sf sort}[d^2]$ at $\beta = 5$, $7$, and $9$. Fig.~\ref{fig:LJ7_FE_confs_sortd2} shows the free energy with respect to the learned CV at $\beta = 9$ and a collection of representative atomic configurations. The diffusion matrix at $\beta = 9$ is visualized in Fig.~\ref{fig:LJ7_DM_sortd2}. The free energies at $\beta = 5$ and $7$ are found in Fig. S1 in the SI. The visualizations of the diffusion matrix at all three values of $\beta$ look similar.

\subsection{Machine learned CVs in sorted coordination numbers}

\begin{figure}[htbp]
    \centering
    \includegraphics[width=0.6\textwidth]{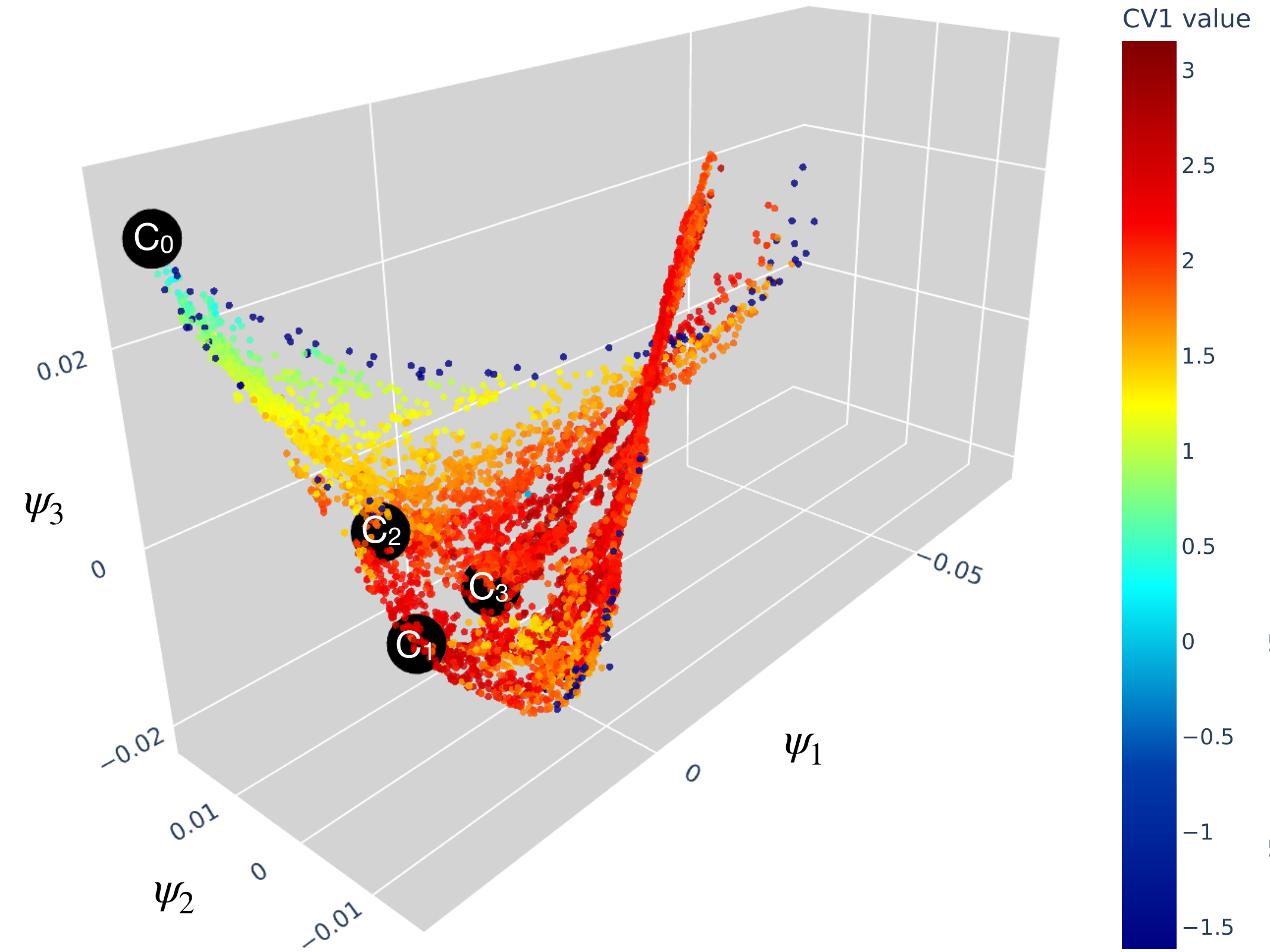}
    \caption{The learned residence manifold in the space spanned by the three dominant eigenvectors of the measure diffusion map for LJ7 in 2D with the ${\sf sort}[c]$ feature map. The manifold is colored according to the first CV learned by Algorithm~\ref{alg:learnCVs}.}
    \label{fig:LJ7_manifold_sortCNum}
\end{figure}

The second feature map considered for LJ7 in 2D is the sorted vector of coordination numbers~\eqref{eq:CNum}, ${\sf sort}[c]$. The dataset of $n = 16,641$ vectors of atomic coordinates $\{x_i\}_{i=1}^n\subset\mathbb{R}^{14}$ generated as described in Section \ref{sec:LJ7data} is mapped to the feature space as $\phi_i = {\sf sort}[c_{1,i},c_{2,i},\ldots,c_{7,i}]$ resulting in $\{\phi_i\}_{i=1}^n\subset\mathbb{R}^{7}$.  The diffusion map with the right-normalized kernel \eqref{eq:K1} rather than \eqref{eq:Kmu}, the sparsification via $k = 4,160$ nearest neighbors and the bandwidth $\epsilon = 1.0$ is used to project the dataset to the span of the three dominant eigenvectors of the diffusion map.
%Using the dataset generated via the \textit{binning method}, we first transform the $x$- and $y$-coordinates into coordination numbers $c \in \mathbb{R}^7$. The residence manifold in 3D is then identified using the eigenvectors of the diffusion map. We set $k = 4160$—one-fourth of the number of data points—to approximate sparsity in the $k$-nearest neighbors graph, and choose a kernel width of $\epsilon = 1.0$.
The resulting 3D manifold is shown in Fig.~\ref{fig:LJ7_manifold_sortCNum}. It is reminiscent of a swimming stingray. The diffusion map is approximated with a diffusion net with three hidden layers with  45, 30, and 25 neurons. Then the confining potential $V_1$ is learned as a neural network with two hidden layers with 30 neurons in each layer.

\begin{figure}[htbp]
    \centering
    \includegraphics[width=0.6\textwidth]{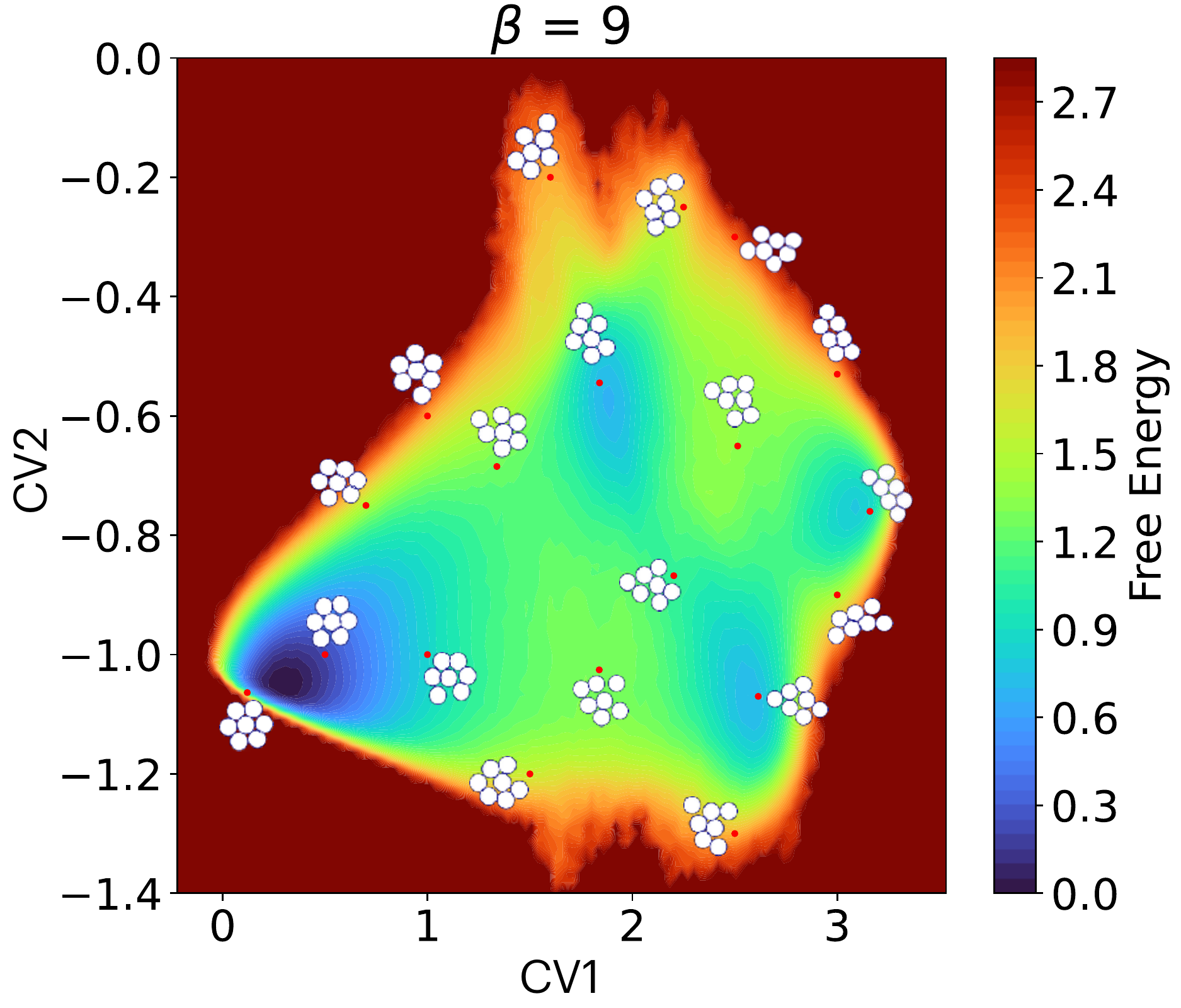}
    \caption{The free energy for LJ7 in 2D with respect to the ML CV ${\tt sort}[c]$ as input. }
    \label{fig:LJ7_FE_confs_sortCNum}
\end{figure}

\begin{figure}[htbp]
    \centering
    \includegraphics[width=0.6\textwidth]{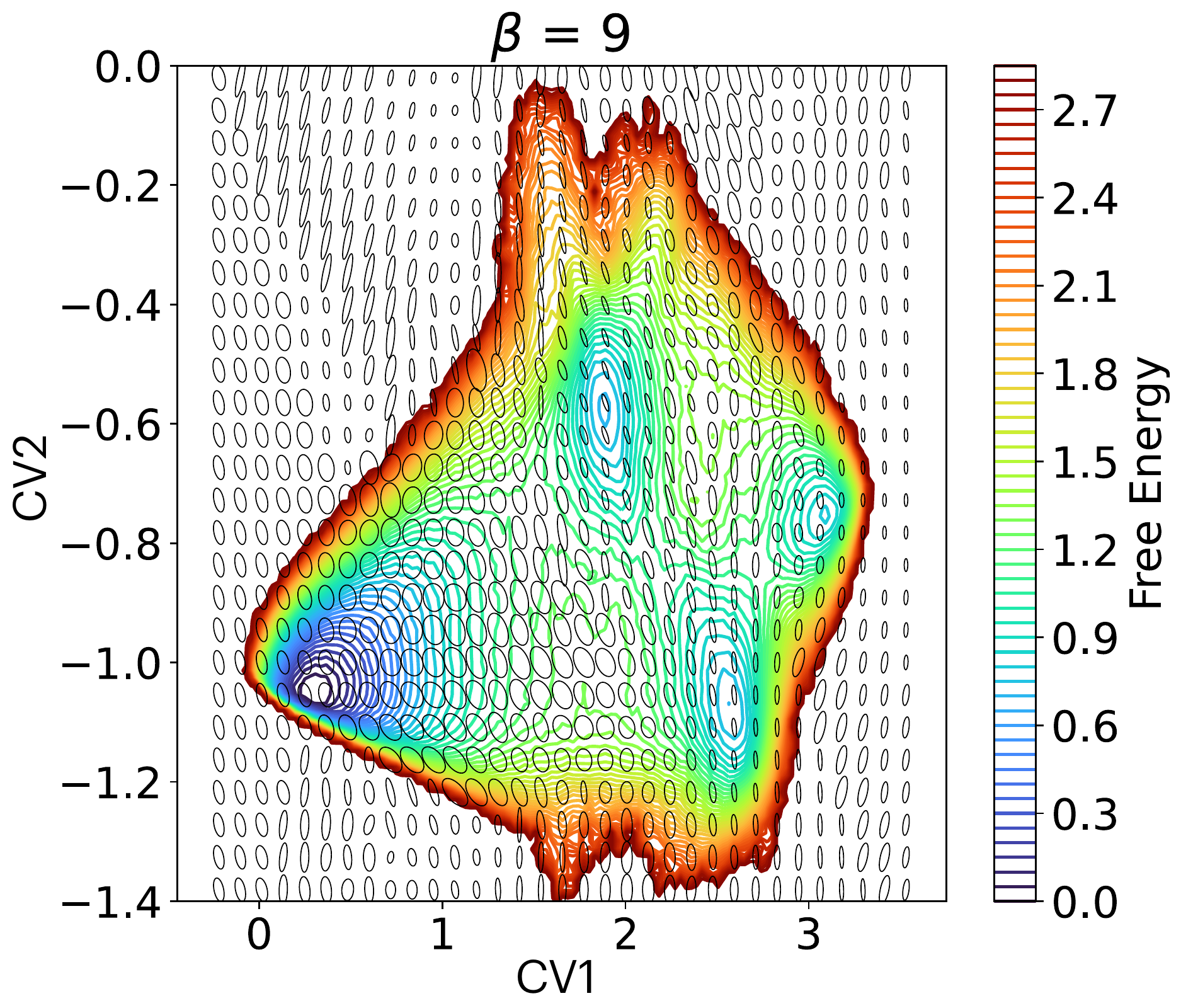}
    \caption{The diffusion matrix  with respect to the learned CVs with the ${\sf sort}[c]$ feature map at $\beta = 9$. }
    \label{fig:LJ7_DM_sortCNum}
\end{figure}

The autoencoder used for learning CVs consists of an encoder and a decoder, each with two hidden layers containing 30 neurons. The ELU function is used as the activation function. The loss function is given by~\eqref{eq:CV_loss}. In contrast with the case with the ${\sf sort}[d^2]$ feature map, a good result was achieved with the loss~\eqref{eq:CV_loss} without the use of the MEP data. 
Figs.~\ref{fig:LJ7_FE_confs_sortCNum} and \ref{fig:LJ7_DM_sortCNum} show the free energy landscape at $\beta = 9$ in the learned CV space with a collection of representative atomic configurations and the diffusion matrix at $\beta = 9$. The free energies at $\beta = 5$ and $7$ are shown in Fig. S1 in the SI.

\subsection{CV space committor as reaction coordinate for forward flux sampling}
\begin{figure}[htbp]
    \centering
\includegraphics[width = 0.9\textwidth]{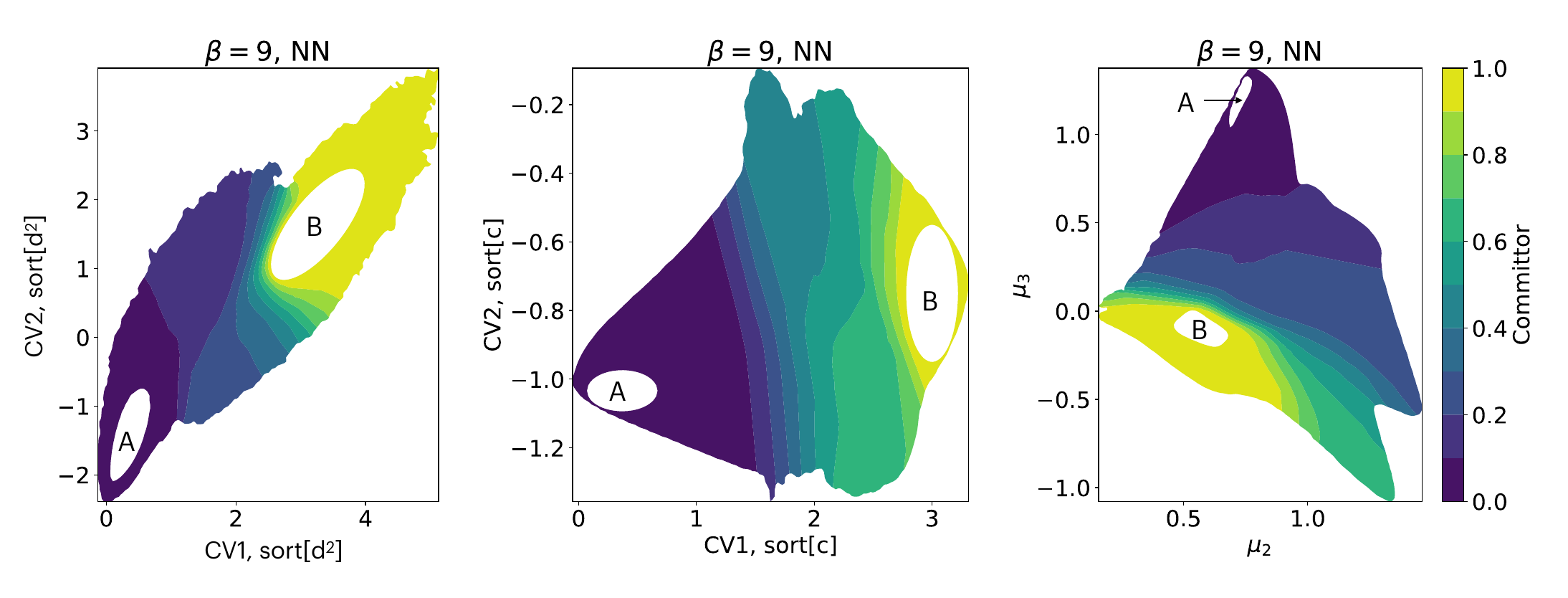}% Here is how to import EPS art
\caption{\label{fig:LJ7comm} The chosen sets $A$ and $B$ surround the hexagon and the trapezoid minima respectively. The committor in the learned CV space is computed using the finite element method at $\beta = 9$ and then approximated by a neural network (NN). The feature maps are: the sorted vector of pairwise distances squared, $\sf{sort}[d^2]$ (left),  the sorted vector of coordination numbers, ${\sf sort}[c]$ (middle), and the second and third central moments of the coordination numbers, $(\mu_2,\mu_3)$ (right).}
\end{figure}

\begin{figure}[htbp]
    \centering
    \includegraphics[width=0.9\textwidth]{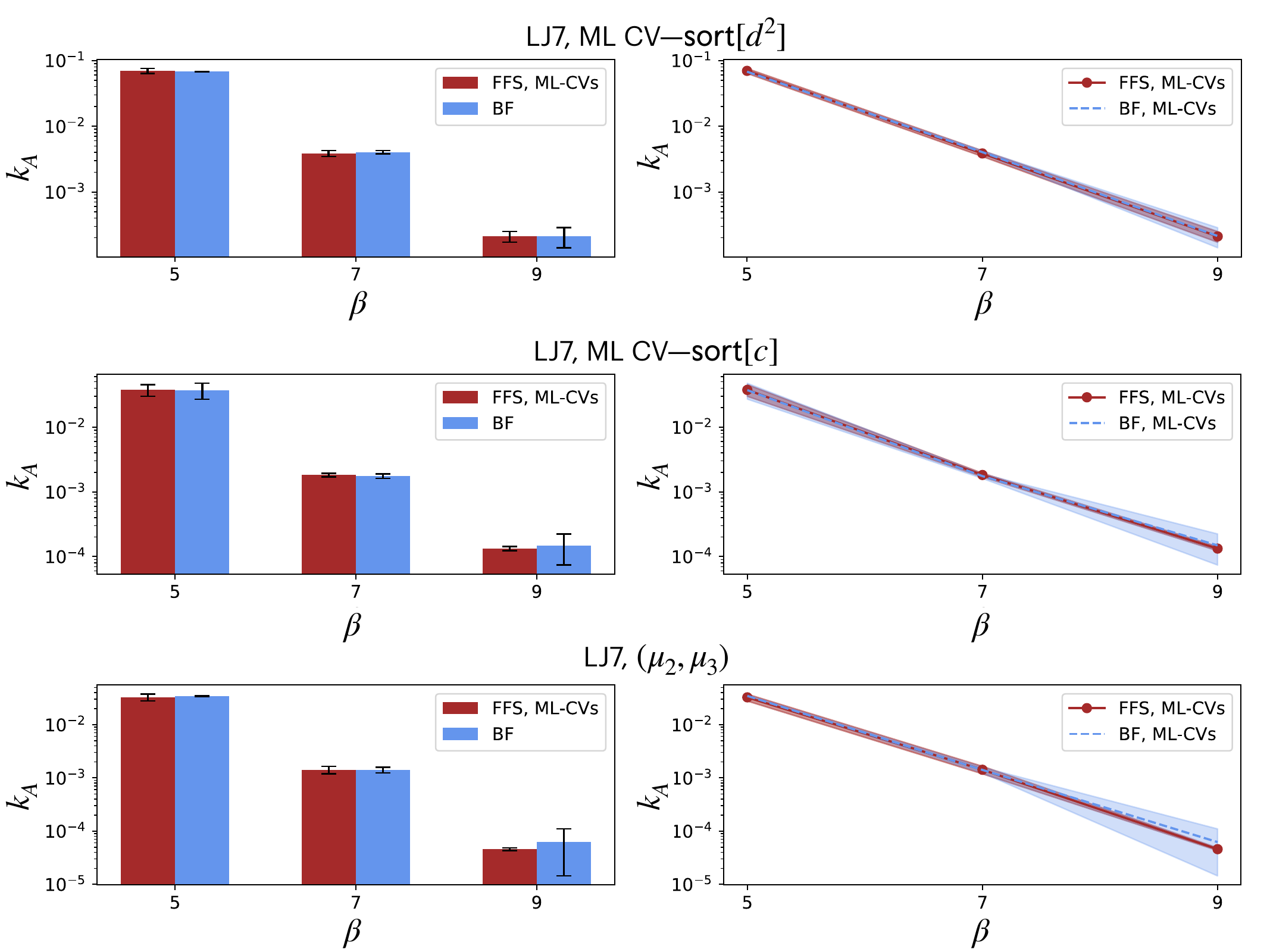}
    \caption{Comparison of FFS escape rate $k_A$ and brute force $k_A$ for three temperatures in log scale. Standard deviations are labeled in figures as error bar (left) or shaded regions (right).}
    \label{fig:LJ7_BF-FFS_all}
\end{figure}

\begin{figure}[htbp]
\includegraphics[width = 0.9\textwidth]{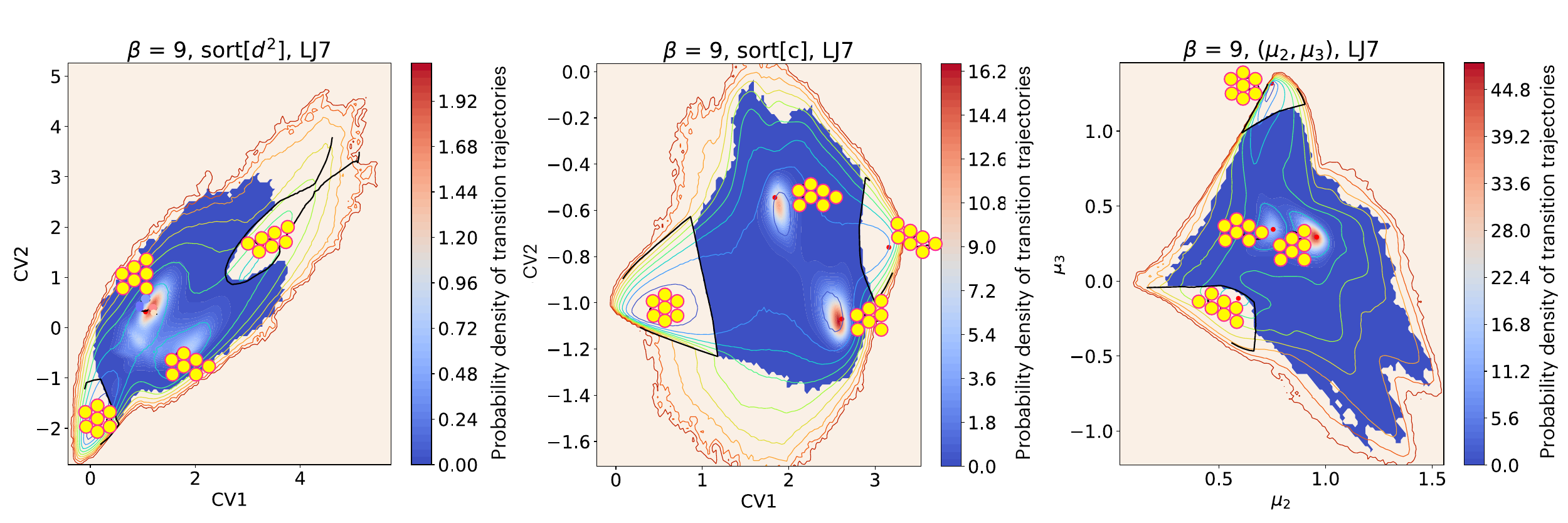}% Here is how to import EPS art
\caption{\label{fig:LJ7probdensity} The probability density of reactive trajectories at $\beta = 9$ obtained using stochastic control in the original $\mathbb{R}^{14}$ space of atomic coordinates and projected onto the learned CV space with feature maps $\sf{sort}[d^2]$ (left), ${\sf sort}[c]$ (middle), and $(\mu_2,\mu_3)$ (right). The sets $A$ and $B$ are determined by level sets of the committor in Fig. \ref{fig:LJ7comm} corresponding to values 0.01 and 0.99, respectively.}
\end{figure}

The orthogonality condition~\eqref{eq:OC} enforced in the loss~\eqref{eq:CV_loss} for learning CVs via autoencoders promotes a small discrepancy between the law of the process in the feature space and the reduced dynamics in the learned CV space. It does not ensure that this discrepancy is small between the process in atomic coordinates and the reduced dynamics in the CV space. Therefore, we cannot expect the transition rates in the all-atom dynamics governed by~\eqref{eq:overdampedLangevin} and the dynamics in CVs~\eqref{eq:dynamics_Maragliano} to be close to each other. Instead, we utilize the learned CVs in the following manner.
\begin{enumerate}
    \item We define the metastable states $A$ and $B$ in the CV space. The precise definitions are found in Section S5 of the SI. The sets $A$ and $B$ defined at this stage for $\beta = 9$ are shown in Fig.~\ref{fig:LJ7comm}.
    \item We calculate the free energy and the diffusion matrix with respect to the CVs and solve the committor problem~\eqref{eq:LgenCV} to find the committor $\tilde{q}(z)$. We approximate $\tilde{q}(z)$ with neural networks. The committors in the machine-learned CVs (ML CV) with the feature maps ${\sf sort}[d^2]$ and ${\sf sort}[c]$ and the CVs $(\mu_2,\mu_3)$ at $\beta =9$ are shown in Fig.~\ref{fig:LJ7comm}. All committors for LJ7 in 2D at $\beta = 5$, $7$, and $9$ are found in Fig. S1 in the SI.
    \item We redefine the sets $A$ and $B$ as level sets of the committor: 
    \begin{align}
        A & : = \{x\in\mathbb{R}^{14}~\mid~\tilde{q}(\xi(\phi(x)) < \epsilon_A\},\label{eq:Adef}\\
        B & : = \{x\in\mathbb{R}^{14}~\mid~\tilde{q}(\xi(\phi(x)) >1- \epsilon_B\},
    \end{align}
    where $\epsilon_A$ and $\epsilon_B$ are small parameters. See Fig.~\ref{fig:LJ7probdensity} for the redefined $A$ and $B$ at $\beta = 9$. The values of $\epsilon_A$ and $\epsilon_B$ chosen for all settings are found in Tables S4--S6 of the SI. 
     This redefinition is advantageous because it helps mitigate the entropic effect if the selected initially $A$ and $B$ are too small for the system to reliably find them while within their respective basins. The redefined $A$ and $B$ will be significantly larger in this case.
     
    \item We use the forward flux sampling (FFS) algorithm~\cite{Allen_2005,Allen_2009} described in Section~\ref{sec:FFS} to find the escape rates $k_A$ and  $k_B$. The milestones for FFS are found in Tables S4--S6 in the SI.  The FFS escape rates $k_A$ from the sets $A$ at $\beta = 5$, $7$, and $9$ are compared to the $k_A$ rates obtained by brute force all-atom unbiased simulations in Fig.~\ref{fig:LJ7_BF-FFS_all}. 
    \item We use $k_A$ and $k_B$ to calculate the transition rate $\nu_{AB}$ and the probabilities $\rho_A$ and $\rho_B$—the probabilities that, at a random time, the system most recently visited $A$ or $B$, respectively, as explained in Section~\ref{sec:rates&control}. The numerical values of all these quantities at $\beta = 5$, $7$, and $9$ are presented together with those obtained by brute force unbiased all-atom simulations in Tables~\ref{table:FFS_result_LJ7_sortd2}, \ref{table:FFS_result_LJ7_MLCV_sortCNum}, and \ref{table:FFS_result_LJ7_mu2mu3} in Appendix \ref{app:tables} for the ML CVs the feature maps ${\sf sort}[d^2]$, ${\sf sort}[c]$, and $(\mu_2,\mu_3)$ respectively.
    \item We use the stochastic control with the controller $2\beta^{-1}\nabla\log\tilde{q}(\xi(\phi(x)))$ as described in Section~\ref{sec:rates&control} to generate transition trajectories and find their probability density. The results for $\beta = 9$ are displayed in Fig.~\ref{fig:LJ7probdensity}. The results for all values of $\beta$ are found in Fig. S8 in the SI.
\end{enumerate}

%============================================================

\section{Test case 2: Lennard-Jones-8 in 3D}
\label{sec:LJ8results}
Unlike the well-studied LJ7 in 2D, the second case study, LJ8 in 3D, is not a standard test problem for coarse-graining and transition rate estimation. The system is governed by the overdamped Langevin dynamics \eqref{eq:overdampedLangevin} simulated using MALA~\cite{Roberts1996ExponentialCO}. Its potential energy landscape contains eight local minima enumerated in the increasing order of their potential energies, two of which, Minima 1 and 2, are significantly deeper than the other ones~\cite{Forman_Cameron_2017} -- see Fig. \ref{fig:LJ8digraph}.  Minimum 1, the capped pentagonal bipyramid, has a point group of order (PGO) 2. Minimum 2 has PGO 8. Minimum 2 can be obtained by taking two pentagonal pyramids and gluing four of their vertices while connecting the fifth pair of their vertices via an edge. If one rotates Minimum 2 by a 90-degree angle around the axis passing through the midpoint of that connecting edge and the opposite vertex, and then rotates the structure by 180 degrees around a particular axis parallel to that connecting edge, Minimum 2 maps into itself. Additionally, Minimum 2 has two orthogonal planes of symmetry. 

\subsection{Settings and CVs}
We primarily consider the problem of estimating the transition rate between Minima 1 and 2. In addition, we also find the escape rate from the union of neighborhoods of Minima 1 and 2 to the union of neighborhoods of Minima 3--8.

The potential energy barrier between Minima 1 and 2 is lower than the lowest potential energy barrier connected to the lowest minimum of LJ7 in 2D. Therefore, we use lower values of temperature corresponding to $\beta = 10$, $15$, and $ 20$.

The MALA time step for simulating dynamics of this system is $5\cdot10^{-5}$ reduced units. The restraining spring force with spring constant $\kappa_r = 100$ turns on if an atom is at a distance greater than 2.5 from the center of mass of the cluster. The parameters for metadynamics are specified in Appendix \ref{app:makedata}.

\begin{figure}[htbp]
    \centering
    \includegraphics[width=0.9\textwidth]{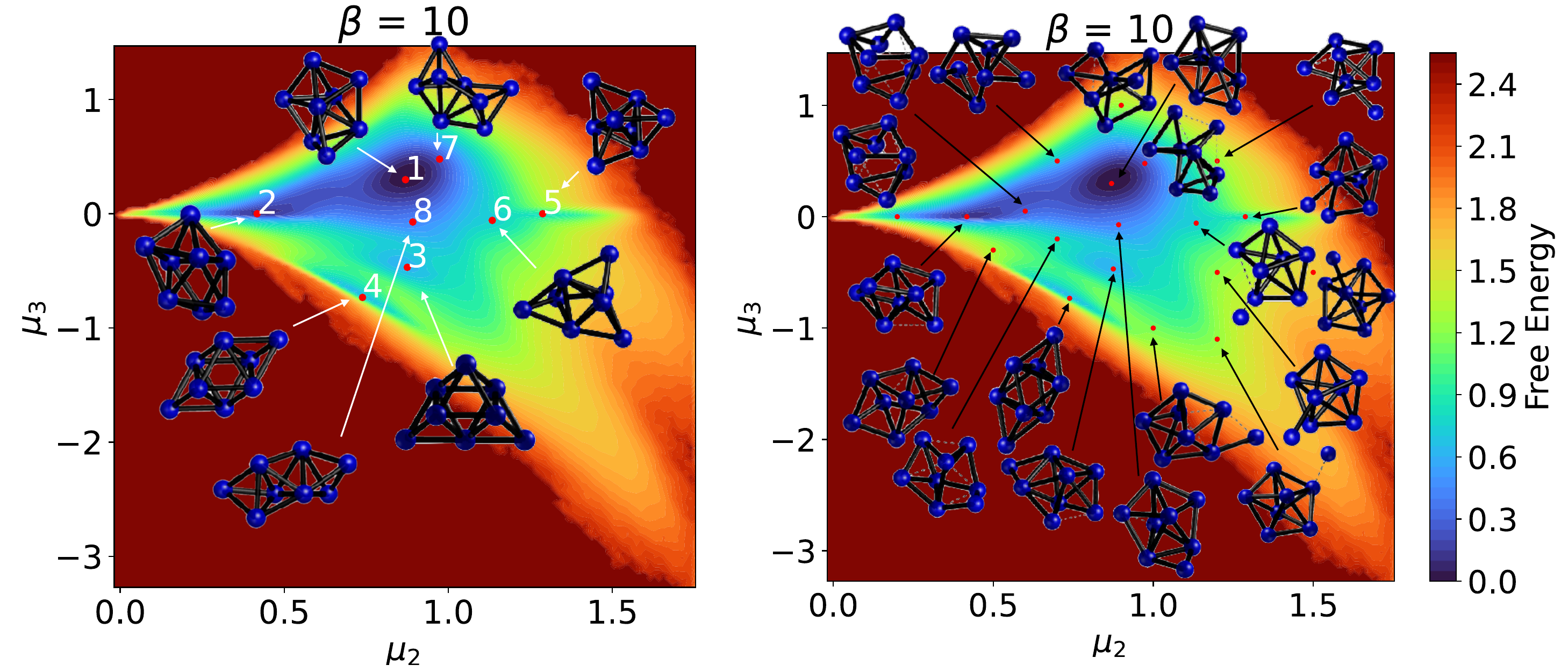}
    \caption{The free energy for LJ8 in 3D in $(\mu_2,\mu_3)$ with the projections of the potential energy minima (left) and representative configurations (right). HEre and in all further ``balls-and-sticks" depictions of the configurations of LJ8 in 3D, any two atoms, $i$, and $j$ are connected by a rod-like edge if and only if the distance between their centers, $r_{i,j}$, differs from $2^{1/6}$ by less than 0.05, and atoms $i$ and $j$ are connected by a dashed edge if and only if $|r_{i,j} - 2^{1/6}|\in[0.05,0.1)$.}
    \label{fig:LJ8_FE_mu2mu3}
\end{figure}

\begin{figure}[htbp]
    \centering
    \includegraphics[width=0.6\textwidth]{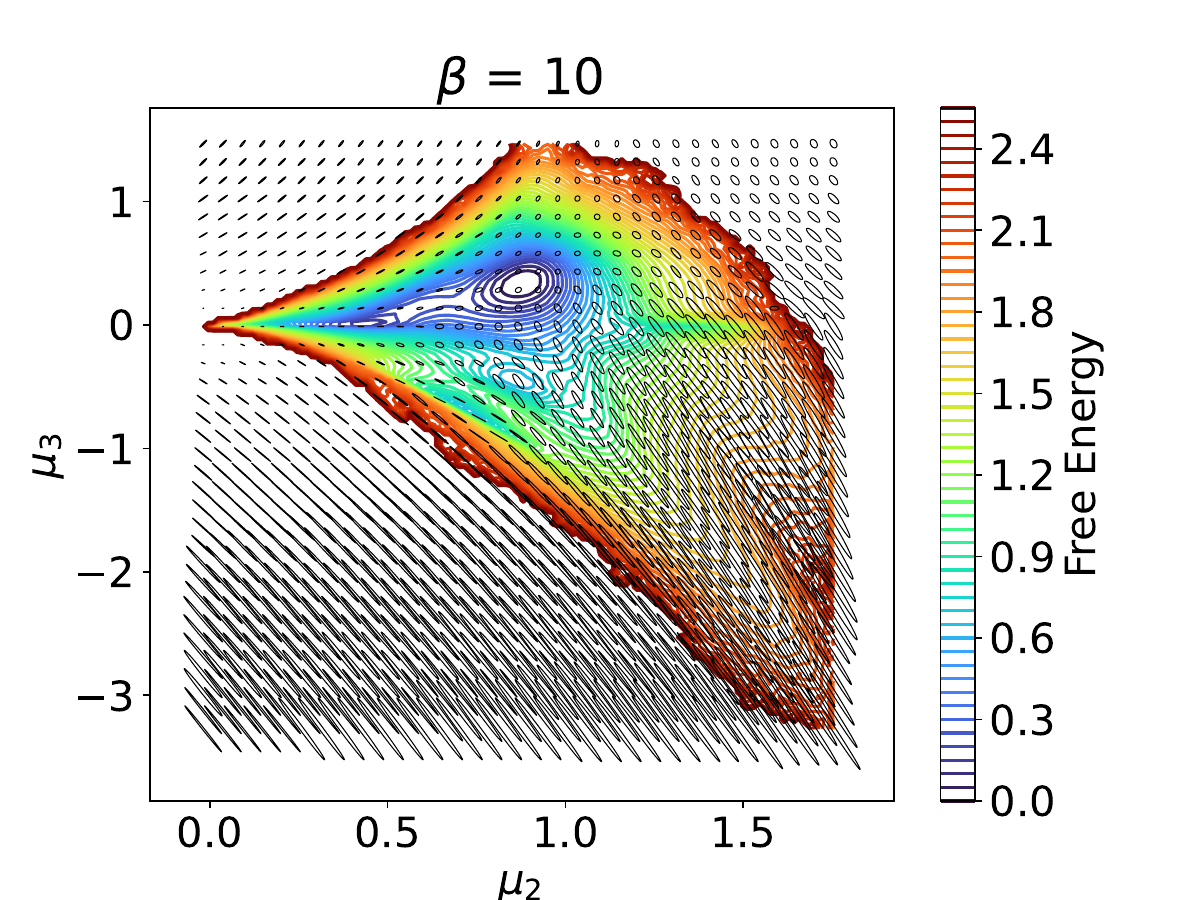}
    \caption{The diffusion matrix for LJ8 in 3D in $(\mu_2,\mu_3)$. }
    \label{fig:LJ8_DM_mu2mu3}
\end{figure}

\begin{figure}[htbp]
    \centering
    \includegraphics[width=0.6\textwidth]{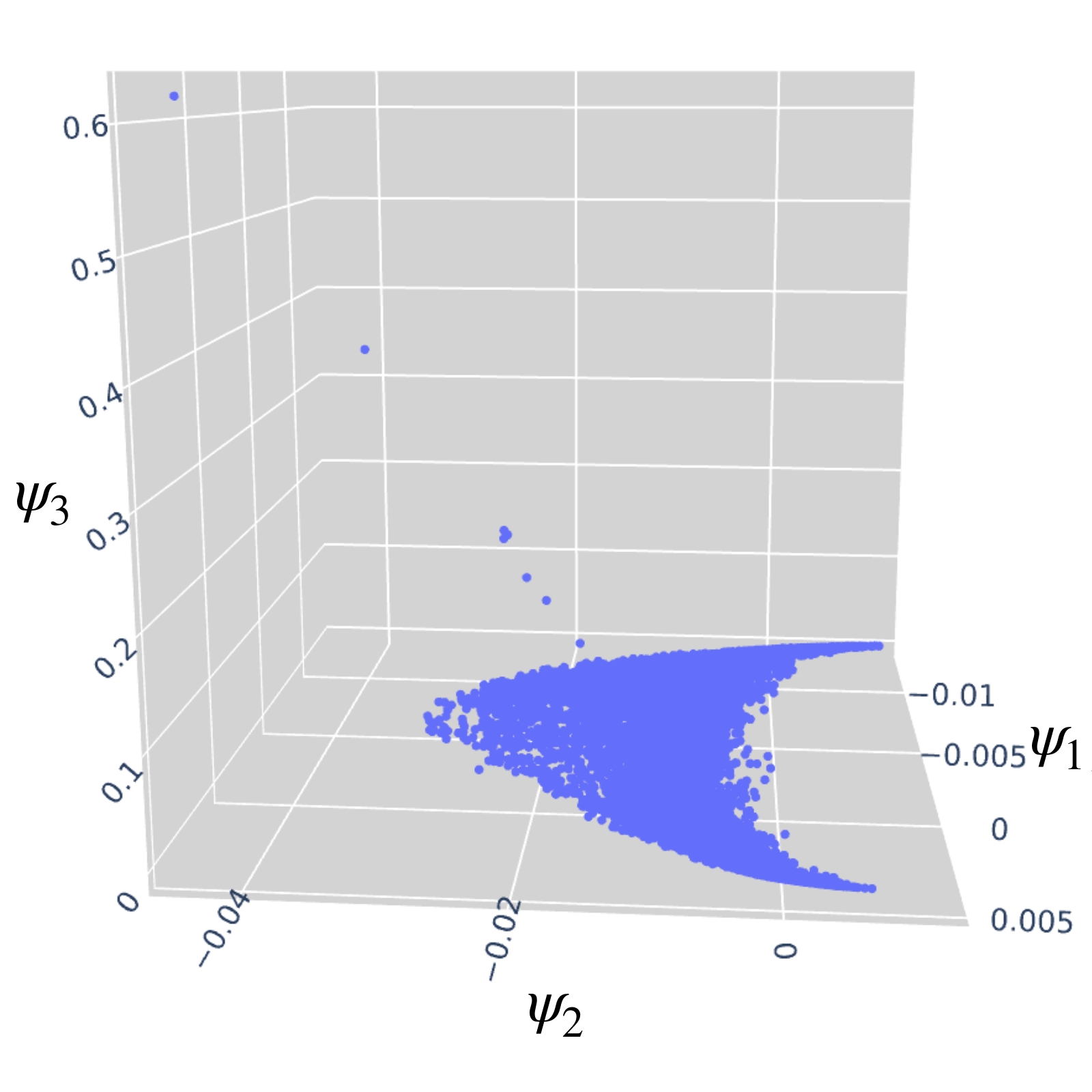}
    \caption{The residence manifold  for LJ8 in 3D with the feature map ${\sf sort}[c]$ obtained using the target measure diffusion map. }
    \label{fig:LJ8manifold}
\end{figure}

\begin{figure}[htbp]
    \centering
    \includegraphics[width=0.9\textwidth]{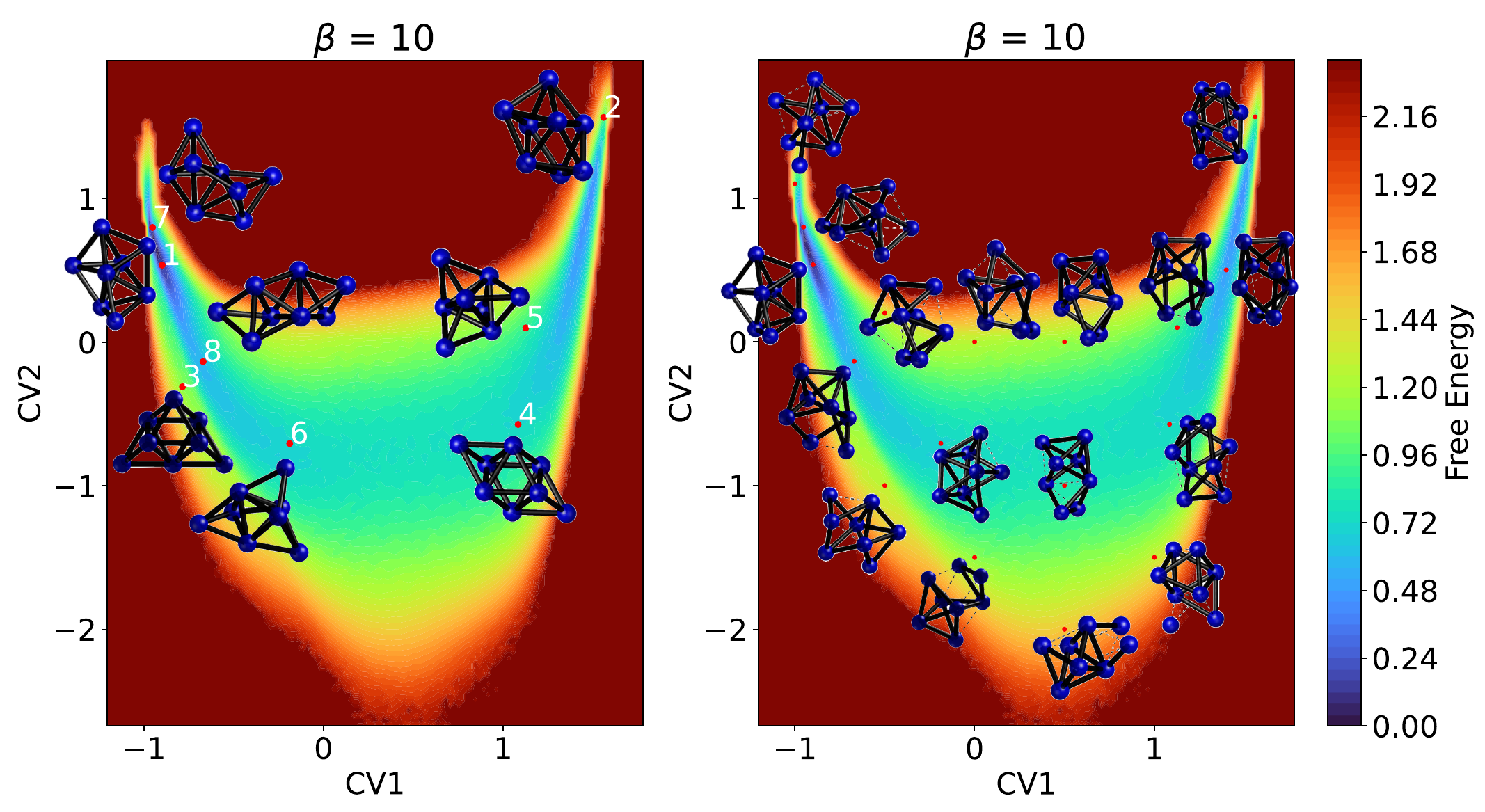}
    \caption{The free energy for LJ8 in 3D with respect to the CV learned by Algorithm~\ref{alg:learnCVs} with the ${\tt sort}[c]$ feature map at $\beta = 10$. The projections of the potential energy minima are displayed on the left while the representative configurations are shown on the right. }
    \label{fig:LJ8_FE_sortCNum}
\end{figure}

\begin{figure}[htbp]
    \centering
    \includegraphics[width=0.6\textwidth]{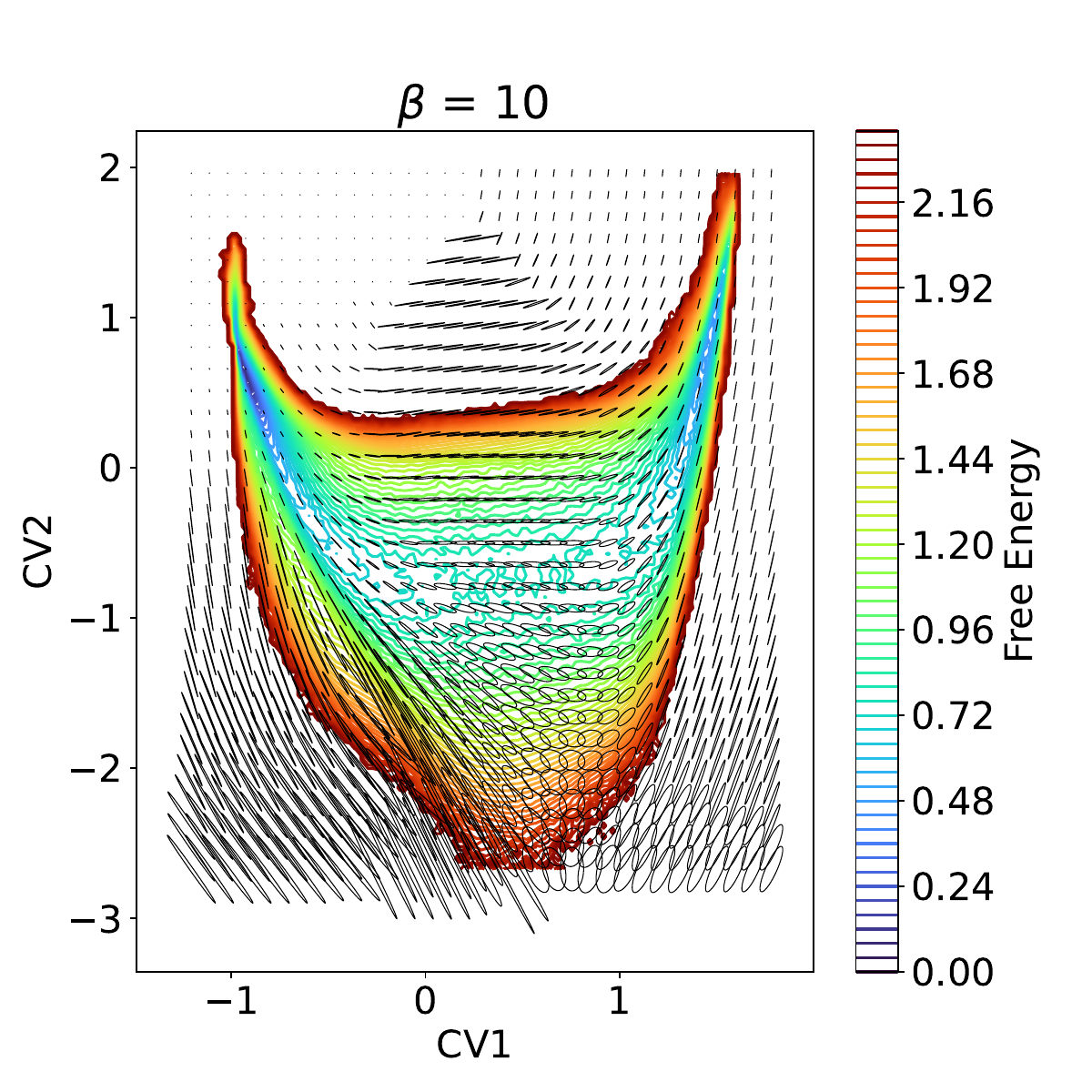}
    \caption{The diffusion matrix  for LJ8 in 3D with respect to the CV learned by Algorithm~\ref{alg:learnCVs} with the ${\tt sort}[c]$ feature map at $\beta = 10$. }
    \label{fig:LJ8_DM_sortCNum}
\end{figure}

\begin{figure}[htbp]
    \centering
    \includegraphics[width=0.6\textwidth]{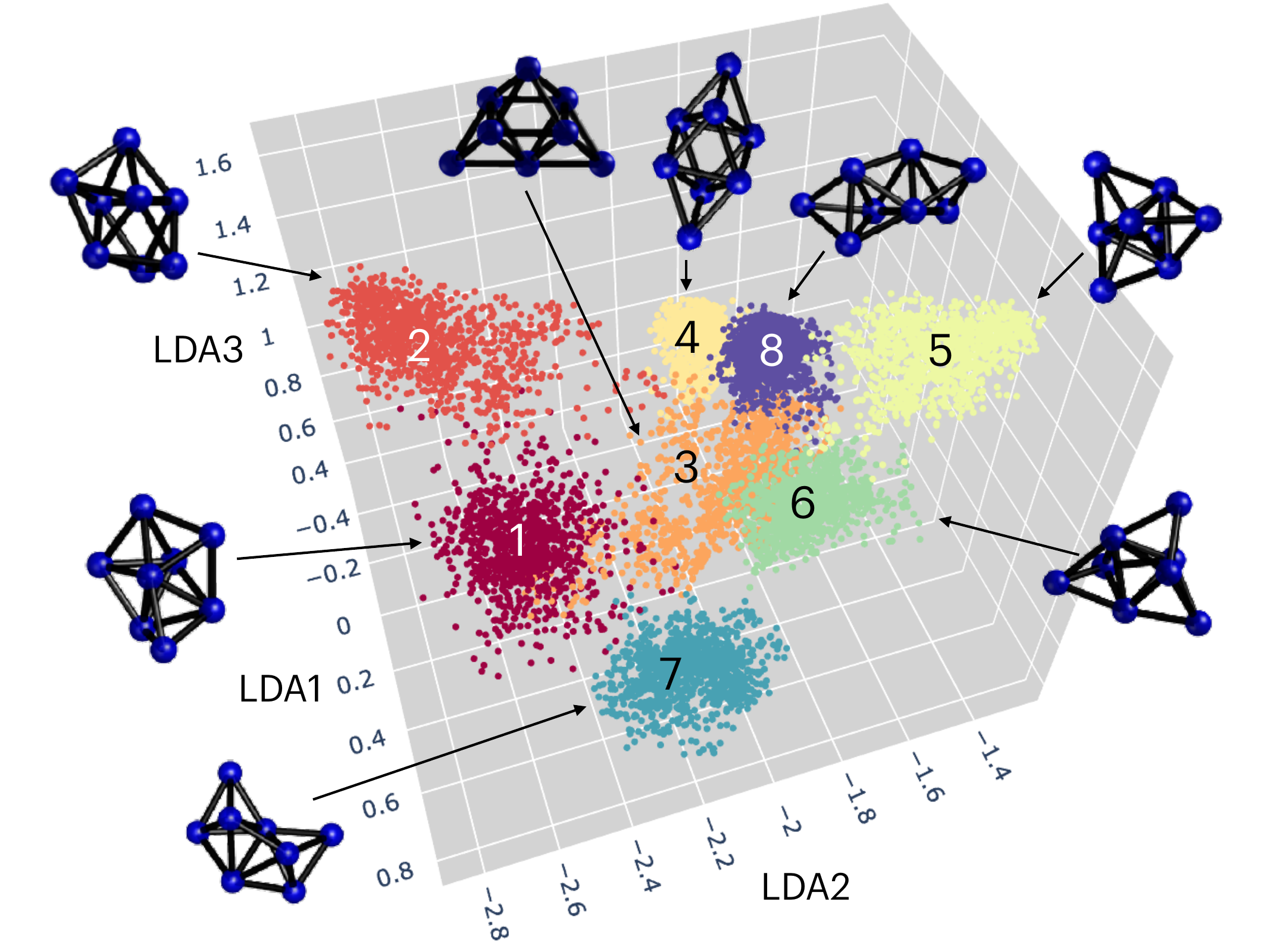}
    \caption{The ${\tt sort}[c]$ data for LJ8 in 3D projected onto the span of the three dominant eigenvectors of the linear discriminant analysis generalized eigenvalue problem. }
    \label{fig:LJ8_LDA}
\end{figure}

\begin{figure}[htbp]
    \centering
    \includegraphics[width=0.9\textwidth]{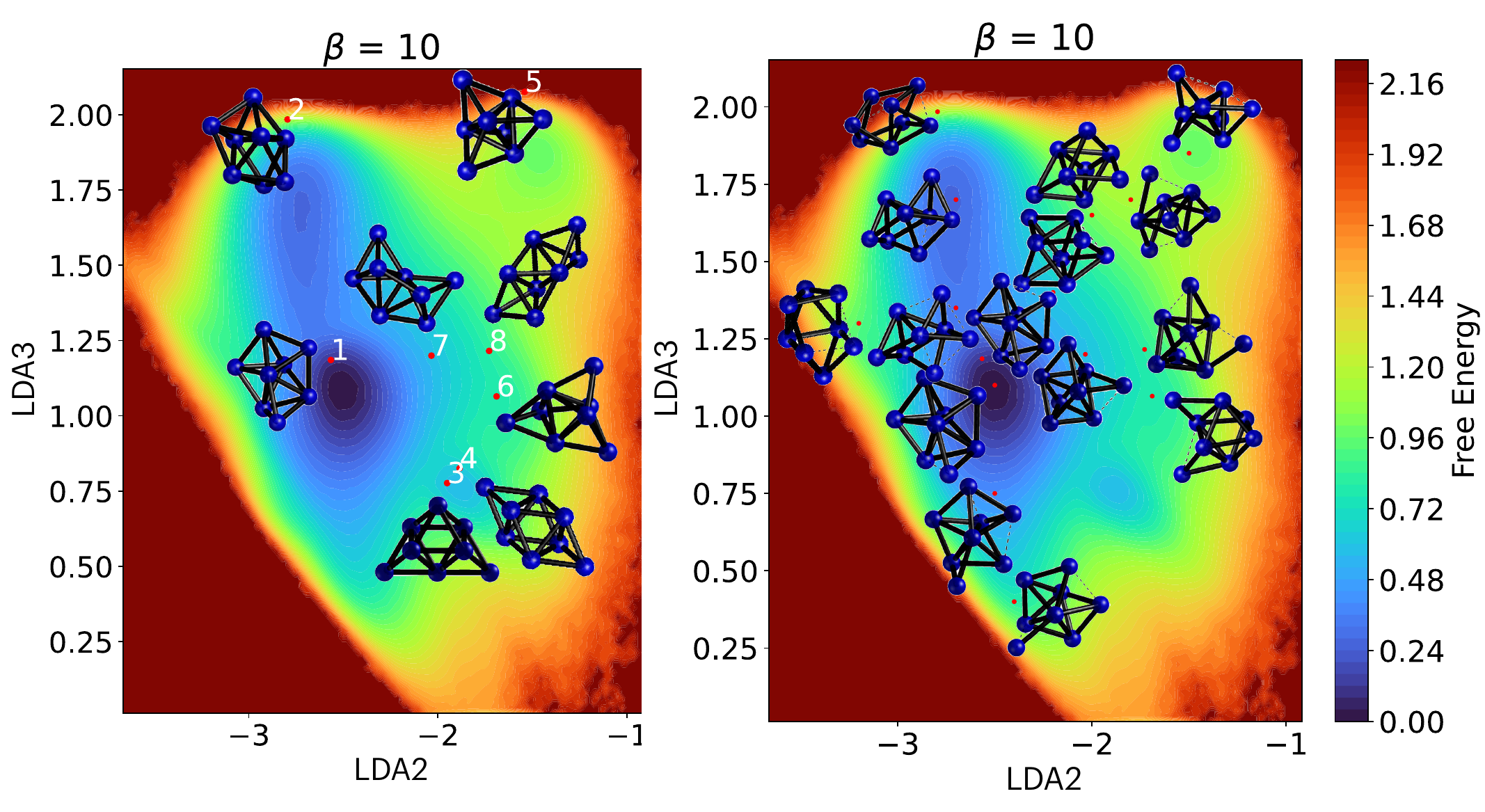}
    \caption{The free energy for LJ8 at $\beta = 10$ in the CV (LDA2,LDA3) with the projections of the potential energy minima (left) and representative configurations (right).}
    \label{fig:LJ8_LDA23_confs}
\end{figure}

\begin{figure}[htbp]
    \centering
    \includegraphics[width=0.9\textwidth]{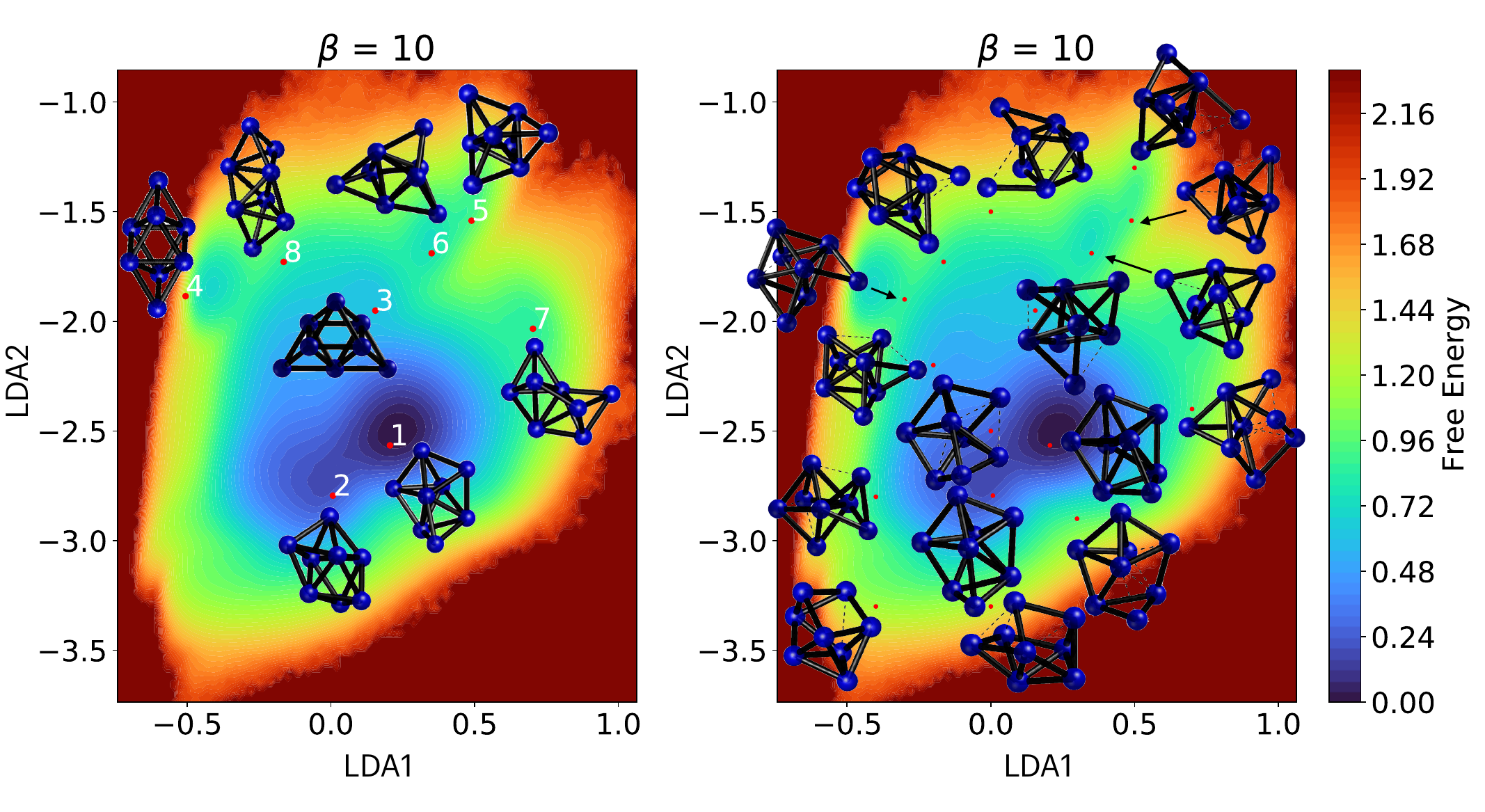}
    \caption{The free energy for LJ8 at $\beta = 10$ in the CV (LDA1,LDA2) with the projections of the potential energy minima (left) and representative configurations (right).}
    \label{fig:LJ8_LDA12_confs}
\end{figure}

\begin{figure}[htbp]
    \centering
    \includegraphics[width=0.9\textwidth]{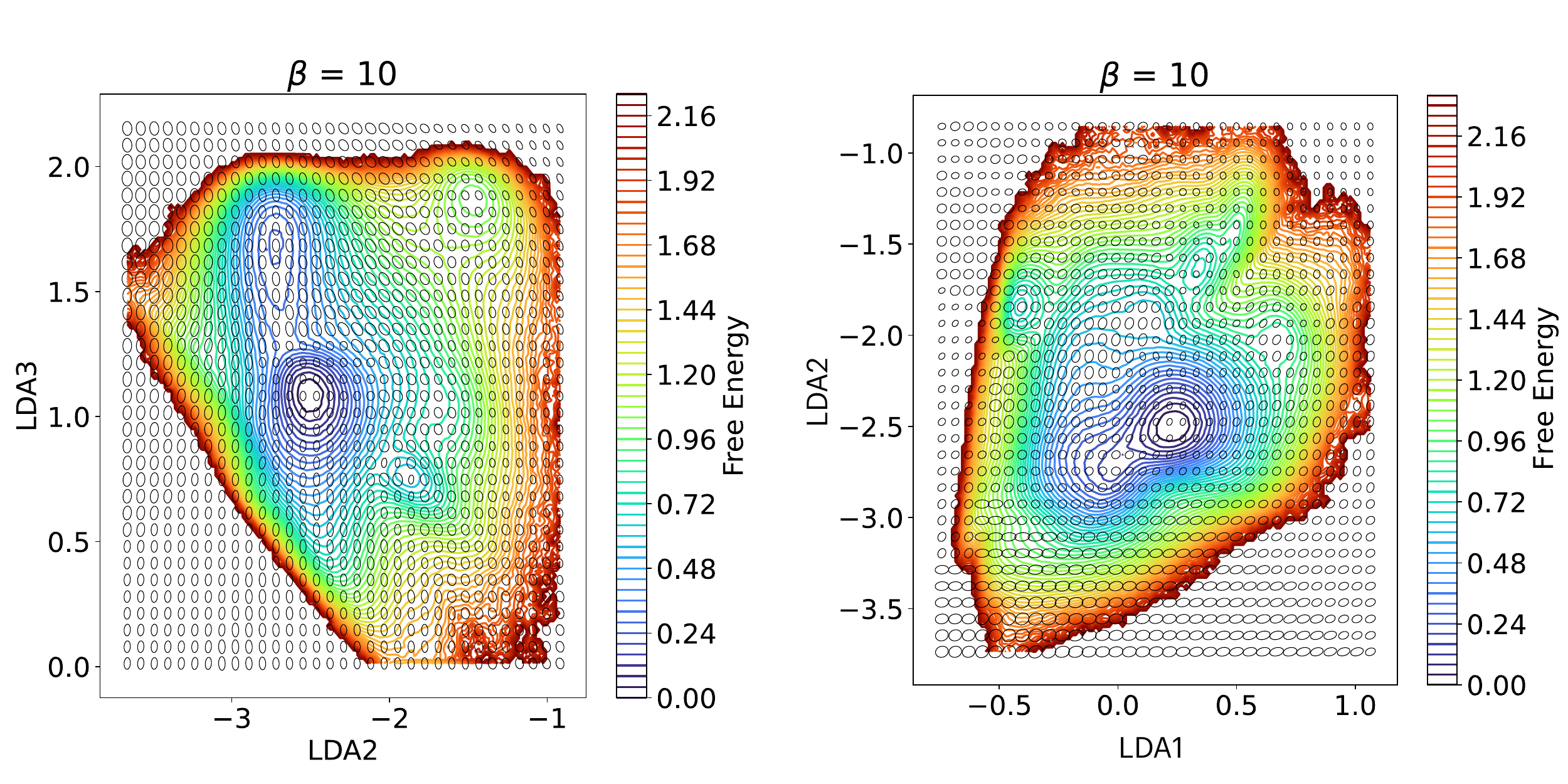}
    \caption{The diffusion matrix  for LJ8 in 3D  at $\beta = 10$ in (LDA2,LDA3) (left) and (LDA1,LDA2) (right).}
    \label{fig:LJ8_LDA_DM}
\end{figure}

We have considered the following five sets of CVs.
\begin{enumerate}
    \item The second and third central moments of the coordination numbers, $(\mu_2,\mu_3)$. The free energy and the diffusion matrix at $\beta = 10$ are shown in Figs.~\ref{fig:LJ8_FE_mu2mu3} and \ref{fig:LJ8_DM_mu2mu3} respectively. The free energy in $(\mu_2,\mu_3)$ has five local minima whose basins include the images of Minima 1, 2, 3, 4, and 5. Minima 6, 7, and 8 are mapped into the basin of Minimum 1. See Fig. \ref{fig:LJ8_FE_mu2mu3} (left). In addition, Fig. \ref{fig:LJ8_FE_mu2mu3} (right) displays a collection of projections of representative configurations at a collection of points in the $(\mu_2,\mu_3)$ space, i.e., the first visited configurations in the mesh cells containing those selected points. The examination of those configurations suggests that the basins of free energy minima containing the images of Minima 1, 2, and 4 are populated with perturbations of the respective minima, while the other two free energy minima may contain images of configurations more similar to Minimum 1 along with images of Minima 3 and 5 and their slight perturbations.

    \item The CVs learned by Algorithm~\ref{alg:learnCVs} with the feature map ${\sf sort}[d^2]$ and the loss function ~\eqref{eq:lossMEP} involving the MEP data for the autoencoder. The resulting free energy landscape has a unique local minimum. We did not proceed with these CVs further.

    \item The CVs learned by Algorithm~\ref{alg:learnCVs} with the feature map ${\sf sort}[c]$. This resulted in a success. The corresponding residence manifold, the free energy, and the diffusion matrix at $\beta = 10$ are shown in Figs.~\ref{fig:LJ8manifold}, \ref{fig:LJ8_FE_sortCNum}, and \ref{fig:LJ8_DM_sortCNum} respectively. The input data for Algorithm \ref{alg:learnCVs} were obtained using $(\mu_2,\mu_3)$. We set up a $129\times 129$ mesh in the $(\mu_2,\mu_3)$-space and saved one configuration per mesh cell visited by a long metadynamics run. The resulting dataset consists of 16,641 vectors of atomic coordinates. We used the target measure diffusion map with the bandwidth $\epsilon = 0.5$ and the kNN sparsification with $k = 4,160$. The residence manifold after removal of a few outliers is a 2D manifold lying in the space $(\psi_1(\phi),\psi_2(\phi))$ spanned by the two dominant eigenvectors of the diffusion map -- see \eqref{eq:Psi123}. Therefore, there is no need to learn the confining potential $V_1$ and the CVs: $\psi_1(\phi(x))$ and $\psi_2(\phi(x))$ are the desired CVs in this case.  The free energy landscape in CVs $(\psi_1,\psi_2)$ has two local minima whose basins contain the images of Minima 1 and 2. The representative configurations visualized in Fig. \ref{fig:LJ8_FE_sortCNum} (right) depict a gradual transition between Minima 1 and 2, which we will study. We will refer to these CVs as ML CV with the feature map ${\sf sort}[c]$.

    \item For comparison, we also applied the linear discriminant analysis (LDA)~\cite{DHS2001} to the ${\sf sort}[c]$ data. Details are found in Appendix \ref{app:LDA}. The projection onto the span of the three dominant eigenvectors denoted by LDA1, LDA2, and LDA3, is shown in Fig.~\ref{fig:LJ8_LDA}. The projection on the span of LDA2 and LDA3 separates the images of Minima 1 and 2 -- see Fig. \ref{fig:LJ8_LDA23_confs}. The representative configurations on the right panel suggest that LDA2 and LDA3 may be suitable CVs for the study of the transition between these two minima. The diffusion matrix at $\beta = 10$ in (LDA2,LDA3) is visualized in Fig. \ref{fig:LJ8_LDA_DM} (left).
    \item 
    The projection onto the span of LDA1 and LDA2 separates the union of the two deepest minima, 1 and 2, from the rest of the minima -- see Fig. \ref{fig:LJ8_LDA12_confs}.  Therefore, we consider the escape problem from the union of the Minima 1 and 2 to the union of the basins of the remaining minima using LDA1 and LDA2 as CVs. The representative configurations in  Fig. \ref{fig:LJ8_LDA12_confs} (right) also suggest (LDA1,LDA2) may be suitable CVs for this purpose.  The diffusion matrix at $\beta = 10$ in (LDA1,LDA2) is shown in Fig. \ref{fig:LJ8_LDA_DM} (right).
\end{enumerate}

\subsection{Results}
\begin{figure}[htbp]
    \centering
    \includegraphics[width=0.9\textwidth]{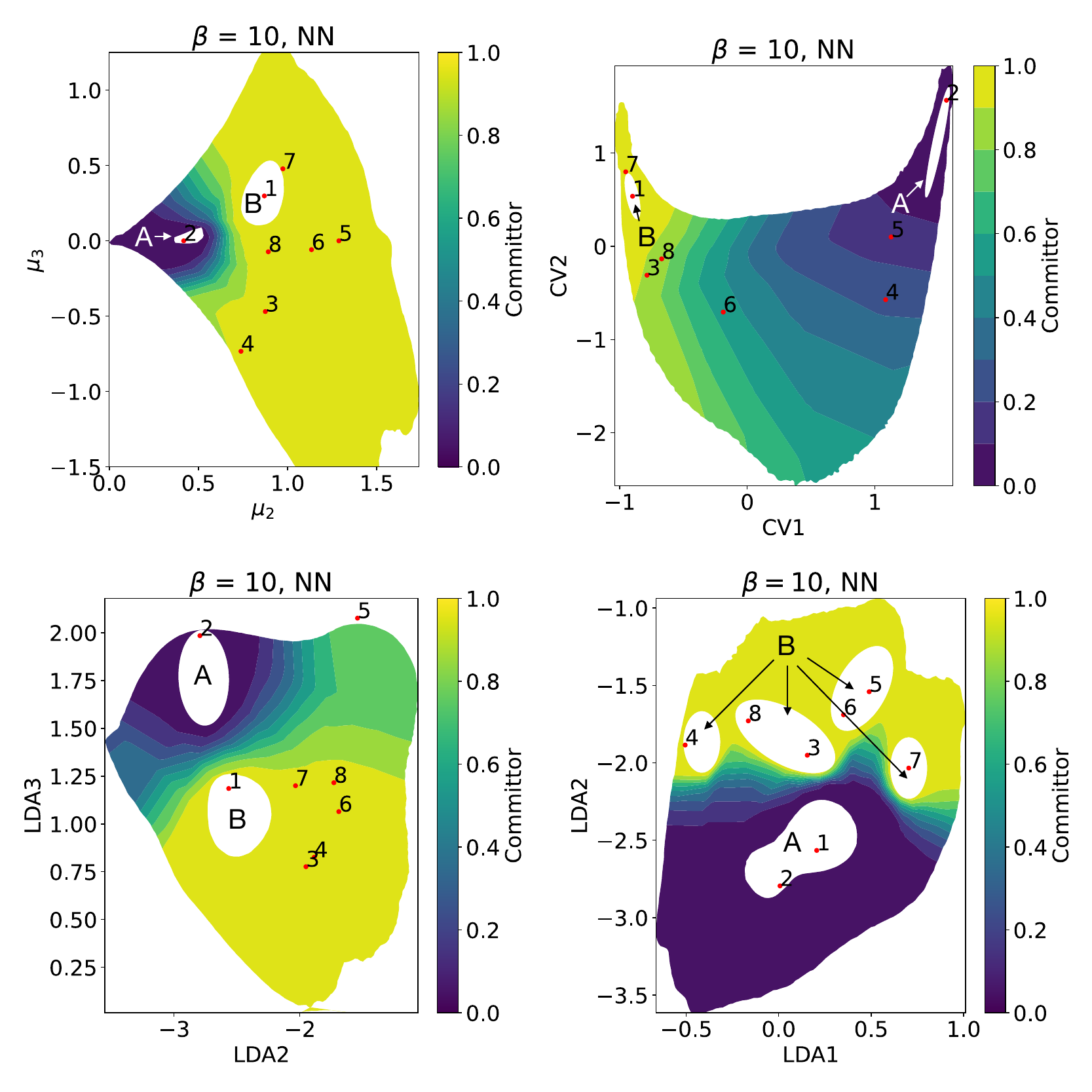}
    \caption{The committors for LJ8 in 3D at $\beta = 10$ in $(\mu_2,\mu_3)$ (top left), CVs learned by Algorithm \ref{alg:learnCVs} (top right), (LDA2,LDA3) (bottom left), and (LDA1,LDA2) (bottom right).}
    \label{fig:LJ8_committor}
\end{figure}

\begin{figure}[htbp]
    \centering
    \includegraphics[width=0.9\textwidth]{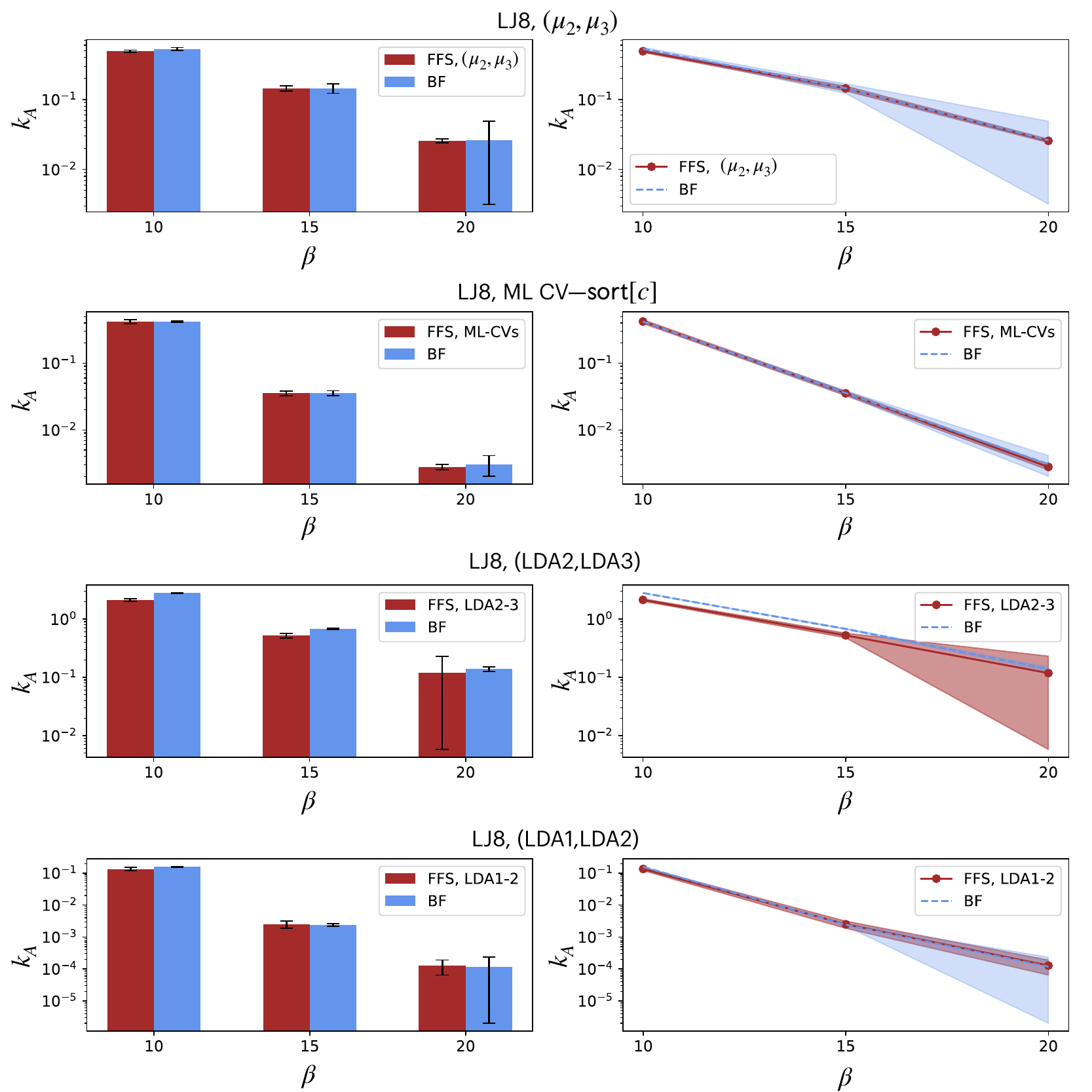}
    \caption{The escape rates $k_A$ from sets $A$ to sets $B$ for LJ8 in 3D obtained using the forward flux sampling (FFS) the committors as the reaction coordinates and brute force sampling (BF). The sets $A$ and $B$ are defined and the committor problem is solved in the following CV spaces: $(\mu_2,\mu_3)$ (row 1), CVs learned by Algorithm \ref{alg:learnCVs} with the feature map ${\sf sort}[c]$ (row 2), LDA2 and LDA3 (row 3), LDA1 and LDA2 (row 4).}
    \label{fig:LJ8_FFS&BF}
\end{figure}

\begin{figure}[htbp]
    \centering
    \includegraphics[width=0.9\textwidth]{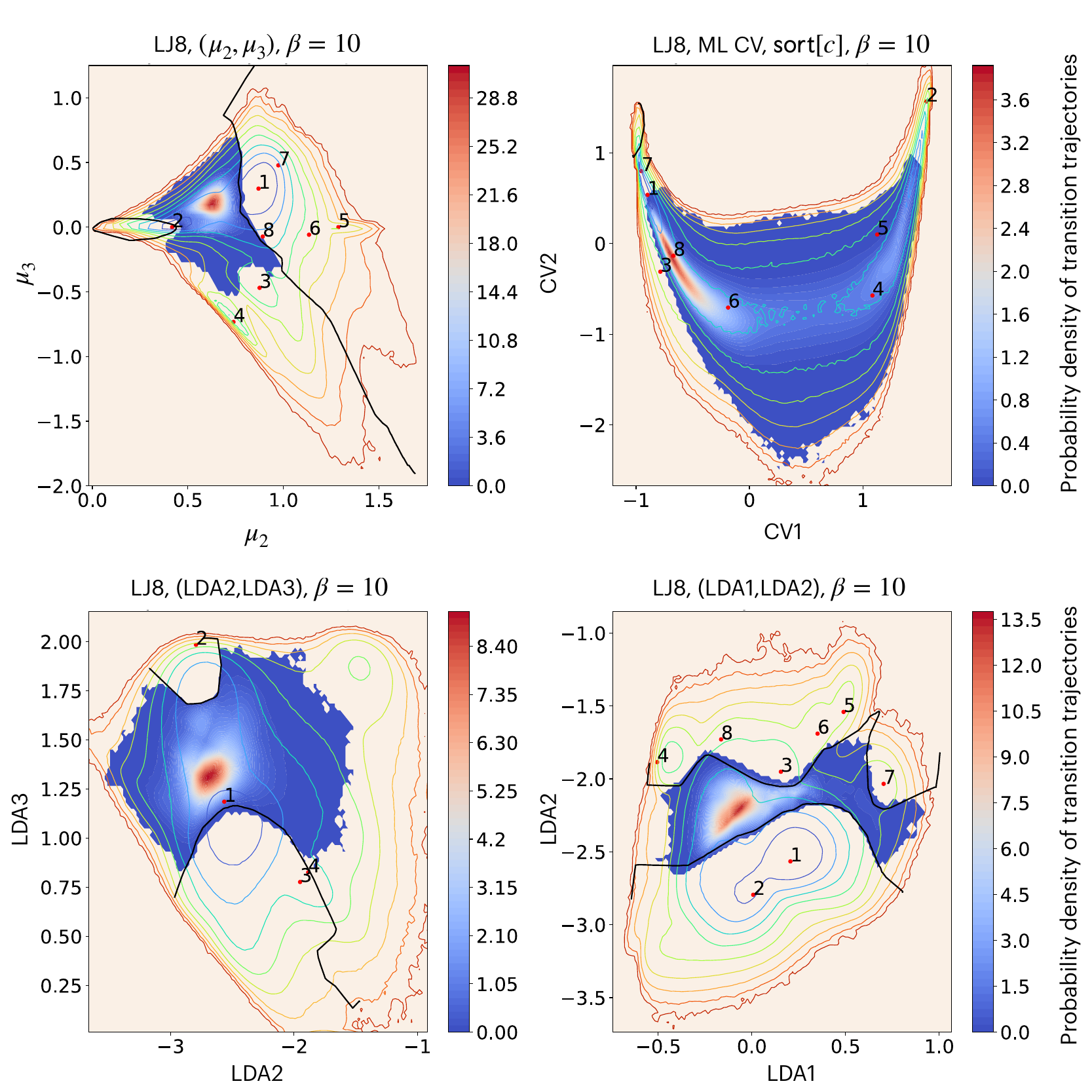}
    \caption{The probability density of transition trajectories for LJ8 in 3D at $\beta = 10$ in $(\mu_2,\mu_3)$ (top left), CVs learned by Algorithm \ref{alg:learnCVs} (top right), (LDA2,LDA3) (bottom left), and (LDA1,LDA2) (bottom right).}
    \label{fig:LJ8_prob_density}
\end{figure}

For each pair of viable CVs, $(\mu_2,\mu_3)$, ML CVwith the ${\sf sort}[c]$ feature map, (LDA2,LDA3), and (LDA1,LDA2), we refined the regions $A$ and $B$ and solved the committor problem \eqref{eq:LgenCV} using the finite element method. Then we approximated the computed committors with neural networks to obtain globally defined differentiable functions. The results are shown in Fig. \ref{fig:LJ8_committor}. Next, we redefine the sets $A$ and $B$ as the level sets of the committor close to zero and one respectively. We use the forward flux sampling (FFS) with the committor as the reactive coordinate to find the rates $k_A$, $k_B$, $\nu_{AB}$, and the probabilities $\rho_A$ and $\rho_B$ and the stochastic control to find the probability density of transition trajectories. The transition rates $k_A$ obtained using FFS and by brute force, i.e., by running a long unbiased trajectory, are presented in Fig. \ref{fig:LJ8_FFS&BF}. The probability densities of transition trajectories at $\beta = 10$ projected onto the corresponding CV spaces are displayed in Fig. \ref{fig:LJ8_prob_density}. 
The numerical values of $k_A$, $k_B$, $\nu_{AB}$, $\rho_A$, and $\rho_B$ at $\beta = 10$, $15$, and $20$ are reported in Tables~\ref{table:FFS_result_LJ8_mu2mu3}, \ref{table:FFS_result_LJ8_MLCV}, \ref{table:FFS_result_LJ8_LDA23} and \ref{table:FFS_result_LJ8_LDA12} in Appendix \ref{app:tables}.
Further details about the four pairs of CVs and associated computations such as the choice of $A$ and $B$, neural network architectures used, and a full set of figures with the free energy and the committor (Figs. S4--S7), and the probability density of transition trajectories (Figs. S9--S10) at $\beta = 10$, $15$, and $20$ for LJ8 in 3D  are found in the SI. The visualizations of the diffusion matrix for each pair of CVs change insignificantly with $\beta$.

\section{Discussion}
\label{sec:discussion}
In this section, we discuss what we have learned about the proposed methodology and the two case studies, LJ7 in 2D and LJ8 in 3D.

\subsection{The proposed framework}
The proposed framework for learning CVs is quite general and admits various implementations of each step. Algorithm \ref{alg:learnCVs} is robust with respect to imperfections in the execution of its steps. For example, the appearance of the residence manifold for LJ8 in Fig. \ref{fig:LJ8manifold} clearly shows that we did not remove outliers. Nonetheless, the rates were estimated accurately. 

The redefinition of the metastable regions $A$ and $B$ as the level sets of the computed committor $\epsilon_A$ and $1-\epsilon_B$ makes a nice autocorrection. If the initially chosen regions $A$ and $B$ are too small so that finding them presents an entropic barrier to the system, the redefined regions tend to enlarge them and remove the entropic effect.

The choice of a feasible feature map is very important for the success of learning CVs that separate the metastable states of interest. The proposed featurization ${\sf sort}[c]$ that maps the vector of atomic coordinates into a sorted vector of coordination numbers led to good CVs and accurate rate estimates for both LJ7 in 2D and LJ8 in 3D. 
In contrast, the featurization ${\sf sort}[d^2]$, the sorted vector of squared pairwise interatomic distances, was successful for LJ7 in 2D, especially when used with a minimum energy path, yielding good collective variables and an accurate rate estimate. However, it failed to separate the basins of Minima 1 and 2 for LJ8 in 3D.
The second and third central moments of the coordination numbers, $(\mu_2,\mu_3)$, make a reasonable 2D CV for both case studies. We used $(\mu_2,\mu_3)$ for obtaining the initial dataset of points by running metadynamics with biasing with respect to them. We also used the results obtained with $(\mu_2,\mu_3)$ for reference.

While this paper was in the final stages of preparation, we learned about other featurizations that may be helpful for larger and more complex systems\footnotemark[1]: 
\footnotetext[1]{Thanks to Dr. J. Rogal.} Geom2vec~\cite{Geom2vec2025} that uses pre-trained graph neural networks with token mixers as geometric featurizers and Behler-Parrinello symmetry functions~\cite{BehLerParrinelloSymFun2007}. Moreover, instead of using the diffusion net algorithm to approximate the derivative of the diffusion map at the training points, one can approximate them by a direct calculation proposed by Meila's group~\cite{Meila2022}.  We leave the exploration of learning CVs with these featurizations and rationalizations for future work.

\subsection{Results for LJ7 in 2D}
For LJ7 in 2D, we have compared three pairs of CVs: CVs learned by Algorithm \ref{alg:learnCVs} with feature maps ${\sf sort}[d^2]$ and ${\sf sort}[c]$ denoted by ML CV--${\sf sort}[d^2]$ and ML CV--${\sf sort}[c]$, respectively, and $(\mu_2,\mu_3)$. ML CV--${\sf sort}[c]$ separates the potential energy minima the best. At all values of $\beta$ considered, $\beta = 5$, $7$, and $9$, the images of the four potential energy minima lie in basins of the corresponding four distinct free energy minima, and these free energy minima are far apart as evident from Fig. \ref{fig:LJ7_FE_confs_sortCNum}.
So is the case for $(\mu_2,\mu_3)$, except that the images of minima $C_1$ and $C_2$ are closer to each other -- see Fig. \ref{fig:LJ7confs}.
ML CV--${\sf sort}[d^2]$ yields a free energy landscape at $\beta = 5$ with only three basins, while with the desired four basins at $\beta = 7$ and $9$ -- see Fig. \ref{fig:LJ7_FE_confs_sortd2} and Fig. S1 in the SI. 

The diffusion matrix is nondegenerate for both ML CV--${\sf sort}[d^2]$ and ML CV--${\sf sort}[c]$ (Figs. \ref{fig:LJ7_DM_sortd2} and \ref{fig:LJ7_DM_sortCNum}, while it is nearly degenerate for $(\mu_2,\mu_3)$ (cf. Fig. 8 in Ref.~\cite{EVANS2023}).

The free energy landscape in $(\mu_2,\mu_3)$ has a ``tail", a large area in the bottom left corresponding to states in which one of the atoms is separated from the cluster. This is a rarely visited region due to the restraining spring force. A nice feature of ML CVs with ${\sf sort}[d^2]$ and ${\sf sort}[c]$ featurizations is that their free energy landscapes do not have such a tail. 

Fig. \ref{fig:LJ7_BF-FFS_all} visualizes the results for the rate $k_A$ done by FFS and brute force. Computational details are given in Appendix \ref{app:tables}. The plots in the log scale show that the estimated $k_A$ rates agree with the Arrhenius law. The estimates of $k_A$ by FFS agree well with those done by brute force, except for the one with $(\mu_2,\mu_3)$ and $\beta = 9$ where the brute force estimate involves a large statistical error due to the rarity of the event. This suggests that the chosen region $B$ is too small in $(\mu_2,\mu_3)$ even after the redefinition a the level set of the committor $1-\epsilon_B$. 

The estimates of the transition rates  $k_A$, $k_B$, and $\nu_{AB}$ and the last hit probabilities $\rho_A$ and $\rho_B$ obtained with the aid ML CV--${\sf sort}[d^2]$, ML CV--${\sf sort}[c]$, and $(\mu_2,\mu_3)$ are reported in Tables \ref{table:FFS_result_LJ7_sortd2}, \ref{table:FFS_result_LJ7_MLCV_sortCNum}, and \ref{table:FFS_result_LJ7_mu2mu3} in Appendix \ref{app:tables}.
These tables show that the FFS rates $k_A$ (escape rate from the hexagon) agree within one standard deviation with the means of the brute force estimates of the corresponding rates. The rates $k_A$ are systematically smaller for the regions defined in the $(\mu_2,\mu_3)$ space than those defined using the ML CVs.
The agreement between the FFS and brute force rates $\nu_{AB}$ for all three CVs is excellent: the means of these estimates are closer than the corresponding standard deviations. The agreement between the FFS and brute force rates $k_B$ is notably worse: it reaches the factor of 2 for ML CVs at $\beta = 7$. It is the best for $(\mu_2,\mu_3)$.  This disagreement might be due to the difficulty of defining the trapezoidal region. The agreement between the probabilities $\rho_A$ is within one standard deviation for all three CVs. The probabilities $\rho_B$ are much smaller than $\rho_A$, and estimates for them are much less accurate.

\subsection{Results for LJ8 in 3D}
The problem of learning a good set of CVs for LJ8 in 3D is much harder than the one for LJ7 in 2D. The reasons for it are the additional degrees of freedom in 3D and the increase in the complexity of the energy landscape. 

An important phenomenon observed in our study of LJ8 in 3D that contrasts it with LJ7 in 2D is that the free energy landscape may totally ignore some of the potential minima. For instance, Fig. \ref{fig:LJ8_FE_sortCNum} shows the images of the local minima in the ML CV--${\sf sort}[c]$ space on the left and a collection of representative configurations on the right. The representative configurations suggest that the free energy landscape represents the two lowest minima of LJ8 and the transition configurations between them, while the other minima do not have their niche in the free energy landscape. Similarly, Minima 6, 7, and 8 are lost in the $(\mu_2,\mu_3)$ free energy landscape (Fig. \ref{fig:LJ8_FE_mu2mu3}). Mimimum 2 is lost in (LDA1,LDA2) free energy landscape (Fig. \ref{fig:LJ8_LDA12_confs}). Therefore, a practical conclusion is that, once a free energy landscape is computed, it is important to visualize representative configurations for a spanning set of points and check what part of the energy landscape it actually represents.

The most adequate sets $A$ and $B$ around Minima 1 and 2 were defined in the ML CV--${\sf sort}[c]$ space. The FFS and brute force rates $k_A$ found with the aid of ML  CV--${\sf sort}[c]$ agree well and follow the Arrhenius law as evident from Fig. \ref{fig:LJ8_FFS&BF}. The $(\mu_2,\mu_3)$ CVs bear a large error in the brute force estimate for $k_A$. (LDA2,LDA3) do not separate minima 1 and 2 well, leaving a low free energy barrier between the corresponding basins --see Fig. \ref{fig:LJ8_LDA23_confs}. This results in a large uncertainty in the FFS rate (Fig. \ref{fig:LJ8_FFS&BF}, row 3). Tables \ref{table:FFS_result_LJ8_mu2mu3}, \ref{table:FFS_result_LJ8_MLCV}, and \ref{table:FFS_result_LJ8_LDA23} show that the FFS and brute force rates for $k_A$ agree within one standard deviation for $(\mu_2,\mu_3)$ and ML CVs--${\sf sort}[c]$, and (LDA2,LDA3). The agreement between the FFS and brute force $k_B$ rates is worse for all these CVs. The agreement for the $\nu_{AB}$ rates is the best for ML CV--${\sf sort}[c]$ and the worst for (LDA2,LDA3).

Figs. \ref{fig:LJ8_LDA12_confs} and \ref{fig:LJ8_prob_density} (bottom right) show that (LDA1,LDA2) separate Minima 1 and 2 from the rest quite well. The agreement between the FFS and brute force rate $k_A$ for the escape from the region containing the images of Minima 1 and 2 to the region containing the images of the rest of the minima is quite good -- within 1 standard deviation for $\beta = 15$ and $20$ -- see Table \ref{table:FFS_result_LJ8_LDA12}. 

Our experience with the LDA-based CVs suggests that, if a representative set of local minima of the system is known, they can be used for the initial exploration of the system and generating the initial dataset.

\section{Conclusion}
\label{sec:conclusion}
We proposed a framework for learning CVs.  Its distinctive components are 
\begin{itemize}
    \item the use of a featurization that maps all states related by a symmetry group to a single point in the feature space and 
    \item the enforcement of the orthogonality condition~\eqref{eq:OC}.  
\end{itemize}

We proposed a feature map ${\sf sort}[c]$, the sorted vector of coordination numbers. This feature map and Algorithm \ref{alg:learnCVs} allowed us to obtain useful CVs for both case studies, LJ7 in 2D and LJ8 in 3D. 

For LJ7 in 2D, ML CVs--${\sf sort}[c]$ gave a free energy landscape separating the potential energy minima better than the standard set of CVs $(\mu_2,\mu_3)$~\cite{Tribell02010LJ7}. For LJ8 in 3D, ML CVs--${\sf sort}[c]$ separated the two lowest potential energy minima, minima 1 and 2, the best -- the free energy barrier between them is comparable to the potential energy barrier.

The CVs learned by Algorithm \ref{alg:learnCVs} with a feature map respecting the permutational symmetry are not expected to yield an accurate transition rate determined by the Transition Path Theory formula~\cite{EVE2006,EVE2010} \eqref{eq:nuABy}. However, the committor computed in the learned CV space is a good reaction coordinate for the FFS~\cite{Allen_2005,Allen_2009}.

An imperative for future work is to apply the proposed framework to more complex systems with permutational symmetry. A famous benchmark system is the Lennard-Jones-38 cluster~\cite{Wales1999} whose potential energy landscape has a double-funnel structure, a deep and narrow face-centered cubic funnel, containing a global minimum with the $O_h$ symmetry group of order 48, and a broader and shallower icosahedral funnel with the second lowest minimum with the point group $C_{5v}$ of order 10 at the bottom. It will be interesting to compare the CVs and the free energy landscapes with those obtained in Ref.~\cite{Tribello2013}.

 \section{Acknowledgements}
 This work was partially supported by AFOSR MURI grant FA9550-20-1-0397 and by 
  NSF REU grant DMS-2149913.

\begin{appendices}
\setcounter{equation}{0}
\renewcommand{\theequation}{\Alph{section}-\arabic{equation}}
    \setcounter{lemma}{0}
    \renewcommand{\thelemma}{\Alph{section}\arabic{lemma}}

%\section*{Appendix}

\section{Generating datasets for learning CVs}
\label{app:makedata}
First, a the well-tempered metadynamics algorithm~\cite{Dama2014WelltemperedMC} is used to deposit Gaussian bumps. The timestep is $5\cdot10^{-5}$. The bumps are deposited after every $N_{\sf skip} = 500$ steps. A total of $N_{\sf bumps} = 50,000$ bumps are deposited. The $k$th Gaussian bump is of the form
\begin{equation}
    G_k(x): = h_k\exp\left(-\frac{\|\xi(\phi(x)) - z_k\|^2}{2\sigma^2}\right),
\end{equation}
where $\xi(\phi(x))$ is the vector of CVs, $z_k$ is the vector of values of CVs at step $N_{\sf skip}k$. The width parameter of the bump is set to $\sigma = 0.02$ for LJ7 in 2D and $\sigma = 0.01$ for LJ8 in 3D.
The height of the $k$th Gaussian bump is
\begin{equation}
    h_k = h_0\exp\left(-\gamma^{-1}\sum_{j=0}^{k-1}G_j(x)\right),
\end{equation}
where $h_0 = 0.02$ for LJ7 in 2D and $h_0 = 0.01$ for LJ8 in 3D, and $\gamma = 1$ for both systems.
After the well-tempered metadynamics run is completed, the bounding box in the CV space is first set using the minimum and maximum values of the centers of the deposited bumps and then expanded by 5 to 10 percent in each direction. The resulting box in the CV space is meshed by $129\times129$ grid. The biasing potential
\begin{equation}
    \label{eq:Vbias}
    V_{\sf WTmetad}(x) = \sum_{k=0}^{N_{\sf bumps}-1}G_k(x)
\end{equation}
is approximated with the standard bicubic spline using the values of $V_{\sf WTmetad}(x)$ at the grid points and its gradients determined by finite differences at the grid points, because this results in a less bumpy biasing potential. The resulting bicubic spline and its gradient are evaluated much faster than the sum of the Gaussian bumps and their gradients. 

Next, a long trajectory is run in the biased potential, and, at each timestep, the grid cell in the CV space is identified. $N_{\sf save}$ vectors of atomic coordinates per grid cell are saved. Our experience shows that it suffices to take $N_{\sf save} = 1$. This collection of atomic coordinates data, $\{x_j\}_{j=1}^n$ is used at the input for Algorithm~\ref{alg:learnCVs}.

\section{Computing the free energy and the diffusion matrix}
\label{app:FE&DM}
While the free energy can be approximated directly from the biasing potential created by a long enough well-tempered metadynamics run~\cite{Dama2014WelltemperedMC}, we have found that it is hard to tune the metadynamics parameters $\sigma$, $h_0$, $\gamma$, $N_{\sf skip}$, and $N_{\sf bumps}$ to obtain satisfactory results. Therefore, we use a binning approach to calculate the free energy. We run a trajectory of $10^9$ steps with timestep $k\cdot10^{-5}$ in the well-tempered metadynamics-biased potential and bin the trajectory in the meshed CV space. Then the free energy grid point $(i,j)$ is found as
\begin{equation}
    \label{eq:FEcalculation}
    F(z_{i,j}) = -\beta^{-1}\log N_{i,j} + V_{\sf bias}(z_{i,j}),
\end{equation}
where $N_{i,j}$ is the number of visits of the long trajectory to the cell centered at $z_{i,j}$. Finally, to make the minimum value of the free energy equal to zero, we subtract $\min_{i,j} F(z_{i,j})$ from the found free energy function. 
%For a fast evaluation, we approximate the grid function with the standard bicubic spline.

The diffusion matrix $M(z)$ in the CV space is computed as described in~\cite{Maragliano}. To compute $M(z_{i,j})$ at the grid point $(i,j)$, we introduce a spring-force potential
\begin{equation}
\label{eq:spring}
    V_{\sf spring}(x) = \frac{\kappa}{2}\|\xi(x) - z_{i,j}\|^2
\end{equation}
with $\kappa = 500.0$, take the vector of atomic coordinates saved at the cell centered at $z_{i,j}$ as the initial configuration, and run a trajectory $N_{\sf DMsteps} = 10^6$ steps with timestep $10^{-5}$ reduced units. Note that this timestep is smaller than the one used for the well-tempered metadynamics and the calculation of the free energy due to the large spring constant in \eqref{eq:spring}.
Then the diffusion matrix at the mesh point $(i,j)$ is defined as
\begin{equation}
    M(z_{i,j}) = \frac{1}{N_{\sf DMsteps}}\sum_{m = 0}^{N_{\sf DMsteps-1}} \nabla \xi(x(t_m))\nabla\xi(x(t_m))^\top.
\end{equation}

\section{LDA for LJ8 in 3D}
\label{app:LDA}
The goal of the \emph{linear discriminant analysis} (LDA), also known as the \emph{multiple discriminant analysis} (MDA)~[Ref.\cite{DHS2001}, Section 3.8.3], is to find an optimal projection of high-dimensional data onto a low-dimensional space such that the data points from the same categories are mapped close to each other, while data points from different categories are mapped far away from each other. 

Let $X\in\mathbb{R}^{n\times d}$ be a dataset where $n$ is the number of data points and $d$ is their dimensionality. The data are subdivided into $c$ categories or classes. Let $\mathcal{I}_j$ be the set of indices of data in category $j$, $1\le j \le c$. Let $m_j$ be the mean of the data in category $j$,  and $m$ be the mean of the combined data of all categories. The \emph{within-class scatter matrix} $S_w$ quantifies the variation of data within the categories. It is defined as 
\begin{equation}
    \label{eq:Sw}
    S_w :=\sum_{j=1}^c S_j,\quad S_j: = \sum_{i\in\mathcal{I}_j}(x_i-m_j)^\top(x_i-m_j) \in \mathbb{R}^{d\times d},
\end{equation}
where $x_i$ is the $i$th row of $X$. The \emph{between-class scatter matrix} $S_b$ characterizes the spread of the centroids of different categories:
\begin{equation}
    \label{eq:Sb}
    S_b :=\sum_{j=1}^c |\mathcal{I}_j|(m_j-m)^\top(m_j-m) \in \mathbb{R}^{d\times d}.
\end{equation}
It is easy to check that the rank of $S_b$ is at most $c-1$. 

Let $W$ be the desired linear mapping from $\mathbb{R}^d$ to $\mathbb{R}^{d_1}$, $1\le d_1\le c-1$. The image of $X$ under the mapping $W$ is $XW\in\mathbb{R}^{n\times d_1}$.
The within- and between-class scatter matrices of the image of the dataset $X$ under the linear transformation $W$ are $\tilde{S}_w = W^\top S_wW$ and $\tilde{S}_b = W^\top S_b W$ respectively. The transformation $W$ is found by maximizing the objective function 
\begin{equation}
    \label{eq:Jw} J(w): = \frac{w^\top S_bw}{w^\top S_w w}, \quad w\in\mathbb{S}^d,
\end{equation}
where $\mathbb{S}^d$ is the $(d-1)$-dimensional sphere embedded into $\mathbb{R}^d$. It suffices to consider $w\in\mathbb{S}^d$ because the function $J(w)$ is invariant under scaling $w\mapsto \alpha w$, $\alpha\neq 0$. $J(w)$ is maximized by setting its gradient to zero, which is equivalent to solving the following generalized eigenvalue problem:
\begin{equation}
    \label{eq:Jweig}
    S_bw = J(w) S_w w.
\end{equation}
The $d_1$ dominant eigenvectors comprise the desired matrix $W$.

We obtained LDA-based CVs for LJ8 in 3D as follows. Starting at each of the eight local potential energy minima, we generated unbiased trajectory of $10^4$ steps each with the time step $5\cdot 10^{-5}$ and saved each tenth configuration. As a result, we obtained 1000 configuration in each of eight categories representing neighborhoods of the minima. We checked by visualizing the configurations in each category that they indeed are near the corresponding minima. For each of these 8000 configurations, we computed the sorted vectors of coordination numbers and used this $8000\times 8$ dataset as the input for the LDA. The dataset mapped into the span of three dominant generalized eigenvectors denoted by LDA1, LDA2, and LDA3 is shown in Fig. \ref{fig:LJ8_LDA}.

\section{Tables with FFS and brute force rates}
\label{app:tables}
Twenty level sets of the committor equispaced between $\epsilon_A$ and $1-\epsilon_B$, where $\epsilon_A$ and $\epsilon_B$ are case-dependent small parameters, were set up as milestones for the forward flux sampling (FFS), i.e., the parameter $m$ in Section \ref{sec:FFS} is equal to 19. The exit flux from the set $A$ was computed by running a long unbiased trajectory starting in $A$ until $N_{\sf exit} = 10^4$ exit events from $A$ were registered. Then the exit flux $\Phi_{A,0}$ is found as the ratio $N_{\sf exit}/T$, where $T$ is the elapsed time. The long trajectory never visited $B$ in our settings. Then the probabilities $\mathbb{P}(\lambda_{i+1}\mid\lambda_i)$, $i = 0,1,\ldots,m-1$, are calculated as described in Section  \ref{sec:FFS}. The trajectories from the interface $\lambda_i$ are launched until $N_{\sf exit} = 10^4$ number of crossings of the interface $\lambda_{i+1}$ is accumulated. 
The whole calculation is repeated $N_{\sf repeat} = 10$ times to obtain the standard deviation of the FFS estimate of $k_A$. The estimate of $k_B$ is obtained likewise.

The brute force estimates of $k_A$, $k_B$, $\nu_{AB}$, $\rho_A$, and $\rho_B$ are obtained by running a long unbiased trajectory consisting of $10^9$ time steps. The time step is $5\cdot10^{-5}$ for both LJ7 in 2D and LJ8 in 3D. 
The calculation is repeated $N_{\sf repeat} = 10$ times to obtain the standard deviations.

Tables \ref{table:FFS_result_LJ7_sortd2}, \ref{table:FFS_result_LJ7_MLCV_sortCNum}, and \ref{table:FFS_result_LJ7_mu2mu3} present the results for LJ7 in 2D. Tables \ref{table:FFS_result_LJ8_mu2mu3}, \ref{table:FFS_result_LJ8_MLCV}, \ref{table:FFS_result_LJ8_LDA23}, and \ref{table:FFS_result_LJ8_LDA12} report the results for LJ8.

%\begin{landscape}

\begin{table*}[ht]
\centering
\scriptsize
 % \footnotesize
% \small
\renewcommand{\arraystretch}{1.5} 
\begin{tabular}{|p{1cm}|p{2.5cm}|p{2.5cm}|p{2.5cm}|p{2.5cm}|p{2.5cm}|}
\hline
& \multicolumn{5}{c|}{\textbf{\color{blue} FFS rate with committtor in ML CVs, ${\sf sort}[d^2]$, LJ7 in 2D}}\\
\cline{2-6}
 \multirow{4}{*}{$\beta = 5$}  & $k_{A}$ & $k_{B}$ & $\rho_A$ & $\rho_B$ & $\nu_{AB}$ \\
 \cline{2-6}
 & $(6.96 \pm 0.59)\cdot10^{-2}$ & $(8.12 \pm 2.25)\cdot10^{-2}$ & $(5.39 \pm 0.83) \cdot 10^{-1}$ & $(4.61 \pm 0.83) \cdot 10^{-1}$ & $(3.75 \pm 0.66)\cdot10^{-2}$ \\
 \cline{2-6}
  & \multicolumn{5}{c|}{\textbf{Brute force rate, A, B defined in ML CVs, ${\sf sort}[d^2]$, LJ7 in 2D}} \\
 \cline{2-6}
 & $k_{A}$ & $k_{B}$ & $\rho_A$ & $\rho_B$ & $\nu_{AB}$ \\
\cline{2-6}
& $(6.76 \pm 0.09)\cdot10^{-2}$ & $(1.09 \pm 0.02)\cdot10^{-1}$ & $(6.16 \pm 0.05) \cdot 10^{-1}$ & $(3.84 \pm 0.053) \cdot 10^{-1}$ & $(4.17 \pm 0.04)\cdot10^{-2}$ \\
\hline
& \multicolumn{5}{c|}{\textbf{\color{blue} FFS rate with committtor in ML CVs, ${\sf sort}[d^2]$, LJ7 in 2D}}\\
\cline{2-6}
 \multirow{4}{*}{$\beta = 7$}  & $k_{A}$ & $k_{B}$ & $\rho_A$ & $\rho_B$ & $\nu_{AB}$ \\
 \cline{2-6}
 & $(3.85 \pm 0.39)\cdot10^{-3}$ & $(3.81 \pm 1.05)\cdot10^{-2}$ & $(9.08 \pm 0.94) \cdot 10^{-1}$ & $(9.18 \pm 9.39) \cdot 10^{-2}$ & $(3.50 \pm 0.50)\cdot10^{-3}$ \\
 \cline{2-6}
  & \multicolumn{5}{c|}{\textbf{Brute force rate, A, B defined in ML CVs, ${\sf sort}[d^2]$, LJ7 in 2D}} \\
 \cline{2-6}
 & $k_{A}$ & $k_{B}$ & $\rho_A$ & $\rho_B$ & $\nu_{AB}$ \\
\cline{2-6}
& $(4.03 \pm 0.22)\cdot10^{-3}$ & $(6.49 \pm 0.61)\cdot10^{-2}$ & $(9.42 \pm 0.05) \cdot 10^{-1}$ & $(5.84 \pm 0.45) \cdot 10^{-2}$ & $(3.79 \pm 0.21)\cdot10^{-3}$ \\
 \hline
 & \multicolumn{5}{c|}{\textbf{\color{blue} FFS rate with committtor in ML CVs, ${\sf sort}[d^2]$, LJ7 in 2D}}\\
\cline{2-6}
 \multirow{4}{*}{$\beta = 9$}  & $k_{A}$ & $k_{B}$ & $\rho_A$ & $\rho_B$ & $\nu_{AB}$ \\
 \cline{2-6}
 & $(2.10 \pm 0.39)\cdot10^{-4}$ & $(1.97 \pm 1.25)\cdot10^{-2}$ & $(9.90 \pm 1.85) \cdot 10^{-1}$ & $(1.05 \pm 18.5) \cdot 10^{-2}$ & $(2.07 \pm 0.55)\cdot10^{-4}$ \\
 \cline{2-6}
  & \multicolumn{5}{c|}{\textbf{Brute force rate, A, B defined in ML CVs, ${\sf sort}[d^2]$, LJ7 in 2D}} \\
 \cline{2-6}
 & $k_{A}$ & $k_{B}$ & $\rho_A$ & $\rho_B$ & $\nu_{AB}$ \\
\cline{2-6}
& $(2.13 \pm 0.72)\cdot10^{-4}$ & $(3.76 \pm 2.20)\cdot10^{-2}$ & $(9.94 \pm 0.03) \cdot 10^{-1}$ & $(5.65 \pm 2.71) \cdot 10^{-3}$ & $(2.12 \pm 0.71)\cdot10^{-4}$ \\
 \hline
\end{tabular}
\caption{LJ7 in 2D. Feature map: ${\sf sort}[d^2]$. The escape rates from $A$ and $B$, $k_A$ and $k_B$, respectively, the probabilities $\rho_A$ and $ \rho_B$ that at a random time $A$ was hit last rather than $B$ and vice versus, respectively, and the transition rates from $A$ to $B$, $\nu_{AB}$, obtained by FFS with the committor in the machine-learned CV with the feature map ${\sf sort}[d^2]$ and calculations described in Section~\ref{sec:rates&control}, are compared to those found by brute force unbiased all-atom simulations for LJ7 in 2D.}
\label{table:FFS_result_LJ7_sortd2}
\end{table*}

\begin{table}[ht]
\centering
\scriptsize
 % \footnotesize
% \small
\renewcommand{\arraystretch}{1.5} 
\begin{tabular}{|p{1cm}|p{2.5cm}|p{2.5cm}|p{2.5cm}|p{2.5cm}|p{2.5cm}|}
\hline
& \multicolumn{5}{c|}{\textbf{\color{blue} FFS rate with committtor in ML CVs, LJ7 in 2D}}\\
\cline{2-6}
 \multirow{4}{*}{$\beta = 5$}  & $k_{A}$ & $k_{B}$ & $\rho_A$ & $\rho_B$ & $\nu_{AB}$ \\
 \cline{2-6}
 & $(6.96 \pm 0.59)\cdot10^{-2}$ & $(8.12 \pm 2.25)\cdot10^{-2}$ & $(5.39 \pm 0.83) \cdot 10^{-1}$ & $(4.61 \pm 0.83) \cdot 10^{-1}$ & $(3.75 \pm 0.66)\cdot10^{-2}$ \\
 \cline{2-6}
  & \multicolumn{5}{c|}{\textbf{Brute force rate, A, B defined in ML CVs, LJ7 in 2D}} \\
 \cline{2-6}
 & $k_{A}$ & $k_{B}$ & $\rho_A$ & $\rho_B$ & $\nu_{AB}$ \\
\cline{2-6}
& $(6.76 \pm 0.09)\cdot10^{-2}$ & $(1.09 \pm 0.02)\cdot10^{-1}$ & $(6.16 \pm 0.05) \cdot 10^{-1}$ & $(3.84 \pm 0.053) \cdot 10^{-1}$ & $(4.17 \pm 0.04)\cdot10^{-2}$ \\
\hline
& \multicolumn{5}{c|}{\textbf{\color{blue} FFS rate with committtor in ML CVs, LJ7 in 2D}}\\
\cline{2-6}
 \multirow{4}{*}{$\beta = 7$}  & $k_{A}$ & $k_{B}$ & $\rho_A$ & $\rho_B$ & $\nu_{AB}$ \\
 \cline{2-6}
 & $(3.85 \pm 0.39)\cdot10^{-3}$ & $(3.81 \pm 1.05)\cdot10^{-2}$ & $(9.08 \pm 0.94) \cdot 10^{-1}$ & $(9.18 \pm 9.39) \cdot 10^{-2}$ & $(3.50 \pm 0.50)\cdot10^{-3}$ \\
 \cline{2-6}
  & \multicolumn{5}{c|}{\textbf{Brute force rate, A, B defined in ML CVs, LJ7 in 2D}} \\
 \cline{2-6}
 & $k_{A}$ & $k_{B}$ & $\rho_A$ & $\rho_B$ & $\nu_{AB}$ \\
\cline{2-6}
& $(4.03 \pm 0.22)\cdot10^{-3}$ & $(6.49 \pm 0.61)\cdot10^{-2}$ & $(9.42 \pm 0.05) \cdot 10^{-1}$ & $(5.84 \pm 0.45) \cdot 10^{-2}$ & $(3.79 \pm 0.21)\cdot10^{-3}$ \\
 \hline
 & \multicolumn{5}{c|}{\textbf{\color{blue} FFS rate with committtor in ML CVs, LJ7 in 2D}}\\
\cline{2-6}
 \multirow{4}{*}{$\beta = 9$}  & $k_{A}$ & $k_{B}$ & $\rho_A$ & $\rho_B$ & $\nu_{AB}$ \\
 \cline{2-6}
 & $(2.10 \pm 0.39)\cdot10^{-4}$ & $(1.97 \pm 1.25)\cdot10^{-2}$ & $(9.90 \pm 1.85) \cdot 10^{-1}$ & $(1.05 \pm 18.5) \cdot 10^{-2}$ & $(2.07 \pm 0.55)\cdot10^{-4}$ \\
 \cline{2-6}
  & \multicolumn{5}{c|}{\textbf{Brute force rate, A, B defined in ML CVs, LJ7 in 2D}} \\
 \cline{2-6}
 & $k_{A}$ & $k_{B}$ & $\rho_A$ & $\rho_B$ & $\nu_{AB}$ \\
\cline{2-6}
& $(2.13 \pm 0.72)\cdot10^{-4}$ & $(3.76 \pm 2.20)\cdot10^{-2}$ & $(9.94 \pm 0.03) \cdot 10^{-1}$ & $(5.65 \pm 2.71) \cdot 10^{-3}$ & $(2.12 \pm 0.71)\cdot10^{-4}$ \\
 \hline
\end{tabular}
\caption{Escape rates from $A$ and $B$ $(k_A, k_B)$, probabilities of hitting $\bar{A}$ and $\bar{B}$ $(\rho_A, \rho_B)$ and transition rates from $A$ to $B$ $(\nu_{AB})$ via FFS and brute force simulation for LJ7 in 2D. Regions of $A$ and $B$ are defined in ML learned CVs space and the reaction coordinate for FFS is the estimated committor function by NN.}
\label{table:FFS_result_LJ7_MLCV}
\end{table}

\begin{table}[h]
\centering
\scriptsize
 % \footnotesize
% \small
\renewcommand{\arraystretch}{1.5} 
\begin{tabular}{|p{1cm}|p{2.5cm}|p{2.5cm}|p{2.5cm}|p{2.5cm}|p{2.5cm}|}
\hline
& \multicolumn{5}{c|}{\textbf{\color{blue} FFS rate with committtor in ML CVs, ${\tt sort}[c]$, LJ7 in 2D}}\\
\cline{2-6}
 \multirow{4}{*}{$\beta = 5$}  & $k_{A}$ & $k_{B}$ & $\rho_A$ & $\rho_B$ & $\nu_{AB}$ \\
 \cline{2-6}
 & $(3.75 \pm 0.77)\cdot10^{-2}$ & $(7.73 \pm 0.97)\cdot10^{-2}$ & $(6.73 \pm 1.41) \cdot 10^{-1}$ & $(3.27 \pm 1.41) \cdot 10^{-1}$ & $(2.52 \pm 0.74)\cdot10^{-2}$ \\
 \cline{2-6}
  & \multicolumn{5}{c|}{\textbf{Brute force rate, A, B defined in ML CVs, ${\tt sort}[c]$, LJ7 in 2D}} \\
 \cline{2-6}
 & $k_{A}$ & $k_{B}$ & $\rho_A$ & $\rho_B$ & $\nu_{AB}$ \\
\cline{2-6}
& $(3.72 \pm 1.02)\cdot10^{-2}$ & $(9.65 \pm 0.34)\cdot10^{-2}$ & $(7.22 \pm 0.07) \cdot 10^{-1}$ & $(2.78 \pm 0.07) \cdot 10^{-1}$ & $(2.68 \pm 0.07)\cdot10^{-2}$ \\
\hline
& \multicolumn{5}{c|}{\textbf{\color{blue} FFS rate with committtor in ML CVs, ${\tt sort}[c]$, LJ7 in 2D}}\\
\cline{2-6}
 \multirow{4}{*}{$\beta = 7$}  & $k_{A}$ & $k_{B}$ & $\rho_A$ & $\rho_B$ & $\nu_{AB}$ \\
 \cline{2-6}
 & $(1.82 \pm 0.12)\cdot10^{-3}$ & $(4.14 \pm 0.74)\cdot10^{-2}$ & $(9.58 \pm 0.65) \cdot 10^{-1}$ & $(4.2 \pm 6.5) \cdot 10^{-2}$ & $(1.74 \pm 0.17)\cdot10^{-3}$ \\
 \cline{2-6}
  & \multicolumn{5}{c|}{\textbf{Brute force rate, A, B defined in ML CVs, ${\tt sort}[c]$, LJ7 in 2D}} \\
 \cline{2-6}
 & $k_{A}$ & $k_{B}$ & $\rho_A$ & $\rho_B$ & $\nu_{AB}$ \\
\cline{2-6}
& $(1.75 \pm 0.14)\cdot10^{-3}$ & $(5.81 \pm 0.84)\cdot10^{-2}$ & $(9.71 \pm 0.03) \cdot 10^{-1}$ & $(2.92 \pm 0.35) \cdot 10^{-2}$ & $(1.70 \pm 0.14)\cdot10^{-3}$ \\
 \hline
 & \multicolumn{5}{c|}{\textbf{\color{blue} FFS rate with committtor in ML CVs, ${\tt sort}[c]$, LJ7 in 2D}}\\
\cline{2-6}
 \multirow{4}{*}{$\beta = 9$}  & $k_{A}$ & $k_{B}$ & $\rho_A$ & $\rho_B$ & $\nu_{AB}$ \\
 \cline{2-6}
 & $(1.33 \pm 0.10)\cdot10^{-4}$ & $(1.48 \pm 0.20)\cdot10^{-2}$ & $(9.91 \pm 0.72) \cdot 10^{-1}$ & $(0.89 \pm 7.19) \cdot 10^{-2}$ & $(1.32 \pm 0.14)\cdot10^{-4}$ \\
 \cline{2-6}
  & \multicolumn{5}{c|}{\textbf{Brute force rate, A, B defined in ML CVs, ${\tt sort}[c]$, LJ7 in 2D}} \\
 \cline{2-6}
 & $k_{A}$ & $k_{B}$ & $\rho_A$ & $\rho_B$ & $\nu_{AB}$ \\
\cline{2-6}
& $(1.49 \pm 0.75)\cdot10^{-4}$ & $(3.01 \pm 2.79)\cdot10^{-2}$ & $(9.95 \pm 0.38) \cdot 10^{-1}$ & $(4.92 \pm 3.83) \cdot 10^{-3}$ & $(1.48 \pm 0.75)\cdot10^{-4}$ \\
 \hline
\end{tabular}
\caption{Escape rates from $A$ and $B$ $(k_A, k_B)$, probabilities of hitting $\bar{A}$ and $\bar{B}$ $(\rho_A, \rho_B)$ and transition rates from $A$ to $B$ $(\nu_{AB})$ via FFS and brute force simulation for LJ7 in 2D. Regions of $A$ and $B$ are defined in ML learned CVs space, ${\tt sort}[c]$ as input, and the reaction coordinate for FFS is the estimated committor function by NN.}
\label{table:FFS_result_LJ7_MLCV_sortCNum}
\end{table} 

\begin{table}[h]
\centering
\scriptsize
 % \footnotesize
% \small
\renewcommand{\arraystretch}{1.5} 
\begin{tabular}{|p{1cm}|p{2.5cm}|p{2.5cm}|p{2.5cm}|p{2.5cm}|p{2.5cm}|}
\hline
& \multicolumn{5}{c|}{\textbf{\color{blue} FFS rate with committtor in $(\mu_2, \mu_3)$, LJ7 in 2D}}\\
\cline{2-6}
 \multirow{4}{*}{$\beta = 5$}  & $k_{A}$ & $k_{B}$ & $\rho_A$ & $\rho_B$ & $\nu_{AB}$ \\
 \cline{2-6}
 & $(3.25 \pm 0.48)\cdot10^{-2}$ & $(7.83 \pm 1.03)\cdot10^{-2}$ & $ 0.707 \pm 0.108 $ & $0.293 \pm 0.108$ & $(2.3 \pm 0.5)\cdot10^{-2}$ \\
 \cline{2-6}
  & \multicolumn{5}{c|}{\textbf{Brute force rate, A, B defined in $(\mu_2, \mu_3)$, LJ7 in 2D}} \\
 \cline{2-6}
 & $k_{A}$ & $k_{B}$ & $\rho_A$ & $\rho_B$ & $\nu_{AB}$ \\
\cline{2-6}
& $(3.46\pm0.07)\cdot10^{-2}$ & $(9.76\pm0.28)\cdot10^{-2}$ & $0.738 \pm 0.006$ & $0.262 \pm 0.006$ & $(2.56 \pm 0.04)\cdot10^{-2}$ \\
 \hline
 & \multicolumn{5}{c|}{\textbf{\color{blue} FFS rate with committtor in $(\mu_2, \mu_3)$, LJ7 in 2D}}\\
\cline{2-6}
 \multirow{4}{*}{$\beta = 7$}  & $k_{A}$ & $k_{B}$ & $\rho_A$ & $\rho_B$ & $\nu_{AB}$ \\
 \cline{2-6}
 & $(1.42 \pm 0.23)\cdot10^{-3}$ & $(4.33 \pm 0.95)\cdot10^{-2}$ & $ (9.68 \pm 1.57) \cdot 10^{-1} $ & $(0.32 \pm 1.57) \cdot 10^{-1}$ & $(1.37 \pm 0.32)\cdot10^{-3}$ \\
 \cline{2-6}
  & \multicolumn{5}{c|}{\textbf{Brute force rate, A, B defined in $(\mu_2, \mu_3)$, LJ7 in 2D}} \\
 \cline{2-6}
 & $k_{A}$ & $k_{B}$ & $\rho_A$ & $\rho_B$ & $\nu_{AB}$ \\
\cline{2-6}
& $( 1.41 \pm 0.17)\cdot10^{-3}$ & $(4.75 \pm 0.83)\cdot10^{-2}$ & $(9.71 \pm 0.04) \cdot 10^{-1}$ & $(2.8 \pm 0.36) \cdot 10^{-2}$ & $(1.37 \pm 0.17)\cdot10^{-3}$ \\
 \hline
  & \multicolumn{5}{c|}{\textbf{\color{blue} FFS rate with committtor in $(\mu_2, \mu_3)$, LJ7 in 2D}}\\
\cline{2-6}
 \multirow{4}{*}{$\beta = 9$}  & $k_{A}$ & $k_{B}$ & $\rho_A$ & $\rho_B$ & $\nu_{AB}$ \\
 \cline{2-6}
 & $(4.58 \pm 0.27)\cdot10^{-5}$ & $(1.23 \pm 0.48)\cdot10^{-2}$ & $ (9.96 \pm 0.60) \cdot 10^{-1} $ & $(3.7 \pm 59.6) \cdot 10^{-3}$ & $(4.56 \pm 0.39)\cdot10^{-5}$ \\
 \cline{2-6}
  & \multicolumn{5}{c|}{\textbf{Brute force rate, A, B defined in $(\mu_2, \mu_3)$, LJ7 in 2D}} \\
 \cline{2-6}
 & $k_{A}$ & $k_{B}$ & $\rho_A$ & $\rho_B$ & $\nu_{AB}$ \\
\cline{2-6}
& $(6.22 \pm 4.78)\cdot10^{-5}$ & $(1.92 \pm 2.46)\cdot10^{-2}$ & $(9.97 \pm 0.03) \cdot 10^{-1}$ & $(3.23 \pm 3.32) \cdot 10^{-3}$ & $(6.20 \pm 4.76)\cdot10^{-5}$ \\
 \hline
\end{tabular}
\caption{Escape rates from $A$ and $B$ $(k_A, k_B)$, probabilities of hitting $\bar{A}$ and $\bar{B}$ $(\rho_A, \rho_B)$ and transition rates from $A$ to $B$ $(\nu_{AB})$ via FFS and brute force simulation for LJ7 in 2D. Regions of $A$ and $B$ are defined in $(\mu_2, \mu_3)$ space and the reaction coordinate for FFS is the estimated committor function by NN.}
\label{table:FFS_result_LJ7_mu2mu3}
\end{table}

\begin{table}[h!]
\centering
\scriptsize
 % \footnotesize
% \small
\renewcommand{\arraystretch}{1.5} 
\begin{tabular}{|p{1cm}|p{2.5cm}|p{2.5cm}|p{2.5cm}|p{2.5cm}|p{2.5cm}|}
\hline
 & \multicolumn{5}{c|}{\color{blue}\textbf{FFS rate with committtor in $(\mu_2, \mu_3)$}} \\
 \cline{2-6}
 \multirow{4}{*}{$\beta = 10$} & $k_{A}$ & $k_{B}$ & $\rho_A$ & $\rho_B$ & $\nu_{AB}$ \\
\cline{2-6}
 & $(4.9 \pm 0.2) \cdot 10^{-1}$ & $(3.7 \pm 1.8)\cdot 10^{-2}$ & $ 0.07 \pm 0.45$ & $0.93 \pm 0.45$ & $(3.5 \pm 1.6)\cdot 10^{-2}$ \\
\cline{2-6}
& \multicolumn{5}{c|}{\textbf{Brute force rate, A, B defined in $(\mu_2, \mu_3)$}}\\
\cline{2-6}
 & $k_{A}$ & $k_{B}$ & $\rho_A$ & $\rho_B$ & $\nu_{AB}$ \\
 \cline{2-6}
 & $(5.3 \pm 0.2) \cdot10^{-1}$ & $ (9.3 \pm 0.3) \cdot 10^{-2}$ & $(1.5 \pm 0.05) \cdot 10^{-1}$ & $(8.5 \pm 0.05) \cdot 10^{-1}$ & $(8.0 \pm 0.3) \cdot 10^{-2}$ \\
 \hline
  & \multicolumn{5}{c|}{\color{blue}\textbf{FFS rate with committtor in $(\mu_2, \mu_3)$}} \\
\cline{2-6}
\multirow{4}{*}{$\beta = 15$} & $k_{A}$ & $k_{B}$ & $\rho_A$ & $\rho_B$ & $\nu_{AB}$ \\
\cline{2-6}
 & $(1.4 \pm 0.1)\cdot 10^{-1}$ & $(2.3\pm 1.3)\cdot 10^{-3}$ & $0.02 \pm 0.009$ & $0.98 \pm 0.009$ & $(2.2 \pm 1.3) \cdot 10^{-3}$ \\
\cline{2-6}
& \multicolumn{5}{c|}{\textbf{Brute force rate, A, B defined in $(\mu_2, \mu_3)$}}\\
\cline{2-6}
 & $k_{A}$ & $k_{B}$ & $\rho_A$ & $\rho_B$ & $\nu_{AB}$ \\
 \cline{2-6}
 & $(1.5 \pm 0.2) \cdot10^{-1}$ & $(1.6 \pm 0.2) \cdot 10^{-2}$ & $(1.0 \pm 0.1) \cdot 10^{-1}$ & $(9.0 \pm 0.1) \cdot 10^{-1}$ & $(1.5 \pm 0.2) \cdot 10^{-2}$ \\
 \hline
   & \multicolumn{5}{c|}{\textbf{\color{blue}FFS rate with committtor in $(\mu_2, \mu_3)$}} \\
\cline{2-6}
\multirow{4}{*}{$\beta = 20$} & $k_{A}$ & $k_{B}$ & $\rho_A$ & $\rho_B$ & $\nu_{AB}$ \\
\cline{2-6}
 & $(2.5 \pm 0.1)\cdot 10^{-2}$ & $(1.3 \pm 0.9)\cdot 10^{-4}$ & $0.05 \pm 0.003$ & $0.995 \pm 0.003$ & $(1.3 \pm 0.8) \cdot 10^{-4}$ \\
\cline{2-6}
& \multicolumn{5}{c|}{\textbf{Brute force rate, A, B defined in $(\mu_2, \mu_3)$}}\\
\cline{2-6}
 & $k_{A}$ & $k_{B}$ & $\rho_A$ & $\rho_B$ & $\nu_{AB}$ \\
 \cline{2-6}
 & $(2.6 \pm 2.3) \cdot10^{-2}$ & $(2.0 \pm 1.2) \cdot 10^{-3}$ & $(7.0 \pm 4.4) \cdot 10^{-2}$ & $(9.3 \pm 0.4) \cdot 10^{-1}$ & $(1.8 \pm 1.1) \cdot 10^{-3}$ \\
 \hline
\end{tabular}
\caption{FFS and brute force rate results for LJ8 in $(\mu_2, \mu_3)$.}
\label{table:FFS_result_LJ8_mu2mu3}
\end{table}

\begin{table}[h!]
\centering
\scriptsize
 % \footnotesize
% \small
\renewcommand{\arraystretch}{1.5} 
\begin{tabular}{|p{1cm}|p{2.5cm}|p{2.5cm}|p{2.5cm}|p{2.5cm}|p{2.5cm}|}
\hline
 & \multicolumn{5}{c|}{\color{blue}\textbf{FFS rate with committtor in ML CVs, LJ8}} \\
 \cline{2-6}
 \multirow{4}{*}{$\beta = 10$} & $k_{A}$ & $k_{B}$ & $\rho_A$ & $\rho_B$ & $\nu_{AB}$ \\
\cline{2-6}
 & $(4.2 \pm 0.3)\cdot 10^{-1}$ & $(3.4 \pm 2.6)\cdot 10^{-2}$ & $ (7.5 \pm 5.25)\cdot 10^{-2} $ & $ 0.93 \pm 0.05$ & $(3.1 \pm 2.2)\cdot 10^{-2}$ \\
\cline{2-6}
& \multicolumn{5}{c|}{\textbf{Brute force rate, A, B defined in ML CVs, LJ8}}\\
\cline{2-6}
 & $k_{A}$ & $k_{B}$ & $\rho_A$ & $\rho_B$ & $\nu_{AB}$ \\
 \cline{2-6}
 & $(4.2 \pm 0.1) \cdot 10^{-1}$ & $(7.2 \pm 0.2) \cdot 10^{-2}$ & $(1.5 \pm 0.03)\cdot 10^{-1}$ & $(8.5 \pm 0.03)\cdot 10^{-1}$ & $(6.1 \pm 0.1)\cdot 10^{-2}$ \\
 \hline
  & \multicolumn{5}{c|}{\color{blue}\textbf{FFS rate with committtor in ML CVs, LJ8}} \\
\cline{2-6}
\multirow{4}{*}{$\beta = 15$} & $k_{A}$ & $k_{B}$ & $\rho_A$ & $\rho_B$ & $\nu_{AB}$ \\
\cline{2-6}
 & $(3.7 \pm 0.2)\cdot 10^{-2}$ & $(2.6 \pm 0.3)\cdot 10^{-3}$ & $(6.72 \pm 6.75) \cdot 10^{-2}$ & $0.93 \pm 0.07$ & $(2.5 \pm 2.4) \cdot 10^{-3}$ \\
\cline{2-6}
& \multicolumn{5}{c|}{\textbf{Brute force rate, A, B defined in ML CVs, LJ8}}\\
\cline{2-6}
 & $k_{A}$ & $k_{B}$ & $\rho_A$ & $\rho_B$ & $\nu_{AB}$ \\
 \cline{2-6}
 & $ (3.5 \pm 0.5) \cdot 10^{-2}$ & $(4.6 \pm 0.4) \cdot 10^{-3}$ & $(1.2 \pm 0.1)\cdot 10^{-1}$ & $(8.8 \pm 0.1)\cdot 10^{-1}$ & $ (4.1 \pm 0.3)\cdot 10^{-3}$ \\
 \hline
   & \multicolumn{5}{c|}{\textbf{\color{blue}FFS rate with committtor in ML CVs, LJ8}} \\
\cline{2-6}
\multirow{4}{*}{$\beta = 20$} & $k_{A}$ & $k_{B}$ & $\rho_A$ & $\rho_B$ & $\nu_{AB}$ \\
\cline{2-6}
 & $(6.0 \pm 0.5)\cdot 10^{-3}$ & $(1.5 \pm 1.2)\cdot 10^{-4}$ & $(2.5 \pm 1.9)\cdot 10^{-2}$ & $ 0.97 \pm 0.02$ & $ (1.5 \pm 1.2) \cdot 10^{-4}$ \\
\cline{2-6}
& \multicolumn{5}{c|}{\textbf{Brute force rate, A, B defined in ML CVs, LJ8}}\\
\cline{2-6}
 & $k_{A}$ & $k_{B}$ & $\rho_A$ & $\rho_B$ & $\nu_{AB}$ \\
 \cline{2-6}
 & $ (6.0 \pm 4.7) \cdot10^{-3}$ & $(5.2 \pm 2.8) \cdot 10^{-4}$ & $(7.9 \pm 4.6) \cdot 10^{-2}$ & $(9.2 \pm 0.5) \cdot 10^{-1}$ & $(4.8 \pm 2.5)\cdot 10^{-4}$ \\
 \hline
\end{tabular}
\caption{FFS rate results for LJ8. The reaction coordinate is the committor function computed in ML CVs space.}
\label{table:FFS_result_LJ8_MLCV}
\end{table} 

\begin{table}[h]
\centering
\scriptsize
 % \footnotesize
% \small
\renewcommand{\arraystretch}{1.5} 
\begin{tabular}{|p{1cm}|p{2.5cm}|p{2.5cm}|p{2.5cm}|p{2.5cm}|p{2.5cm}|}
\hline
 & \multicolumn{5}{c|}{\color{blue}\textbf{FFS rate with committtor in LDA2-3}} \\
 \cline{2-6}
 \multirow{4}{*}{$\beta = 10$} & $k_{A}$ & $k_{B}$ & $\rho_A$ & $\rho_B$ & $\nu_{AB}$ \\
\cline{2-6}
 & $2.03 \pm 0.13$ & $(2.7 \pm 1.8)\cdot 10^{-2}$ & $ (1.32 \pm 0.88) \cdot 10^{-2}$ & $0.99 \pm 0.0088$ & $(2.7 \pm 1.8) \cdot 10^{-2}$ \\
\cline{2-6}
& \multicolumn{5}{c|}{\textbf{Brute force rate, A, B defined in LDA2-3}}\\
\cline{2-6}
 & $k_{A}$ & $k_{B}$ & $\rho_A$ & $\rho_B$ & $\nu_{AB}$ \\
 \cline{2-6}
 & $ 2.6 \pm 0.1$ & $(3.3 \pm 0.1) \cdot 10^{-1}$ & $(1.1 \pm 0.04) \cdot 10^{-1}$ & $ (8.9 \pm 0.04) \cdot 10^{-1}$ & $(5.8 \pm 0.4) \cdot 10^{-2}$ \\
 \hline
  & \multicolumn{5}{c|}{\color{blue}\textbf{FFS rate with committtor in LDA2-3}} \\
\cline{2-6}
\multirow{4}{*}{$\beta = 15$} & $k_{A}$ & $k_{B}$ & $\rho_A$ & $\rho_B$ & $\nu_{AB}$ \\
\cline{2-6}
 & $(6.5 \pm 0.4) \cdot 10^{-1}$ & $(8.2 \pm 13.4)\cdot 10^{-4}$ & $(1.32 \pm 0.88) \cdot 10^{-2}$ & $0.987 \pm 0.009$ & $(2.7 \pm 1.8) \cdot 10^{-2}$ \\
\cline{2-6}
& \multicolumn{5}{c|}{\textbf{Brute force rate, A, B defined in LDA2-3}}\\
\cline{2-6}
 & $k_{A}$ & $k_{B}$ & $\rho_A$ & $\rho_B$ & $\nu_{AB}$ \\
 \cline{2-6}
 & $(8.0 \pm 1.3) \cdot10^{-1}$ & $(8.2 \pm 0.9) \cdot 10^{-2}$ & $(9.3 \pm 1.0) \cdot 10^{-2}$ & $(9.1 \pm 0.1) \cdot 10^{-1}$ & $(7.4 \pm 0.8) \cdot 10^{-2}$ \\
 \hline
   & \multicolumn{5}{c|}{\textbf{\color{blue}FFS rate with committtor in LDA2-3}} \\
\cline{2-6}
\multirow{4}{*}{$\beta = 20$} & $k_{A}$ & $k_{B}$ & $\rho_A$ & $\rho_B$ & $\nu_{AB}$ \\
\cline{2-6}
 & $(1.8 \pm 0.1) \cdot 10^{-1}$ & $(1.1 \pm 2.6)\cdot 10^{-13}$ & $0.0$ & $ 1.0$ & $(1.1 \pm 2.6) \cdot 10^{-13}$ \\
\cline{2-6}
& \multicolumn{5}{c|}{\textbf{Brute force rate, A, B defined in LDA2-3}}\\
\cline{2-6}
 & $k_{A}$ & $k_{B}$ & $\rho_A$ & $\rho_B$ & $\nu_{AB}$ \\
 \cline{2-6}
 & $ (2.1 \pm 1.9) \cdot10^{-1}$ & $(1.6 \pm 1.0) \cdot 10^{-3}$ & $(6.8 \pm 4.3) \cdot 10^{-2}$ & $(9.3 \pm 0.4) \cdot 10^{-1}$ & $(1.5 \pm 0.9) \cdot 10^{-2}$ \\
 \hline
\end{tabular}
\caption{FFS and brute force rate results for LJ8. The reaction coordinate is the committor function computed in (LDA2, LDA3) space.}
\label{table:FFS_result_LJ8_LDA23}
\end{table}

\begin{table}[h!]
\centering
% \scriptsize
\small
\renewcommand{\arraystretch}{1.5} 
\begin{tabular}{|p{0.7cm}|p{7cm}|p{7cm}|}
\hline
$\beta$ & Brute force $k_A$, A, B defined in (LDA1, LDA2) & FFS $k_A$, committtor in (LDA1, LDA2) \\
\hline 
10 & $(1.6 \pm 0.03) \times 10^{-1}$ & $(1.3 \pm 0.1) \times 10^{-1}$\\
\hline
15 & $(2.4\pm 0.2)\times 10^{-3}$ & $(2.5 \pm 0.6) \times 10^{-3}$\\
\hline
20 & $(1.2 \pm 1.2) \times 10^{-4}$  & $(1.3 \pm 0.6) \times 10^{-4}$ \\
\hline
\end{tabular}
\caption{Transition rate computed via FFS and brute force simulation with reaction coordinates the committor function in LDA1 and LDA2.}
\label{table:FFS_result_LJ8_LDA12}
\end{table}

%\end{landscape}
\end{appendices}

\section*{Supplementary Information}

  \setcounter{table}{0}
  \renewcommand{\thetable}{S\arabic{table}}%
  \setcounter{figure}{0}
  \renewcommand{\thefigure}{S\arabic{figure}}%
\renewcommand{\thepage}{S\arabic{page}}
 \setcounter{page}{1}
 
\renewcommand{\theequation}{S\arabic{equation}}
  \setcounter{equation}{0}
\renewcommand{\thesection}{S\arabic{section}}
\setcounter{section}{0}
%}

%
%\author{Jiaxin Yuan}
%\email{jyuan98@umd.edu}
%\affiliation{Department of Mathematics, University of Maryland, College Park, MD 20742, USA}
%\author{Shashank Sule}
%\email{ssule25@umd.edu}
%\affiliation{Department of Mathematics, University of Maryland, College Park, MD 20742, USA}
%\author{Yeuk Yin Lam}
%\email{lam00185@umn.edu}
%\affiliation{School of Mathematics, University of Minnesota, Twin Cities, MN 55455, USA}
%\author{Maria Cameron}
%\email{mariakc@umd.edu}
%\affiliation{Department of Mathematics, University of Maryland, College Park, MD 20742, USA}

The codes developed for this work are available in the following GitHub repositories.

The C codes for running the well-tempered metadynamics algorithm~\cite{Dama2014WelltemperedMC}, computing  free energy and diffusion matrix, as well as forward flux sampling~\cite{Allen_2009}, brute force sampling, and sampling transition trajectories are available at \\  \href{https://github.com/mar1akc/LJ7in2D_LJ8in3D_learningCVs/blob/main/README.md}{https://github.com/mar1akc/LJ7in2D\_LJ8in3D\_learningCVs/blob/main/\\README.md}

    The Python codes for learning CVs are available at \\   \href{https://github.com/margotyjx/OrthogonalityCVLearning/tree/main/LrCV_permsym}{https://github.com/margotyjx/OrthogonalityCVLearning/tree/main/\\LrCV\_permsym}

%\tableofcontents

\section{Neural network architectures and training details}
\label{SIsec:architectures}
Architectures and training details for all neural networks used are provided in Table \ref{table:NN_details}. The activation function {\sf ELU}, the exponential linear unit, is defined as
\begin{equation}
    \label{ELUdef}
    \sf{ELU}(x) = 
        \begin{cases}
            x & \text{if } x > 0\\
%            \alpha (\exp(x) - 1) & \text{if } x \leq 0.
            \exp(x) - 1 & \text{if } x \leq 0       \end{cases}.
\end{equation}

\section{The form of collective variables}
\label{SI}
The CVs $\xi(\phi(x))$ are represented by the encoder part, $\xi(\phi(x)) = {\sf Encoder}(\phi(x))$, of the autoencoder neural network of the form ${\sf Decoder}\odot{\sf Encoder}(\phi(x))$, where $\phi(x)$ is the feature map. We used two feature maps, $\phi(x) = {\sf sort}[d^2](x)$, sorted pairwise distance squared, and $\phi(x) = {\sf sort}[c](x)$, sorted vector of coordination numbers. The encoder neural networks used have two or three hidden layers -- see Table \ref{table:NN_details}. The ML CVs for LJ7 are of the form
\begin{equation}
    \label{CV2hl}
    \xi(x) = A_3{\sf ELU}\left(A_2{\sf ELU}\left(A_1\phi(x) + b_1\right)+b_2\right)+b_3.
\end{equation}
The ML CV for LJ8 is of the form
\begin{equation}
    \label{CV3hl}
    \xi(x) = A_4{\sf ELU}\left(A_3{\sf ELU}\left(A_2{\sf ELU}\left(A_1\phi(x) + b_1\right)+b_2\right)+b_3\right)+b_4.
\end{equation}

The sizes of the matrices $A_j$ and bias vectors $b_j$ for are given in Table  \ref{table:NN_details}. For example, for LJ7 with the feature map ${\sf sort}[c]$, $A_1$ is $30\times 7$, $A_2$ is $30\times 30$, and $A_3$ is $2\times 30$. 

%\begin{landscape}
\begin{table}[htbp]
\centering
% \scriptsize
 \tiny
% \small
\renewcommand{\arraystretch}{2.5} 
\begin{tabular}{p{1.2cm} p{1.2cm} p{3.0cm} p{2.2cm} p{1.3cm} p{1.5cm} p{1.5cm} }
\hline
\textbf{Test case} & \textbf{CVs} & \textbf{Model} & \textbf{\# neurons} & \textbf{Activation} & \textbf{Optimizer, learning rate} & \textbf{Training epochs}\\
\hline
 \multirow{9}{*}{LJ7 in 2D} & \multirow{6}{*}{\shortstack{MLCV, \\ ${\tt sort}[d^2]$}} & Diffusion net & $[21,45,25,3]$ & ELU() & Adam, 1e-3 & 1000\\
\cline{3-7}
 &  & Manifold & $[3, 45, 30, 1]$ & Tanh() & Adam, 1e-3 & 1000 \\
\cline{3-7}
 &  & Encoder & $[21,30, 30, 2]$ & ELU() & Adam, 1e-3 & 1500 \\
\cline{3-7}
 &  & Decoder & $[2,30, 30, 3]$ & ELU() & Adam, 1e-3 & 1500 \\
\cline{3-7}
 &  & Committor, $\beta = 5$ & $[2,10,10,10,1]$ & ReLU() & Adam, 1e-3 & 1000 \\
 \cline{3-7}
 &  & Committor, $\beta = 7$ & $[2,25,25,25,1]$ & ReLU() & Adam, 1e-3 & 1000 \\
 \cline{3-7}
 &  & Committor, $\beta = 9$ & $[2,40,40,40,1]$ & ReLU() & Adam, 1e-3 & 1000 \\
\cline{2-7}
& \multirow{4}{*}{\shortstack{MLCV, \\ ${\tt sort}[c]$}} & Diffusion net & $[7,45,30,25,3]$ & ELU() & Adam, 5e-3 & 2000 \\
\cline{3-7}
 &  & Manifold & $[3,45,30,1]$ & Tanh() & Adam, 1e-3 & 1000\\
\cline{3-7}
 &  & Encoder & $[7,30, 30, 2]$ & ELU() & Adam, 1e-3 & 1500\\
 \cline{3-7}
&  & Decoder & $[2,30, 30, 3]$ & ELU() & Adam, 1e-3 & 1500\\
\cline{3-7}
  &  & {\shortstack{Committor, \\ $\beta = 5, 7, 9$}}  & $[2,40,40,40,1]$ & ReLU() & Adam, 1e-3 & 1000 \\
\cline{2-7}
& \multirow{3}{*}{\shortstack{$(\mu_2, \mu_3)$}} & Committor, $\beta = 5$ & $[2,10,10,1]$ & Tanh() & Adam, 1e-3 & 1000 \\
 \cline{3-7}
 &  & Committor, $\beta = 7,9$ & $[2,40,40,40,1]$ & ReLU() & Adam, 1e-3 & 1000 \\
 \hline
\multirow{8}{*}{LJ8 in 3D} & \multirow{4}{*}{\shortstack{MLCV, \\ ${\tt sort}[c]$}} & Diffusion net & $[8,45,45,3]$ & ELU() & Adam, 5e-3 & 1000 \\
\cline{3-7}
 &  & Manifold & $[3,45,30,1]$ & Tanh() & Adam, 1e-3 & 1000\\
\cline{3-7}
 &  & Encoder & $[7,45,30, 25, 2]$ & ELU() & Adam, 5e-3 & 1500\\
\cline{3-7}
&  & Decoder & $[2,25,30, 45, 3]$ & ELU() & Adam, 5e-3 & 1500\\
\cline{3-7}
 &  & Committor, $\beta = 10,15$ & $[2,10,10,1]$ & ReLU() & Adam, 1e-3 & 1000 \\
 \cline{3-7}
 &  & Committor, $\beta = 20$ & $[2,25,25,1]$ & ReLU() & Adam, 1e-3 & 1000 \\
\cline{2-7}
& \multirow{1}{*}{\shortstack{LDA1-2}} & Committor, $\beta = 10,15$ & $[2,25,25,1]$ & ReLU() & Adam, 1e-3 & 1000 \\
\cline{2-7}
& \multirow{2}{*}{\shortstack{LDA2-3}} & Committor, $\beta = 10,15$ & $[2,25,25,1]$ & ReLU() & Adam, 1e-3 & 1000 \\
\cline{3-7}
 &  & Committor, $\beta = 20$ & $[2,40,40,1]$ & ReLU() & Adam, 1e-3 & 1000 \\
\cline{2-7}& \multirow{1}{*}{\shortstack{$(\mu_2, \mu_3)$}} & {\shortstack{Committor, \\ $\beta = 10, 15, 20$}} & $[2,25, 25,1]$ & ReLU() & Adam, 1e-3 & 1000 \\
 \hline
\end{tabular}
\caption{Neural network architectures and training configurations for the feed-forward networks used. The “\# neurons” column specifies the network structure in the format [input dimension, hidden layer dimensions, output dimension]. For example, [2,10,10,1] denotes a network with two hidden layers, each containing 10 neurons, taking input of dimension 2 and producing output of dimension 1. 
% The “lr” column indicates the learning rate used during training.
}
\label{table:NN_details}
\end{table}

%\end{landscape}

%%%%%%%%%%%%%%%%%%
\section{Computing the Jacobians of neural networks}
\label{sec: CV form}
Python's package PyTorch provides tools for automatic differentiation. In our C codes, we wrote routines for taking gradients of neural networks. The gradients of ML CVs were required in codes running metadynamics, free energy estimation, and diffusion tensor estimation. The gradient of the committor neural network is needed in the code for generating transition trajectories using stochastic control. 
Consider a neural network function with $l$ hidden layers of the form
\begin{equation}
    \label{NNfunc}
    \mathcal{N}(x) = A_{l+1}\sigma\left(A_l\sigma\ldots\left(\sigma\left(A_1f(x) + b_1\right)\ldots\right)+b_1\right) + b_{l+1},
\end{equation}
where $\sigma(\cdot)$ is an activation function acting element-wise and $f(x)$ is a nonlinear function. If $\mathcal{N}(x)$ is an ML CV, then $f(x)$ is the feature map $\phi(x)$. If $\mathcal{N}(x)$ is the committor, then $f(x)$ is the CV $\xi(x)$.
The gradient $\nabla\mathcal{N}(x)$ can be calculated as follows.
Let 
\begin{equation}
    \label{Jf}
    J_0(x) = \left[\frac{\partial f}{\partial{x}}\right]
\end{equation}
be the Jacobian matrix of the function $f(x)$. Set
\begin{equation}
    \label{w1}
    w_1: = A_1 f(x) + b_1.
\end{equation}
Then
\begin{align*}
    {\sf for}~j &= 1, 2,\ldots, l\\
   & D_j : = {\sf diag}\left[\sigma'(w_j)\right], ~\text{a diagonal matrix with $\sigma'(w_j)$ along its diagonal} \\
   & J_j: = D_j A_j J_{l-1}\\ 
   & w_{j+1}: = A_{j+1}\sigma(w_j) + b_{j+1}\\
    {\sf end}~&{\sf for}
\end{align*}
The last iteration of the for-loop yields $\mathcal{N}(x) = w_{l+1}$. The Jacobian matrix of $\mathcal{N}(x)$ is given by
\begin{equation}
    \label{JacNN}
    \left[\frac{\partial \mathcal{N}}{\partial x}\right] = A_{l+1} J_l.
\end{equation}

\section{Computing the derivatives of feature maps ${\sf sort}[d^2]$ and ${\sf sort}[c]$}
In this section, we detail how to compute the Jacobian \eqref{Jf} of $\phi(x)={\sf sort}[d^2] $ and $\phi(x) = {\sf sort}[c]$.
Since we used the feature map $\phi(x)={\sf sort}[d^2] $ only for LJ7, we describe only this case. The vector of pairwise distances squared is given by
\begin{align}
    [d^2] = \{d^2_{i,j}: = (x_i-x_j)^2 + (y_i-y_j)^2~|
     1\le i,j\le 7,~i < j\}\in\mathbb{R}^{21}.
\end{align}
Here, ${\tt sort}[d^2]\in\mathbb{R}^{21}$ is the vector of pairwise distances sorted in increasing order. The activation function {\sf ELU} is given by \eqref{ELUdef}. 
The Jacobian $J_{{\sf sort}[d^2]}\in\mathbb{R}^{21\times 14}$ of ${\sf sort}[d^2]$ with respect to $x_i$ and $y_i$, $1\le i\le 7$ is computed as
\begin{align}
    J_{{\sf sort}[d^2]}(j,k):=\frac{\partial [{\sf sort}[d^2]]_j }{\partial x_k} &= \begin{cases}2(x_k - x_l),& {\sf sort}[d^2]_j = d_{k,l}^2,\\
    -2(x_i - x_k),& {\sf sort}[d^2]_j = d_{i,k}^2,\\
    0,&{\rm otherwise},\end{cases}\\
     J_{{\sf sort}[d^2]}(j,k+7):=\frac{\partial [{\sf sort}[d^2]]_j }{\partial y_k} &= \begin{cases}2(y_k - y_l),& {\tt sort}[d^2]_j = d_{k,l}^2,\\
    -2(y_i - y_k),& {\sf sort}[d^2]_j = d_{i,k}^2,\\
    0,&{\rm otherwise}.\end{cases}
\end{align}

Now we write out the formulas for the feature map ${\sf sort}[c]$ for LJ8 in 3D. The modification of these formulas for LJ7 in 2D is simple, and we do not include it here.
The vector of coordination numbers is
\begin{equation}
    \label{CNum_vec}
    c_i = \sum_{\substack{j=1\\j\neq i}}^{N_{\sf atoms}}\frac{1 -\left( \frac{r_{i,j}}{1.5}\right)^8}{1 -\left( \frac{r_{i,j}}{1.5}\right)^{16}} =: \sum_{\substack{j=1\\j\neq i}}^{N_{\sf atoms}} g(r_{i,j}) ,\quad i = 1,2,\ldots,N_{\sf atoms}.
\end{equation}
The derivatives of the coordination numbers are:
\begin{align}
    a_{i,j}&:= \frac{-4 r_{i,j}^6 + 8 g(r_{i,j})r^{14}}{1 -\left( \frac{r_{i,j}}{1.5}\right)^{16}}, \quad i,j = 1,\ldots,8, ~~j\neq i\label{auxgder}\\
    \frac{\partial c_i}{\partial x_i} &= \sum_{j=1}^8 2a_{i,j}\frac{x_i - x_j}{1.5^2},\quad i = 1,\ldots,8\label{cix_i}\\
   \frac{\partial c_i}{\partial y_i} &= \sum_{j=1}^8 2a_{i,j}\frac{y_i - y_j}{1.5^2}\quad i = 1,\ldots,8\label{ciy_i}\\
   \frac{\partial c_i}{\partial z_i} &= \sum_{j=1}^8 2a_{i,j}\frac{z_i - z_j}{1.5^2}\quad i = 1,\ldots,8\label{ciz_i}\\
    \frac{\partial c_i}{\partial x_j} &= -2a_{i,j}\frac{x_i - x_j}{1.5^2},\quad i,j = 1,\ldots,8, ~~j\neq i\label{cix_j}\\
    \frac{\partial c_i}{\partial y_j} &= -2a_{i,j}\frac{y_i - y_j}{1.5^2},\quad i,j = 1,\ldots,8, ~~j\neq i\label{ciy_j}\\
    \frac{\partial c_i}{\partial z_j} &= -2a_{i,j}\frac{z_i - z_j}{1.5^2},\quad i,j = 1,\ldots,8, ~~j\neq i\label{ciz_j}.   
\end{align}

The Jacobian $J_{{\sf sort}[c]}\in\mathbb{R}^{8\times 24}$ of ${\sf sort}[c]$ with respect to $x_i$, $y_i$, and $z_i$, $1\le i\le 8$, is computed as
\begin{align}
    J_{{\sf sort}[c]}(j,k)&:=\frac{\partial [{\sf sort}[c]]_j }{\partial x_k} = \frac{\partial c_l}{\partial x_k},\\
     J_{{\sf sort}[c]}(j,k+8)&:=\frac{\partial [{\sf sort}[c]]_j }{\partial y_k} = \frac{\partial c_l}{\partial y_k}\\
     J_{{\sf sort}[c]}(j,k+16)&:=\frac{\partial [{\sf sort}[c]]_j }{\partial z_k} = \frac{\partial c_l}{\partial z_k}\\
    {\rm where}&~~l = {\sf isort}(j),~~\text{i.e. $l$ is the $j$th smallest coordination number. } 
\end{align}

\section{Definitions of the domain $\Omega$ and  regions $A$ and $B$ in the CV space}
\label{sec:omega_A_B}
Once the free energy and the diffusion matrix in a given pair of CVs are computed, we select the domain $\Omega$ in the CV space and the metastable regions $A$ and $B$ near the images of the potential energy minima, transitions between which we want to study. Typically, the domain $\Omega$ is defined as a high enough level set of the free energy. If the images of the potential energy minima of interest lie in nice "bowls" of the free energy landscape, we define $A$ and $B$ as appropriate free energy level sets. Otherwise, $A$ and/or $B$ are defined as ellipses surrounding the images of the potential energy minima. 

The free energy and the diffusion matrix slightly change with the temperature. Therefore, we need to adjust our choice of $\Omega$, $A$, and $B$ for each value of $\beta$. Below we list the definitions of $\Omega$, $A$, and $B$ for all cases studied in this work.

\subsection*{Lennard-Jones 7 in 2D}
\begin{figure}[htbp]
    \centering
    \includegraphics[width=0.9\textwidth]{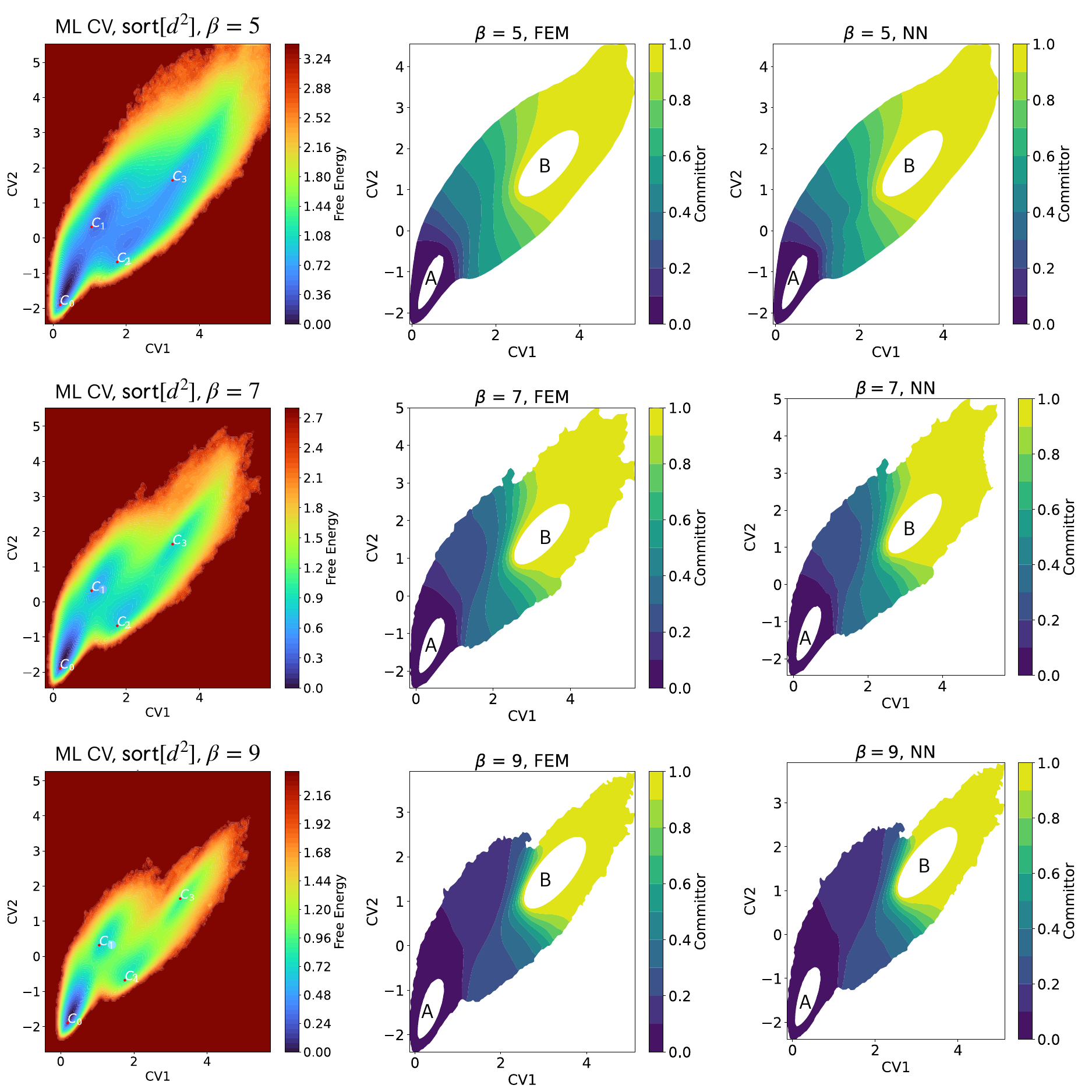}
    \caption{LJ7 in 2D. The free energy landscapes in ML CV with the feature map ${\sf sort}[d^2]$ (column 1), the chosen sets $A$ and $B$ and the committor computed by the finite element method (FEM) (column 2), and its approximation with a neural network (column 3) at $\beta = 5$ (row 1), $\beta = 7$ (row 2), and $\beta = 9$ (row 3). }
    \label{fig:LJ7_sortd2_FE_FEM_NN}
\end{figure}

\begin{figure}[htbp]
    \centering
    \includegraphics[width=0.9\textwidth]{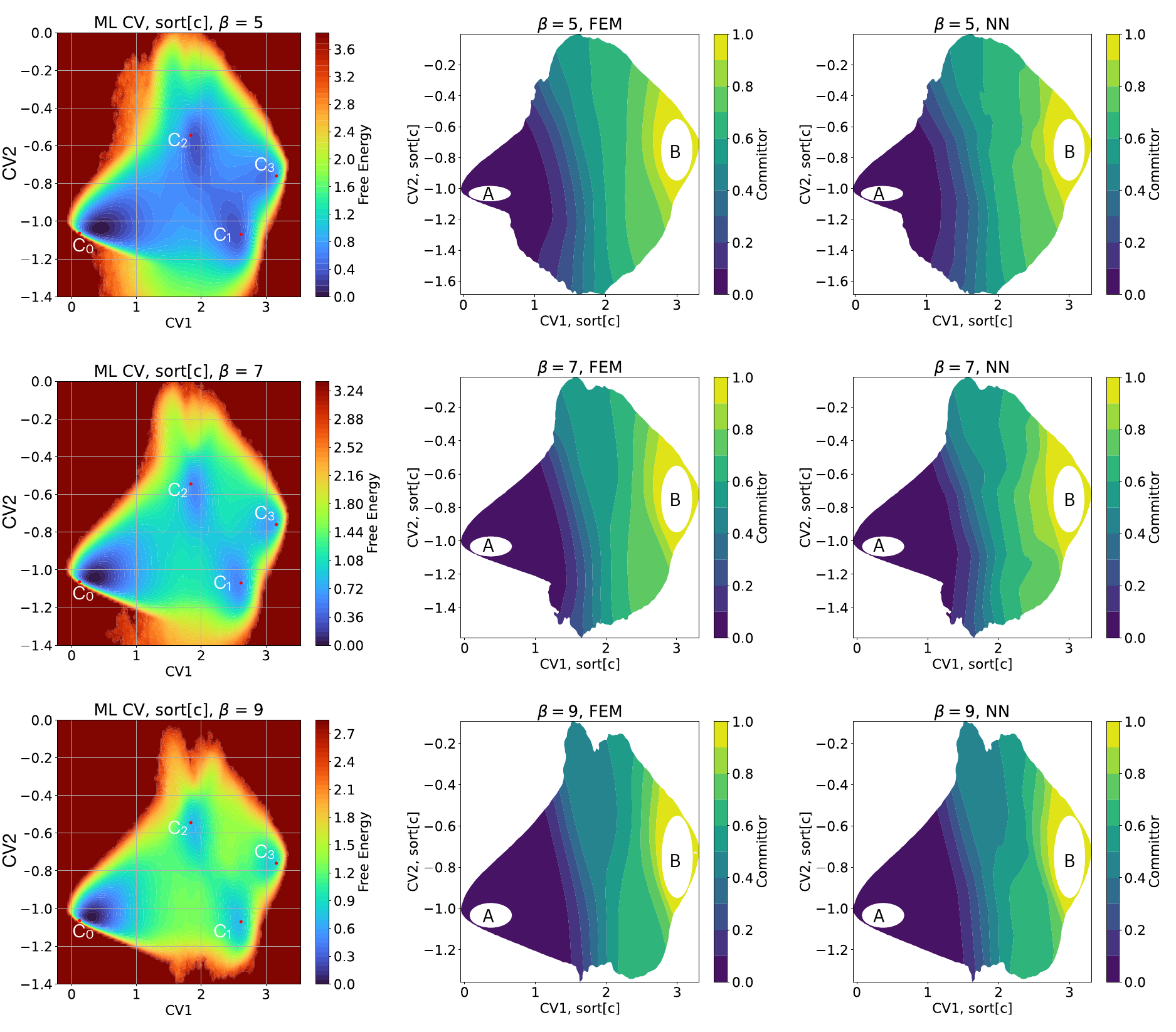}
    \caption{LJ7 in 2D. The free energy landscapes in ML CV with the feature map ${\sf sort}[c]$ (column 1), the chosen sets $A$ and $B$ and the committor computed by the finite element method (FEM) (column 2), and its approximation with a neural network (column 3) at $\beta = 5$ (row 1), $\beta = 7$ (row 2), and $\beta = 9$ (row 3). }
    \label{fig:LJ7_sortCNum_FE_FEM_NN}
\end{figure}

Figure \ref{fig:LJ7_sortd2_FE_FEM_NN} displays the free energy in ML CVs with the feature map ${\sf sort}[d^2]$, the chosen sets $A$ and $B$, the committor found by the finite element method (the FEM committor), and its approximation by a neural network (the NN committor) at $\beta = 5$, $7$, and $9$. Figures \ref{fig:LJ7_sortCNum_FE_FEM_NN}, and  \ref{fig:LJ7_mu2mu3_FE_FEM_NN} do the same for ML CVs with the feature map ${\sf sort}[c]$ and CVs $(\mu_2,\mu_3)$ respectively.  
% \begin{enumerate}
     \subsubsection*{ ML CV with the feature map ${\sf sort}[d^2]$}
    \begin{itemize}
        \item $\beta = 5: ~\Omega = \{z\in\mathbb{R}^2~|~F(z) \le 2\}, ~A = \{z \in \mathbb{R}^2 | F(z) \leq 0.4 \}$
        \item $\beta = 7,9: ~\Omega = \{z\in\mathbb{R}^2~|~F(z) \le 2.75\}, ~A = \{z \in \mathbb{R}^2 | F(z) \leq 0.6 \}$
    \end{itemize}
    For $\beta = 5,~7$, and $9$, the region $B$ is defined as the ellipse:
    \begin{align*}
        B = \Big\{ (x,y) \in \mathbb{R}^2 ~\vline~ &\frac{[0.65*(x - 3.26) + 0.76 * (y - 1.64)]^2}{1.008^2} +\\
        &\frac{[0.76*(x - 3.26) - 0.65 * (y - 1.64)]^2}{0.4^2} \leq 1 \Big\}.
    \end{align*}

    \subsubsection*{ ML CV with the feature map ${\sf sort}[c]$}
    \begin{itemize}
        \item $\beta = 5: ~\Omega = \{z\in\mathbb{R}^2~|~F(z) \le 2.27\}$
        \item $\beta = 7: ~\Omega = \{z\in\mathbb{R}^2~|~F(z) \le 2.65\}$
        \item $\beta = 9: ~\Omega = \{z\in\mathbb{R}^2~|~F(z) \le 2.82\}$.
    \end{itemize}
    For $\beta = 5$, $7$, and $9$, the regions $A$ and $B$ are defined as  ellipses
    \begin{align*}
    A&:=\left\{ (x,y) \in \mathbb{R}^2 ~\vline~  \frac{(x - 0.37)^2}{0.3^2} + \frac{(y + 1.64)^2}{0.06^2} \leq 1\right\}, \\
    B&:=\left\{ (x,y) \in \mathbb{R}^2 ~\vline~  \frac{(x - 2.99)^2}{0.3^2} + \frac{(y + 0.75)^2}{0.06^2} \leq 1\right\}.
\end{align*}

    \subsubsection*{ CVs $(\mu_2, \mu_3)$}
    \begin{itemize}
        \item $\beta = 5: ~\Omega = \{z\in\mathbb{R}^2~|~F(z) \le 4.73\}$,\\  $A = \{z \in \mathbb{R}^2 | F(z) \leq 0.7 \}, ~B = \{z \in \mathbb{R}^2 | F(z) \leq 1.05 \}$
        \item $\beta = 7: ~\Omega = \{z\in\mathbb{R}^2~|~F(z) \le 4.73\}$, \\ $A = \{z \in \mathbb{R}^2 | F(z) \leq 0.7 \}, ~B = \{z \in \mathbb{R}^2 | F(z) \leq 1.2 \}$
        \item $\beta = 9: ~\Omega = \{z\in\mathbb{R}^2~|~F(z) \le 4.73\}$, \\ $A = \{z \in \mathbb{R}^2 | F(z) \leq 0.8 \}, ~B = \{z \in \mathbb{R}^2 | F(z) \leq 1.4 \}$
    \end{itemize}
% \end{enumerate}

\begin{figure}[htbp]
    \centering
    \includegraphics[width=0.9\textwidth]{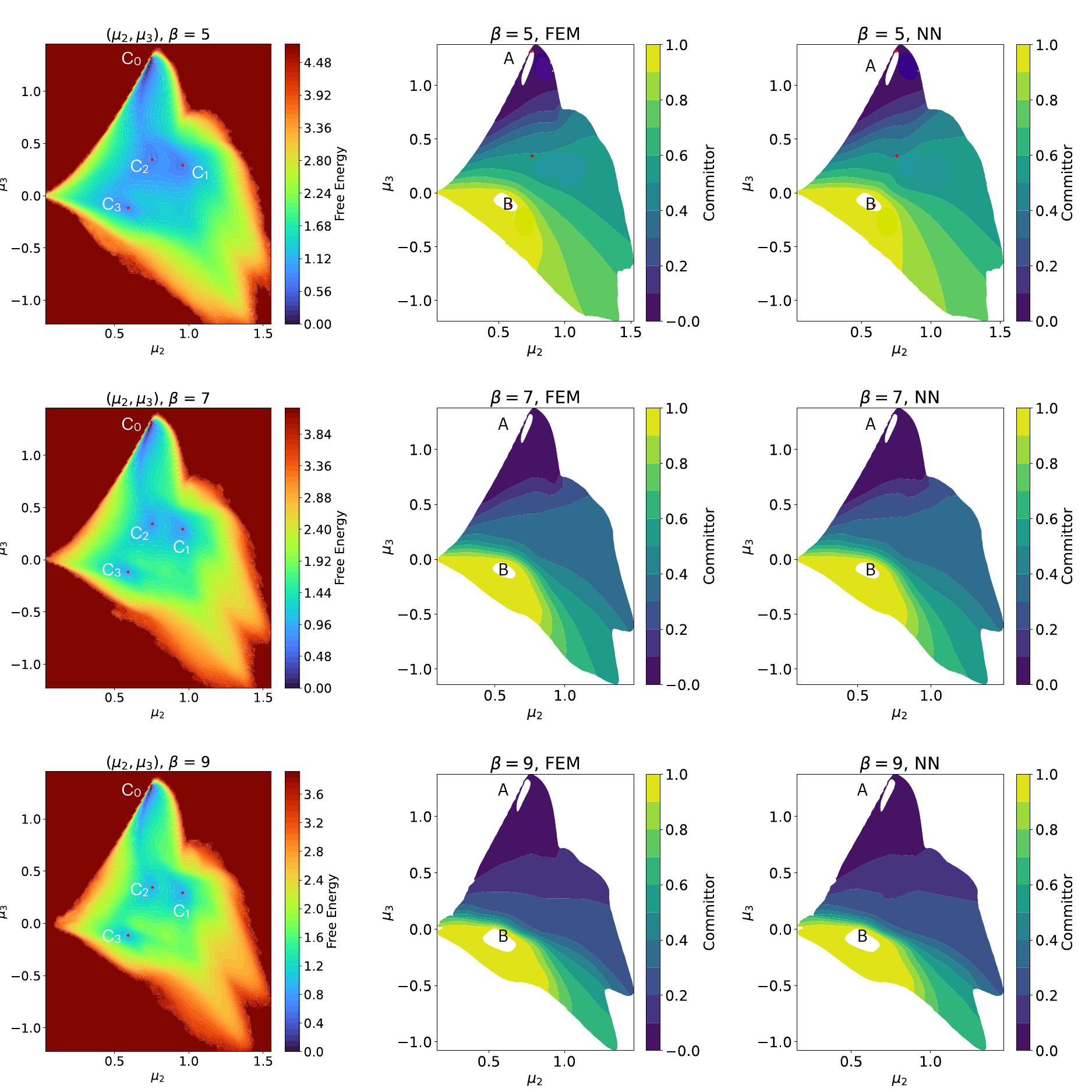}
    \caption{LJ7 in 2D. The free energy landscapes in CVs $(\mu_2,\mu_3)$ (column 1), the chosen sets $A$ and $B$ and the committor computed by the finite element method (FEM) (column 2), and its approximation with a neural network (column 3) at $\beta = 5$ (row 1), $\beta = 7$ (row 2), and $\beta = 9$ (row 3). }
    \label{fig:LJ7_mu2mu3_FE_FEM_NN}
\end{figure}

\subsection*{Lennard-Jones 8 in 3D}
\begin{figure}[htbp]
    \centering
    \includegraphics[width=0.9\textwidth]{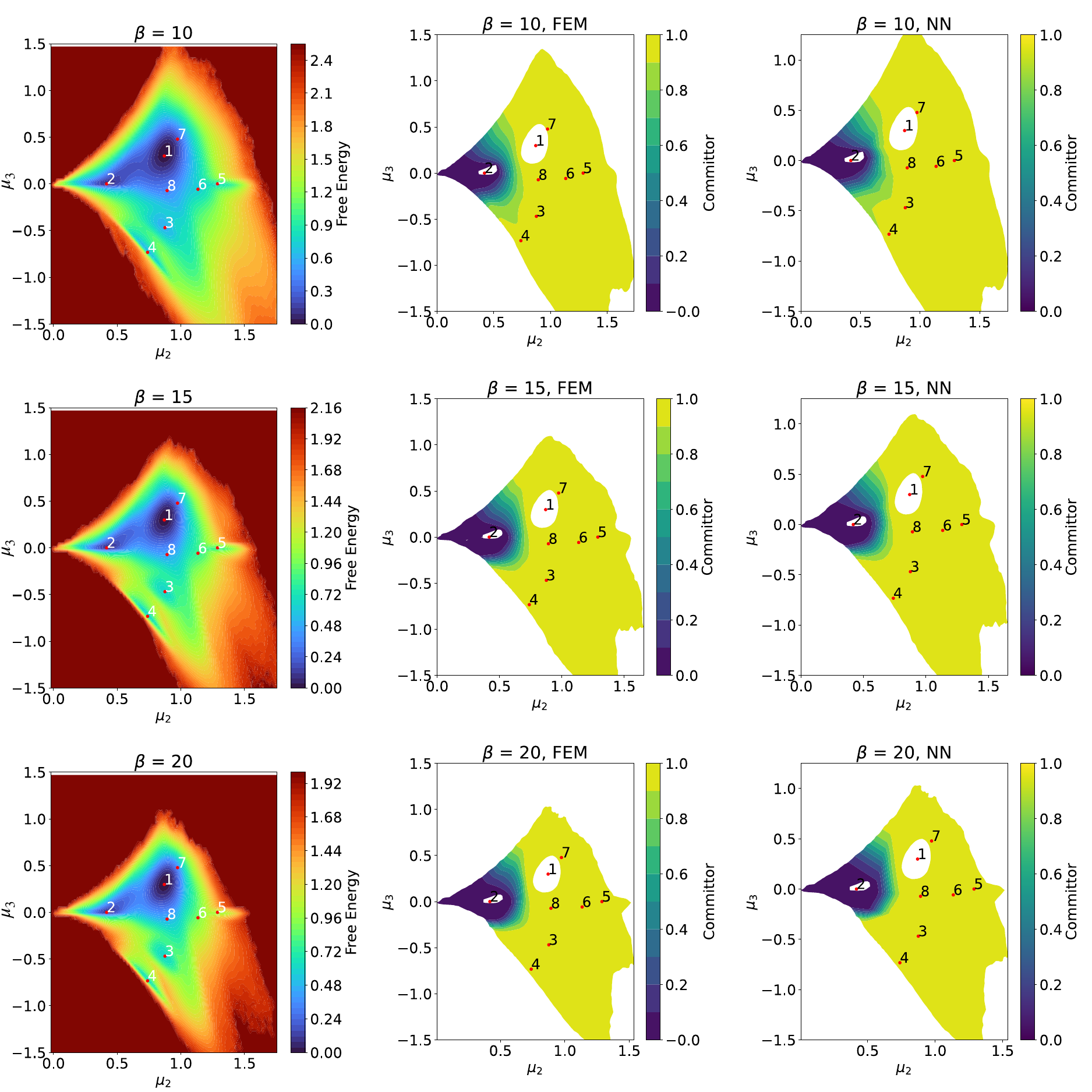}
    \caption{LJ8 in 3D. The free energy landscapes in CVs $(\mu_2,\mu_3)$ (column 1), the chosen sets $A$ and $B$ and the committor computed by the finite element method (FEM) (column 2), and its approximation with a neural network (column 3) at $\beta = 10$ (row 1), $\beta = 15$ (row 2), and $\beta = 20$ (row 3). }
    \label{fig:LJ8_mu2mu3_FE_FEM_NN}
\end{figure}

The free energy in $(\mu_2,\mu_3)$ for LJ8 in 3D, the chosen sets $A$ and $B$, and the FEM and NN committors are shown in Fig. \ref{fig:LJ8_mu2mu3_FE_FEM_NN}. The data in ML CV with the feature map ${\sf sort}[c]$, and CVs (LDA2,LDA3), and (LDA1,LDA2) are displayed in Figs. \ref{fig:LJ8_sortCNum_FE_FEM_NN}, \ref{fig:LJ8_LDA23_FE_FEM_NN}, and \ref{fig:LJ8_LDA12_FE_FEM_NN} respectively.

% \begin{enumerate}
     \subsubsection*{ CVs  $(\mu_2, \mu_3)$}

    \begin{itemize}
        \item $\beta = 10: ~\Omega = \{z\in\mathbb{R}^2~|~F(z) \le 1.85\},$
        \item $\beta = 15: ~\Omega = \{z\in\mathbb{R}^2~|~F(z) \le 1.78\},$ %$, ~A = \{z \in \mathbb{R}^2 | F(z) \leq 0.7 \}, ~B = \{z \in \mathbb{R}^2 | F(z) \leq 1.2 \}$ 
        \item $\beta = 20: ~\Omega = \{z\in\mathbb{R}^2~|~F(z) \le 1.68\}.$ %$, ~A = \{z \in \mathbb{R}^2 | F(z) \leq 0.8 \}, ~B = \{z \in \mathbb{R}^2 | F(z) \leq 1.4 \}$
    \end{itemize}
    For all values of $\beta$, the regions $A$ and $B$ are defined as 
    \begin{equation*}
        A = \text{connected component of }\{z \in \mathbb{R}^2 | F(z) \leq 0.2\} \text{ that contains minimum 2}, 
    \end{equation*}
    \begin{equation*}
        B = \text{connected component of }\{z \in \mathbb{R}^2 | F(z) \leq 0.2\} \text{ that contains minimum 1}.
    \end{equation*}

    \subsubsection*{ ML CV with the feature map ${\sf sort}[c]$}
    The domain in this case is defined as   
       \begin{equation}
       \label{Fdef}
\Omega = \{z\in\mathbb{R}^2~|~F(z) \leq F_{\Omega}\}.
\end{equation}
The regions $A$ and $B$ are ellipses of the form \begin{equation}
\label{eldef}
    \frac{[(x-x_0)*v_{x_0} + (y - y_0)*v_{y_0}]^2}{r_{x_0}^2} +\frac{[(x-x_0)*v_{y_0} - (y - y_0)*v_{x_0}]^2}{r_{y_0}^2} \leq 1. 
\end{equation}
The values of the parameters $F_\Omega$ in \eqref{Fdef} and  $x_0$, $y_0$, $v_{x_0}$, $v_{y_0}$, $r_{x_0}$, and $r_{y_0}$ in \eqref{eldef} are listed in  Table~~\ref{table:LJ8_MLCV_committor_detail}.
\begin{table*}[h]
    \centering
    \tiny
    \renewcommand{\arraystretch}{2.5} 
    \begin{tabular}{|c|c|c|c|c|c|c|c|c|c|c|c|c|c|}
    \hline
    & \multicolumn{6}{c|}{Region $B$} & \multicolumn{6}{c|}{Region $A$} & Domain $\Omega$ \\
    \hline
   $\beta$ & $x_0$ & $y_0$ & $v_{x_0}$ & $v_{y_0}$ & $r_{x_0}$ & $r_{y_0}$  & $x_0$ & $y_0$ & $v_{x_0}$ & $v_{y_0}$ & $r_{x_0}$ & $r_{y_0}$ & $F_{\Omega}$\\
    \hline
    $\beta = 10$
     & 0.54 & -0.9 & -0.15 & 1.0 & 0.25 & 0.05 & \multirow{3}{*}{1.49} & \multirow{3}{*}{1.11} & \multirow{3}{*}{0.15} & \multirow{3}{*}{1} & \multirow{3}{*}{0.6} & \multirow{3}{*}{0.04} & 2.1 \\
    
     \cline{1-7} \cline{14-14}
    $\beta = 15$
     & 0.55 & -0.93 & -0.2 & 1.0 & 0.25 & 0.06 & & & & & &  & 1.6\\
    \cline{1-7} \cline{14-14}
    $\beta = 20$
     & 0.565 & -0.95 & -0.23 & 1.0 & 0.27 & 0.04 & & & & & & & 1.3\\
    \hline
    \end{tabular}
    \caption{The parameters for the definitions of the domain $\Omega$ and the regions $A$ and $B$ in the ML CV space with the feature map ${\sf sort}[c]$ for LJ8 in 3D. See equations \eqref{Fdef} and \eqref{eldef}.  }
    \label{table:LJ8_MLCV_committor_detail}
\end{table*}

\begin{figure}[htbp]
    \centering
    \includegraphics[width=0.9\textwidth]{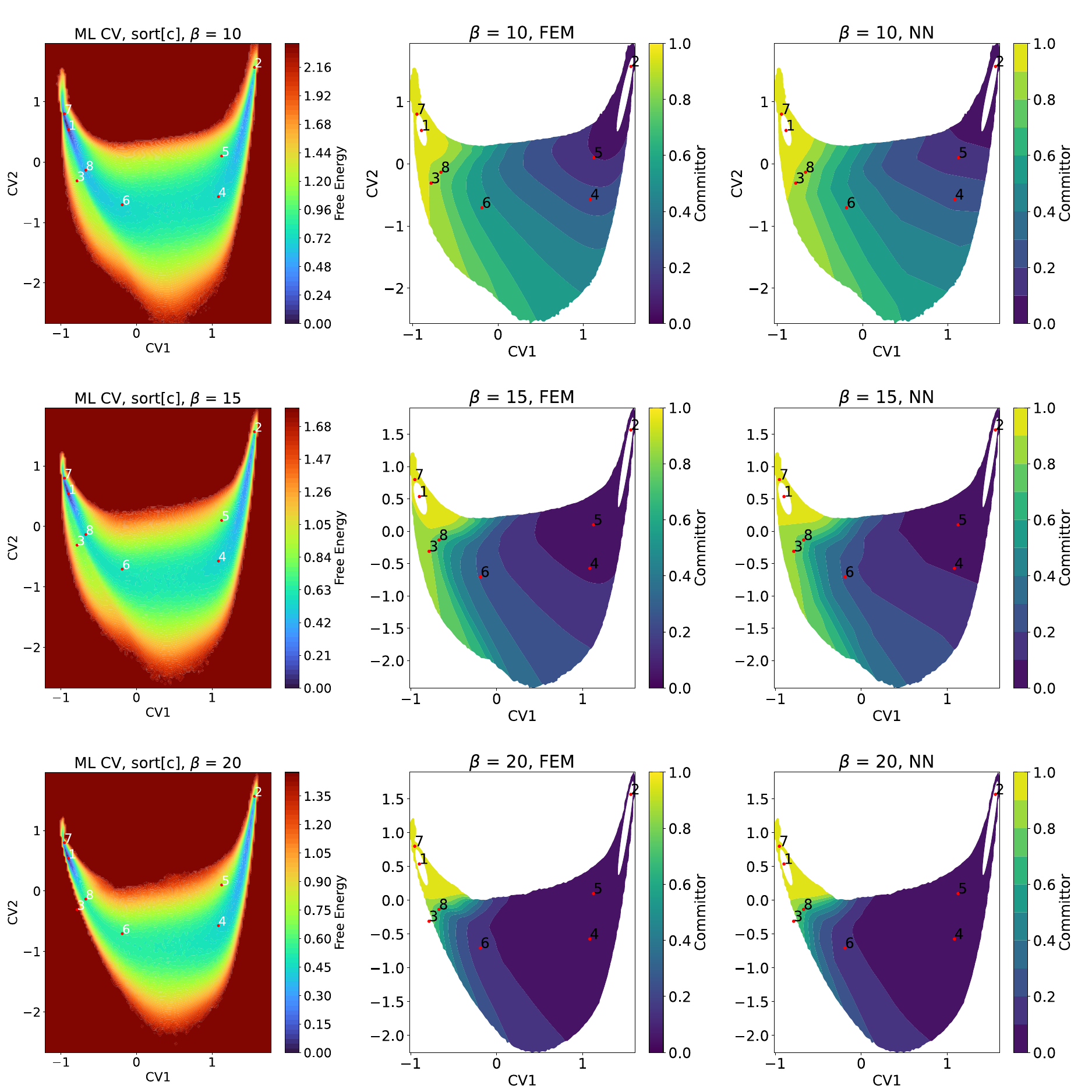}
    \caption{LJ8 in 3D. The free energy landscapes in  ML CV with the feature map ${\sf sort}[c]$ (column 1), the chosen sets $A$ and $B$ and the committor computed by the finite element method (FEM) (column 2), and its approximation with a neural network (column 3) at $\beta = 10$ (row 1), $\beta = 15$ (row 2), and $\beta = 20$ (row 3). }
    \label{fig:LJ8_sortCNum_FE_FEM_NN}
\end{figure}

 \subsubsection*{CVs (LDA2,LDA3)} 
\begin{itemize}
    \item $\beta = 10, ~\Omega = \{z\in\mathbb{R}^2~|~F(z) \le 1.55\}$,
    \item $\beta = 15, ~\Omega = \{z\in\mathbb{R}^2~|~F(z) \le 1.45\}$,
    \item $\beta = 20, ~\Omega = \{z\in\mathbb{R}^2~|~F(z) \le 1.32\}$.
\end{itemize}
For $\beta = 10$, $15$, and $20$, 
\begin{equation*}
    B = \text{connected component of }\{z \in \mathbb{R}^2~ | ~F(z) \leq 0.3\} \text{ that contains minimum 1},
\end{equation*}
\begin{equation*}
    A = \left\{ (x,y) \in \mathbb{R}^2~\vline~\frac{[-0.1*(x - 2.76) + (y - 1.76)]^2}{0.25^2} + \frac{[(x - 2.76) + 0.1 * (y - 1.76)]^2}{0.2^2} \leq 1. \right\}.
\end{equation*}

\begin{figure}[htbp]
    \centering
    \includegraphics[width=0.9\textwidth]{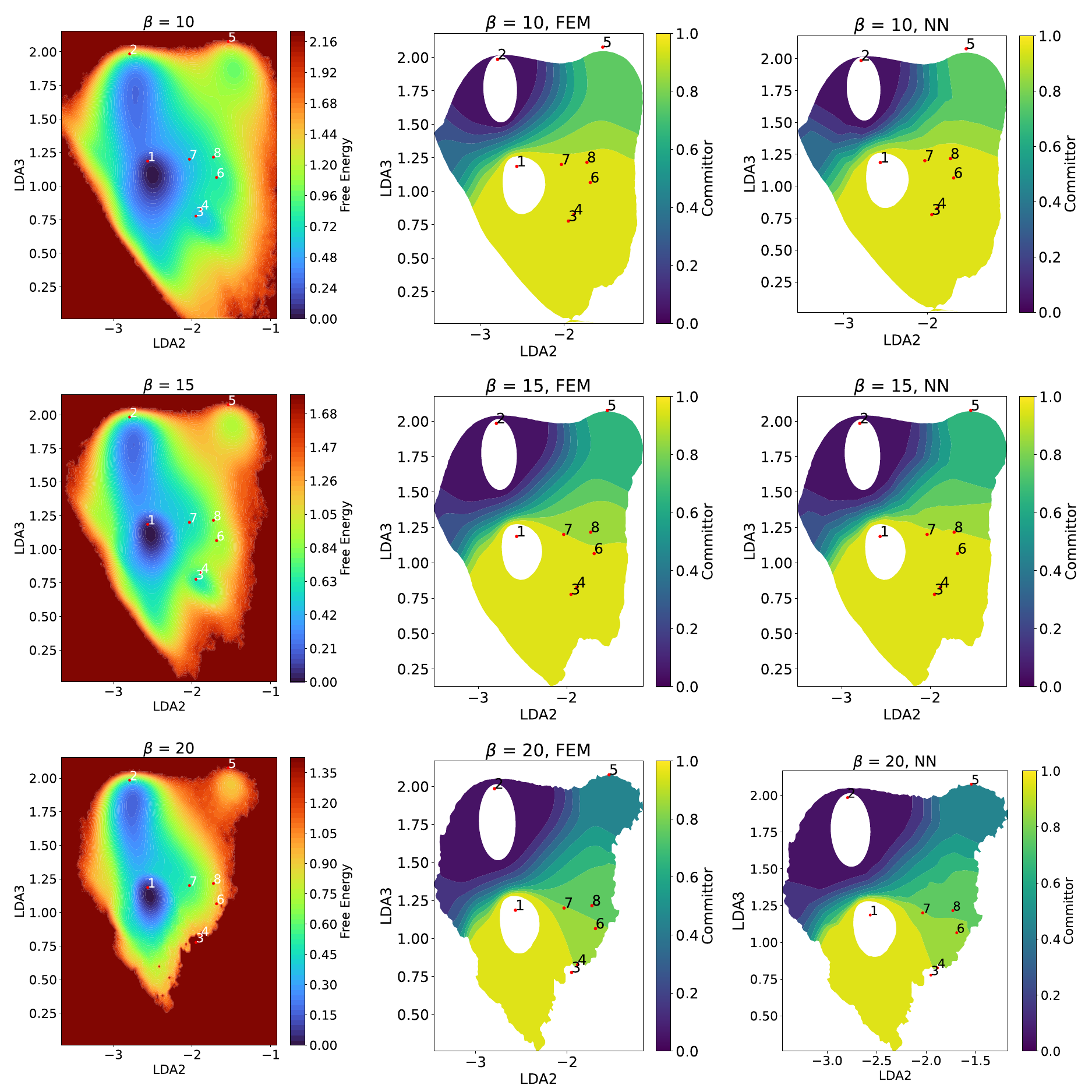}
    \caption{LJ8 in 3D. The free energy landscapes in CVs (LDA2,LDA3) (column 1), the chosen sets $A$ and $B$ and the committor computed by the finite element method (FEM) (column 2), and its approximation with a neural network (column 3) at $\beta = 10$ (row 1), $\beta = 15$ (row 2), and $\beta = 20$ (row 3). }
    \label{fig:LJ8_LDA23_FE_FEM_NN}
\end{figure}

 \subsubsection*{ CVs (LDA1,LDA2)} 
\begin{itemize}
    \item $\beta = 10, ~\Omega = \{z\in\mathbb{R}^2~|~F(z) \le 1.75\}$,
    \item $\beta = 15, ~\Omega = \{z\in\mathbb{R}^2~|~F(z) \le 1.55\}$.
\end{itemize}
The well-tempered metadynamics run never reached neighborhoods of the images of of minimum 4. Therefore, we did not compute the committor in (LDA1,LDA2) at $\beta = 20$. Instead, we used the NN committor at $\beta = 15$ as a reaction coordinate for forward flux sampling and to construct the stochastic control to sample the transition trajectories.

For $\beta = 10$ and $15$, 
\begin{equation*}
    A = \{z \in \mathbb{R}^2~ |~ F(z) \leq 0.25 \}.
\end{equation*}
The region $B$ is defined as a union of disjoint ellipses of the form
\begin{equation}
\label{eldef_LDA12}
    \frac{[(x-x_0)*v_{x_0} + (y - y_0)*v_{y_0}]^2}{r_{x_0}^2} +\frac{[(x-x_0)*v_{y_0} - (y - y_0)*v_{x_0}]^2}{r_{y_0}^2} \leq 1, 
\end{equation}
containing the images of minima 3, 4, 5, 6, 7, and 8. The parameters $x_0$, $y_0$, $v_{x_0}$, $v_{y_0}$, $r_{x_0}$, and $r_{y_0}$  in \eqref{eldef_LDA12} are listed in Table~\ref{table:FFS_LJ8_LDA12_committor_detail}.
\begin{table*}[h]
    \centering
%    \tiny
    \begin{tabular}{|c|c|c|c|c|c|c|c|}
    \hline
    $\beta$ & Minimum & $x_0$ & $y_0$ & $v_{x_0}$ & $v_{y_0}$ & $r_{x_0}$ & $r_{y_0}$ \\
    \hline
    \multirow{4}{*}{$\beta = 10$} 
     & 4 & -0.42 & -1.87 & 1.0 & 0 & 0.1 & 0.2\\
     & 3, 8 & 0.036 & -1.83 & 0.8 & 1.0 & 0.13 & 0.25\\
     & 7 & 0.70 & -2.03 & 0 & 1.0 & 0.2 & 0.1 \\
     & 5, 6 & 0.45 & -1.52 & 0.4 & 1.0 & 0.27 & 0.13 \\
     \hline
    \multirow{4}{*}{$\beta = 15$} 
     & 4 & -0.41 & -1.87 & 1.0 & 0 & 0.12 & 0.2 \\
     & 3, 8 & 0.036 &  -1.83 & 0.85 & 1.0 & 0.1 & 0.25 \\
     & 7 & 0.70 & -2.03 & 0 & 1.0 & 0.2 & 0.1 \\
     & 5, 6 & 0.43 & -1.52 & 0.4 & 1.0 & 0.25 & 0.13 \\
    \hline
    \end{tabular}
    \caption{The parameters for the set $B$ in the CVs (LDA1,LDA2) defined as the union of ellipses of the form \eqref{eldef_LDA12}. }
    \label{table:FFS_LJ8_LDA12_committor_detail}
\end{table*}

\begin{figure}[htbp]
    \centering
    \includegraphics[width=0.9\textwidth]{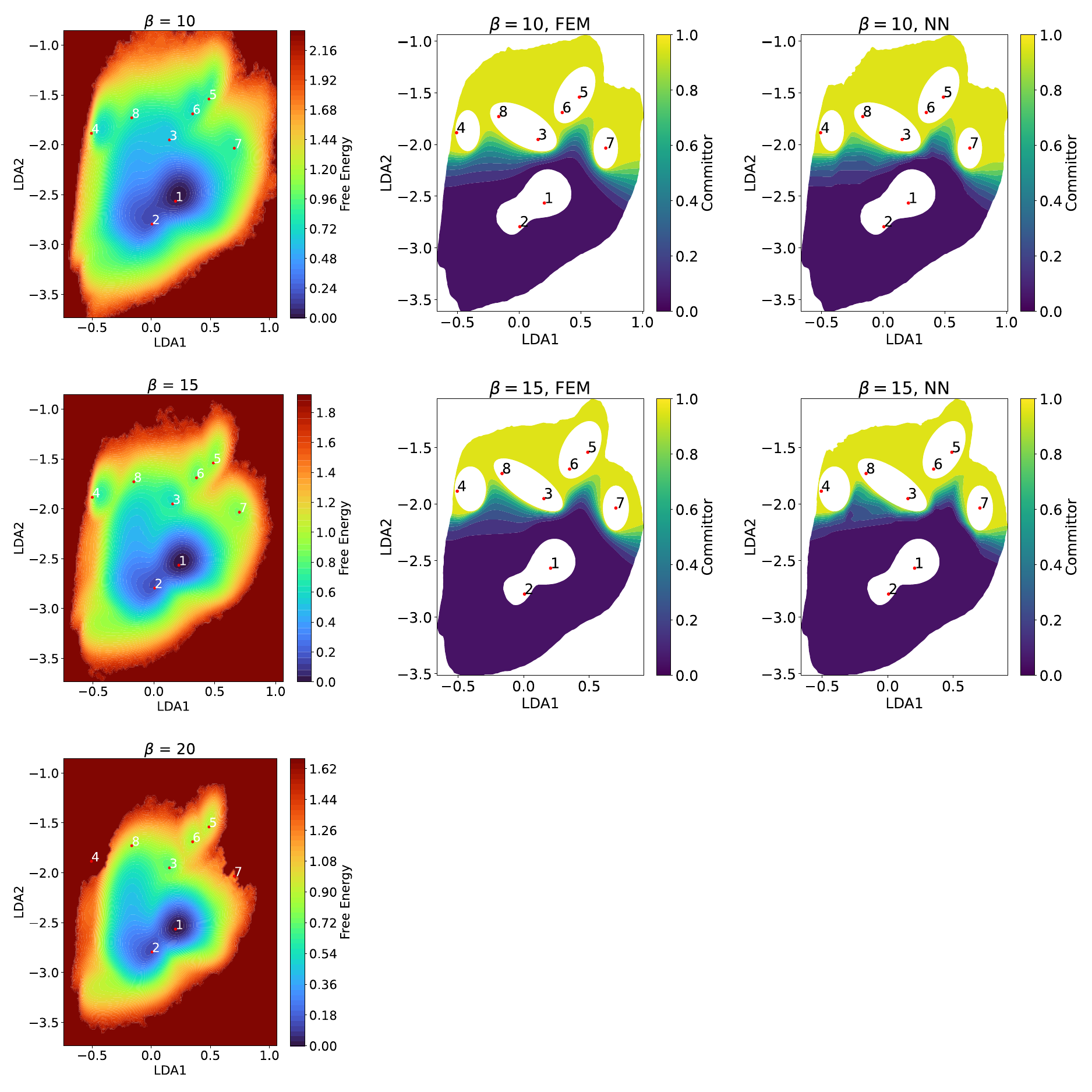}
    \caption{LJ8 in 3D. The free energy landscapes in CVs (LDA1,LDA2) (column 1), the chosen sets $A$ and $B$ and the committor computed by the finite element method (FEM) (column 2), and its approximation with a neural network (column 3) at $\beta = 10$ (row 1), $\beta = 15$ (row 2), and $\beta = 20$ (row 3). The committor was not computed at $\beta = 20$.}
    \label{fig:LJ8_LDA12_FE_FEM_NN}
\end{figure}

% \end{enumerate}

\section{Approximation of committor function via NN}
We use the committor computed in the CV space as the reaction coordinate for the forward flux sampling (FFS). We find it easiest to solve the committor problem using the finite element method (FEM) and then approximate it with a neural network (NN) to create a globally defined and easily evaluated function.

For committor functions at different temperatures and across various collective variables (CVs), we employ neural networks comprising two to three hidden layers. The number of neurons in each layer ranges from 10 to 40, depending on the specific case.  The hidden layers use the \textit{Rectified Linear Unit} ({\sf ReLU})  or {\sf tanh} activation function, and the output layer uses a sigmoid activation function, ${\sf sigmoid}(x) := (1 + \exp(-x))^{-1}$, to ensure that the network's output lies within the interval $[0,1]$. The resulting NN committor is of the form
\begin{equation}
    \label{qNN2hl}
    q_{\sf NN}(z) = {\sf sigmoid}\left(A_3{\sf ReLU}\left(A_2{\sf ReLU}\left(A_1z + b_1\right)+b_2\right)+b_3\right),
\end{equation}
or
\begin{equation}
    \label{qNN2hl_tanh}
    q_{\sf NN}(z) = {\sf sigmoid}\left(A_3{\sf tanh}\left(A_2{\sf tanh}\left(A_1z + b_1\right)+b_2\right)+b_3\right),
\end{equation}

or
\begin{equation}
    \label{qNN3hl}
    q_{\sf NN}(z) = {\sf sigmoid}\left(A_4{\sf ReLU}\left(A_3{\sf ReLU}\left(A_2{\sf ReLU}\left(A_1z + b_1\right)+b_2\right)+b_3\right)+b_4\right),
\end{equation}
where $z = \xi(\phi(x))\in\mathbb{R}^2$.

The network is trained by minimizing the mean squared error between its output and the FEM-computed committor function, using the Adam optimizer with a learning rate of $ 10^{-3}$. The FEM mesh points are used as the training points. Detailed network architectures for each case are provided in Table~\ref{table:NN_details}.

The side-by-side comparisons between the FEM and NN committors are provides in Figures~\ref{fig:LJ7_mu2mu3_FE_FEM_NN} -- \ref{fig:LJ8_LDA12_FE_FEM_NN}.

\section{ The computation of the transition rates using FFS and brute force}
The committor in collective variables at the corresponding values of $\beta$ was used as the reaction coordinate for the forward flux sampling (FFS) for all cases except CVs (LDA1,LDA2) at $\beta = 20$. We did not compute the committor in (LDA1,LDA2) at $\beta = 20$ and used the committor in (LDA1,LDA2) at $\beta = 15$ as the reaction coordinate at $\beta = 20$ instead.

\subsection*{Milestones for FFS}
The first step of the estimation of the transition rates is the redefinition of the sets $A$ and $B$ as the level sets of the NN committor close to zero and one, respectively. The boundaries of the new $A$ and $B$ become the first and the last milestones, respectively.
The milestones for FFS are defined as the level sets of the NN committor.  The milestones for all cases are given in Tables \ref{table:FFS_LJ7_detail_sortD} -- \ref{table:FFS_LJ8_LDA12_detail}.

We repeat each FFS run ten times to estimate statistical errors of the $k_A$ and $k_B$, the escape rates for $A$ and $B$ respectively, and the quantities derived from them:
\begin{equation}
    \label{estimands}
        \nu_{AB} = \frac{1}{\frac{1}{k_A}+\frac{1}{k_B}}\quad
    \rho_A = \frac{1}{1 + k_A/k_B},\quad 
    \rho_B = \frac{1}{1 + k_B/k_A}.  
\end{equation}

The confidence intervals are computed through error propagation with first-order Taylor expansion \cite{ku1966propagation}.
For a function $f(x_1,x_2,...,x_n)$ of multiple variables, its standard deviation $\sigma_f$ can be approximated using the quadrature sum of partial derivatives:
\begin{equation}
    \sigma_f = \sqrt{\left(\frac{\partial f}{\partial x_1} \sigma_{x_1}\right)^2 + ... + \left(\frac{\partial f}{\partial x_n} \sigma_{x_n}\right)^2}.
\end{equation}
For $\rho_B = \frac{1}{1 + k_{B}/k_{A}}$, the partial derivatives are
\begin{equation*}
    \frac{\partial \rho_B}{\partial k_{B}} = -\frac{k_{A}}{(k_{A} + k_{B})^2}, \quad \frac{\partial \rho_B}{\partial k_{A}} =\frac{k_{B}}{(k_{A} + k_{B})}.
\end{equation*}
Hence, given the estimated standard deviation for $k_{A}$ and $k_{B}$, the propagated error is 
\begin{equation*}
    \sigma_{\rho_B} = \sqrt{\left(\frac{\partial \rho_B}{\partial k_{B}} \sigma_{k_{B}}\right)^2 + \left(\frac{\partial \rho_B}{\partial k_{A}} \sigma_{k_{A}}\right)^2}.
\end{equation*}
Since $\rho_A = 1 - \rho_B$, their statistical errors are the same: $\sigma_{\rho_A} = \sigma_{\rho_B}$.

To estimate the standard deviation for the product $\nu_{AB} = \rho_A k_{A}$, we apply the same procedure.
Consequently, we can compute the confidence interval as 
\begin{equation*}
    \sigma_{\nu_{AB}} = \sqrt{\left(k_{A} \sigma_{\rho_A}\right)^2 + \left(\rho_A \sigma_{k_{A}}\right)^2}.
\end{equation*}

\begin{table*}[h]
    \centering
    \begin{tabular}{|c|c|c|p{4cm}|}
    \hline
    $\beta$ & Boundary of $A$ ($\lambda_0$) & Boundary of $B$ ($\lambda_n$) & Milestones \\
    \hline
    $\beta = 5, 7$ & $q = 10^{-3}$  & $q = 1-10^{-3}$ & {\footnotesize[0.54, 1.06, 1.59, 2.11, 2.64, 3.16, 3.69, 4.21, 4.74, 5.26, 5.79, 6.31, 6.84, 7.36, 7.89, 8.41, 8.94, 9.46] $\times$ $10^{-1}$} \\
    \hline
    $\beta = 9$ & $q=10^{-4}$ & $q =1-10^{-4}$ & {\footnotesize [0.53, 1.05, 1.58, 2.11, 2.633, 3.16, 3.68, 4.21, 4.74, 5.26, 5.79, 6.32, 6.84, 7.37, 7.89, 8.42, 8.95, 9.47] $\times$ $10^{-1}$} \\
    % $\beta = 7$ &  $q^7 =$ 1e-3  & $q^7 =$ 1-1e-3  & \\
    \hline
    
    \end{tabular}
    \caption{FFS setting for ML CV, ${\tt sort}[d^2]$, LJ7}
    \label{table:FFS_LJ7_detail_sortD}
\end{table*}

\begin{table*}[h]
    \centering
    \begin{tabular}{|c|c|c|p{4cm}|}
    \hline
    $\beta$  & Boundary of $A$ ($\lambda_0$) & Boundary of $B$ ($\lambda_n$) & Milestones \\
    \hline
    $\beta = 5, 7, 9$ & $q =10^{-2}$  & $q =1-10^{-2}$ & {\footnotesize [0.62, 1.13, 1.65, 2.16, 2.68, 3.19, 3.71, 4.23, 4.74, 5.26, 5.77, 6.29, 6.81, 7.32, 7.84, 8.35, 8.87, 9.38] $\times$ $10^{-1}$} \\
    \hline
    \end{tabular}
    \caption{FFS setting for Ml CV, ${\tt sort}[c]$, LJ7}
    \label{table:FFS_LJ7_detail_sortCNum}
\end{table*}

\begin{table*}[h]
    \centering
    \begin{tabular}{|c|c|c|p{4cm}|}
    \hline
    $\beta$  & Boundary of $A$ ($\lambda_0$) & Boundary of $B$ ($\lambda_n$) & Milestones \\
    \hline
    $\beta = 5, 7, 9$ & $q =10^{-2}$   & $q =1-10^{-2}$ & {\footnotesize[0.62, 1.13, 1.65, 2.16, 2.68, 3.19, 3.71, 4.23, 4.74, 5.26, 5.77, 6.29, 6.81, 7.32, 7.84, 8.35, 8.87, 9.38] $\times$ $10^{-1}$} \\
    \hline
    \end{tabular}
    \caption{FFS setting for $(\mu_2, \mu_3)$, LJ7}
    \label{table:FFS_LJ7_detail_mu2mu3}
\end{table*}

\begin{table*}[h!]
    \centering
    \begin{tabular}{|c|c|c|p{4cm}|}
    \hline
    $\beta$  & Boundary of $A$ ($\lambda_0$) & Boundary of $B$ ($\lambda_n$) & Milestones \\
    \hline
    $\beta = 10$ & $q =3\cdot 10^{-2}$ & $q =1-3\cdot 10^{-2}$ &
    {\footnotesize [0.79, 1.29, 1.78, 2.28, 2.77, 3.27, 3.76, 4.26, 5.25, 5.74, 6.24, 6.73, 7.23, 7.72, 8.22, 8.71, 9.21] $\times$ $10^{-1}$} \\
    \hline
    $\beta = 15$ & $q =2\cdot10^{-2}$ & $q =1-2\cdot10^{-2}$ &
    {\footnotesize [0.71, 1.21, 1.72, 2.22, 2.72, 3.23, 3.74, 4.24, 5.25, 5.76, 6.26, 6.77, 7.27, 7.78, 8.28, 8.79, 9.29] $\times$ $10^{-1}$} \\
    \hline
    $\beta = 20$ & $q =10^{-2}$ & $q =1-10^{-2}$ &
    {\footnotesize [0.62, 1.13, 1.65, 2.16, 2.68, 3.19, 3.71, 4.23, 4.74, 5.26, 5.77, 6.29, 6.80, 7.32, 7.84, 8.35, 8.87, 9.38] $\times$ $10^{-1}$} \\
    \hline    
    \end{tabular}
    \caption{FFS details for transition rates under three different temperatures for ML CVs of sorted coordination number for LJ8 in 3D. }
    \label{table:FFS_LJ8_MLCV_detail}
\end{table*}

\begin{table*}[h!]
    \centering
    \begin{tabular}{|c|c|c|p{4cm}|}
    \hline
    $\beta$  & Boundary of $A$ ($\lambda_0$) & Boundary of $B$ ($\lambda_n$) & Milestones \\
    \hline
    $\beta = 10, 15$ &$q =10^{-2}$ & $q =1-10^{-2}$ &
    {\footnotesize [0.62, 1.13, 1.65, 2.16, 2.68, 3.19, 3.71, 4.23, 4.74, 5.26, 5.77, 6.29, 6.81, 7.32, 7.84, 8.35, 8.87, 9.38] $\times$ $10^{-1}$} \\
    \hline
    $\beta = 20$ & $q =2\cdot10^{-2}$  & $q =1-2\cdot10^{-2}$ &
    {\footnotesize [0.71, 1.21, 1.72, 2.22, 2.72, 3.23, 3.74, 4.24, 5.25, 5.76, 6.26, 6.77, 7.27, 7.78, 8.28, 8.79, 9.29] $\times$ $10^{-1}$} \\
    \hline
    \end{tabular}
    \caption{FFS details for transition rates under three different temperatures in $(\mu_2, \mu_3)$ for LJ8 in 3D. }
    \label{table:FFS_LJ8_mu2mu3_detail}
\end{table*}

\begin{table*}[h]
    \centering
    \begin{tabular}{|c|c|c|p{4cm}|}
    \hline
    $\beta$  & boundary of $A$ ($\lambda_0$) & boundary of $B$ ($\lambda_n$) & milestones \\
    \hline
    $\beta = 10, 15,20$ & $q =10^{-2}$ & $q =1-10^{-2}$ & {\scriptsize [0.62, 1.13, 1.65, 2.16, 2.68, 3.19, 3.71, 4.23, 4.74, 5.26, 5.77, 6.29, 6.81, 7.32, 7.84, 8.35, 8.87, 9.38] $\times$ $10^{-1}$} \\
    \hline
    
    \end{tabular}
    \caption{FFS details for transition rates under three different temperatures in (LDA2,LDA3) for LJ8 in 3D.}
    \label{table:FFS_LJ8_LDA23_detail}
\end{table*}

\begin{table*}[h!]
    \centering
    \begin{tabular}{|c|c|c|p{4cm}|}
    \hline
    $\beta$  & Boundary of $A$ ($\lambda_0$) & Boundary of $B$ ($\lambda_n$) & Milestones \\
    \hline
    $\beta = 10$ &$q =2\cdot10^{-2}$  & $q =1-2\cdot10^{-2}$ & 
    {\scriptsize [0.71, 1.21, 1.72, 2.22, 2.73, 3.23, 3.74, 4.24, 4.75, 5.25, 5.76, 6.26, 6.77, 7.27, 7.78, 8.28, 8.79, 9.29] $\times$ $10^{-1}$} \\
    \hline
    $\beta = 15, 20$ & $10^{-2}$ & $1-10^{-2}$ &
    {\scriptsize [0.1, 0.62, 1.13, 1.65, 2.16, 2.68, 3.19, 3.71, 4.23, 4.74, 5.26, 5.77, 6.29, 7.32, 7.84, 8.35, 8.87, 9.38, 9.9] $\times$ $10^{-1}$} \\
    \hline
    
    \end{tabular}
    \caption{FFS details for transition rates under three different temperatures in (LDA1, LDA2) for LJ8 in 3D. }
    \label{table:FFS_LJ8_LDA12_detail}
\end{table*}

\subsection*{Rates from brute force simulations}
\label{sec:brute_force_rates}
We use brute-force all-atom simulations to compare the results with the FFS rates. The brute-force rates might involve a large statistical error if the transition of interest is rare. 

For each case, we performed 10 unbiased all-atom simulations. Each simulation had $10^9$ time steps of size $\Delta t = 5\cdot 10^{-5}$. During each simulation, we counted the following quantities:
\begin{itemize}
    \item the total time $T_A$ during which the system last entered $A$ rather than $B$,
    \item the total time $T_B$ during which the system last entered $B$ rather than $A$,
    \item $N_{AB}$, the number of times the system entered $B$  while it last hit of $A$ rather than $B$.
\end{itemize}
Using these data, we find the quantities of interest. Let $T = N_{\sf steps}\cdot\Delta t = 5\cdot 10^{4}$ be the total elapsed time.
Then the probabilities $\rho_{A}$ and $\rho_B$ that the system last hit $A$ rather than $B$ and, respectively, the other way around, are 
\begin{equation}
    \rho_A = \frac{T_A}{T}, \quad \rho_B = \frac{T_B}{T}.
\end{equation}
The escape rates $k_{A}$ and $ k_B$ are
\begin{equation}
    k_A = \frac{N_{AB}}{T_A}, \quad k_B = \frac{N_{AB}}{T_B}.
\end{equation}
The transition rate $\nu_{AB}$ is
\begin{equation}
    \nu_{AB} = \frac{N_{AB}}{T}.
\end{equation}

Repeating each brute-force run ten times allows us to estimate the statistical errors of the estimated quantities $\rho_A$, $\rho_B$, $k_A$, $k_B$, and $\nu_{AB}$.

\section{The probability density of transition trajectories}
\label{sec:RTraj}
The NN committor is used to design a stochastic control for generating transition trajectories and estimating their probability density. The following control is added to the drift of the governing SDE:
\begin{equation}
    \label{stochcontrol}
    2\beta^{-1}\nabla\log q_{\sf NN}(\xi(\phi(x))),
\end{equation}
where $\xi$ is the collective variable, $\phi$ is the feature map, and $x$ is the vector of atomic coordinates.
If the stochastic control were optimal, the controlled process would generate stochastic trajectories whose law is exactly the same as the law of the transition trajectories from $A$ to $B$ of the original process. However, the control \eqref{stochcontrol} is not optimal. Due to the redefinition of $A$ as level set of $q_{\sf NN}$ close to zero, we avoid the singularity of the stochastic control at the boundary of $A$. On the downside, we allow some trajectories starting at the boundary of $A$ to return to $A$. 

Therefore, we estimate the probability density of the transition trajectories as follows. 
\begin{enumerate}
    \item We mesh the CV space into $129\times129$ cells. We initialize a set of bins, {\tt bins}, centered at the mesh points, with zeros.
    \item We run an unbiased all-atom simulation starting in $A$ until we collect $N_{\sf exit} = 10^4$ exit events from $A$. For each exit event, we record the atomic configurations. 
    \item We initialize a temporary set of bins, {\tt bins\_tmp}, with zeros.
    \item We randomly pick an exit configuration from $A$ and launch a controlled trajectory from it. We record its visits to the mesh cells in  {\tt bins\_tmp}. If the trajectory reaches $B$, we update {\tt bins}: {\tt bins = bins + bins\_tmp}. Otherwise, we discard {\tt bins\_tmp}.
    \item We repeat steps 3 and 4 until we accumulate a total of $N_{\sf traj} =10^4$ transition trajectories from $A$ to $B$.
    \item The variable {\tt bins} is approximately proportional to the probability density of transition trajectories.
\end{enumerate} 

\begin{figure}[htbp]
    \centering
    \includegraphics[width=0.9\textwidth]{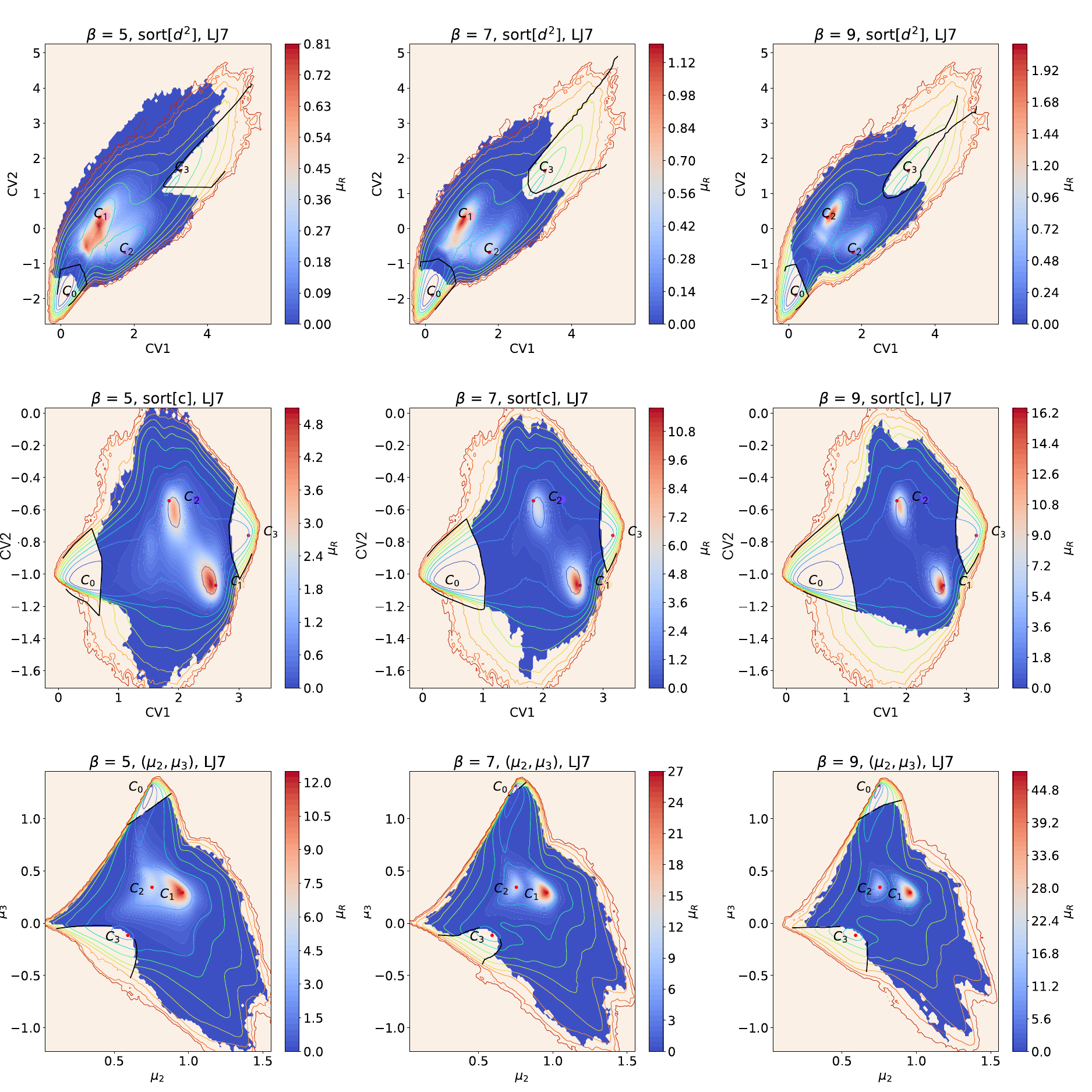}
    \caption{LJ7 in 2D. The estimated probability density of transition trajectories projected onto the ML CV space with the feature map ${\sf sort}[d^2]$ (row 1), ${\sf sort}[c]$ (row 2), and the CVs $(\mu_2,\mu_3)$ (row 3), at $\beta = 5$ (column 1), $\beta = 7$ (column 2), and $\beta = 9$ (column 3).}
    \label{fig:LJ7_ProbDensityRTraj}
\end{figure}

\begin{figure}[htbp]
    \centering
    \includegraphics[width=0.9\textwidth]{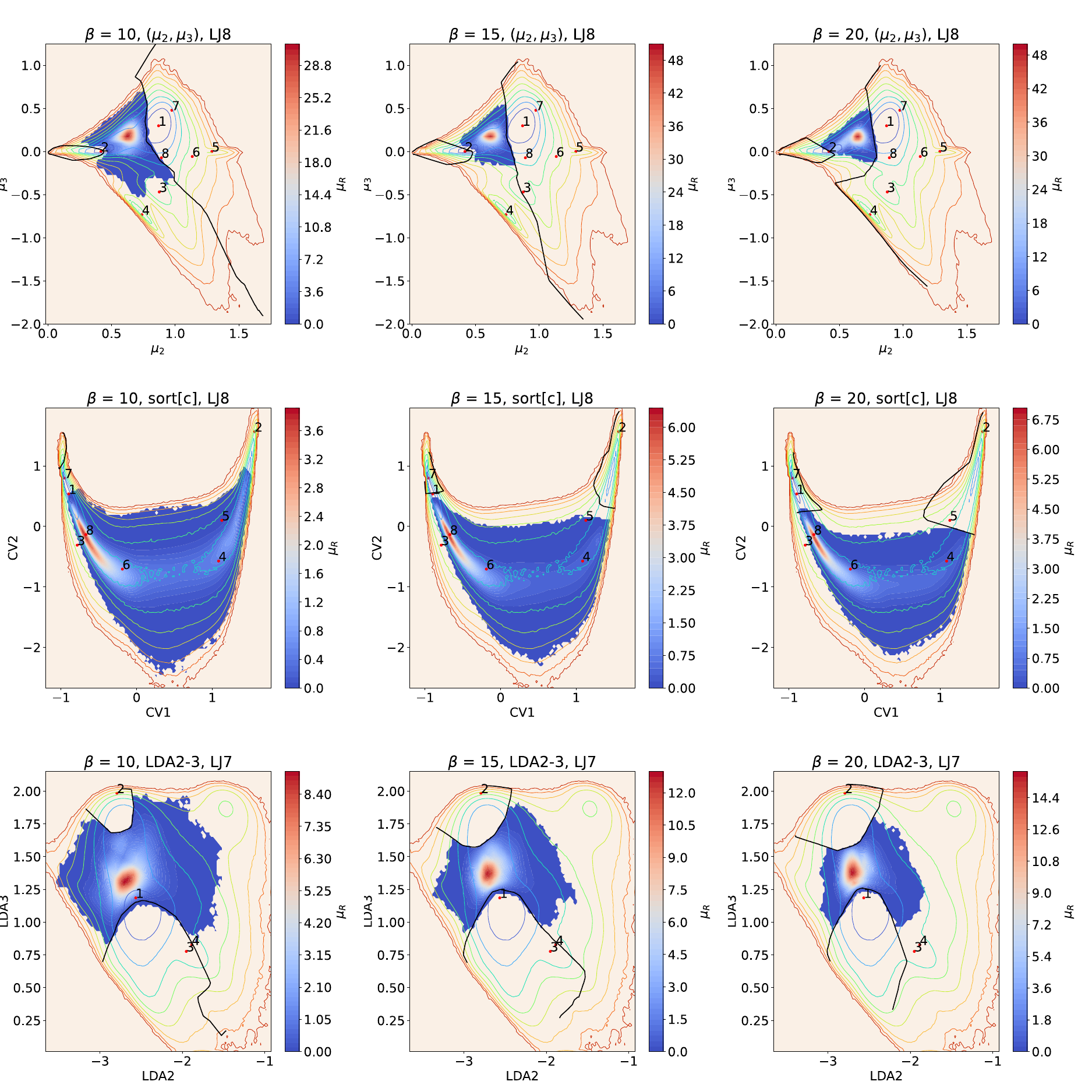}
    \caption{LJ8 in 3D. The estimated probability density of transition trajectories projected onto the CV space $(\mu_2,\mu_3)$ (row 1), ML CV with the feature map ${\sf sort}[c]$ (row 2), and (LDA2,LDA3) (row 3) at $\beta = 10$ (column 1), $\beta = 15$ (column 2), and $\beta = 20$ (column 3).}
    \label{fig:LJ8_ProbDensityRTraj}
\end{figure}

\begin{figure}[htbp]
    \centering
    \includegraphics[width=0.9\textwidth]{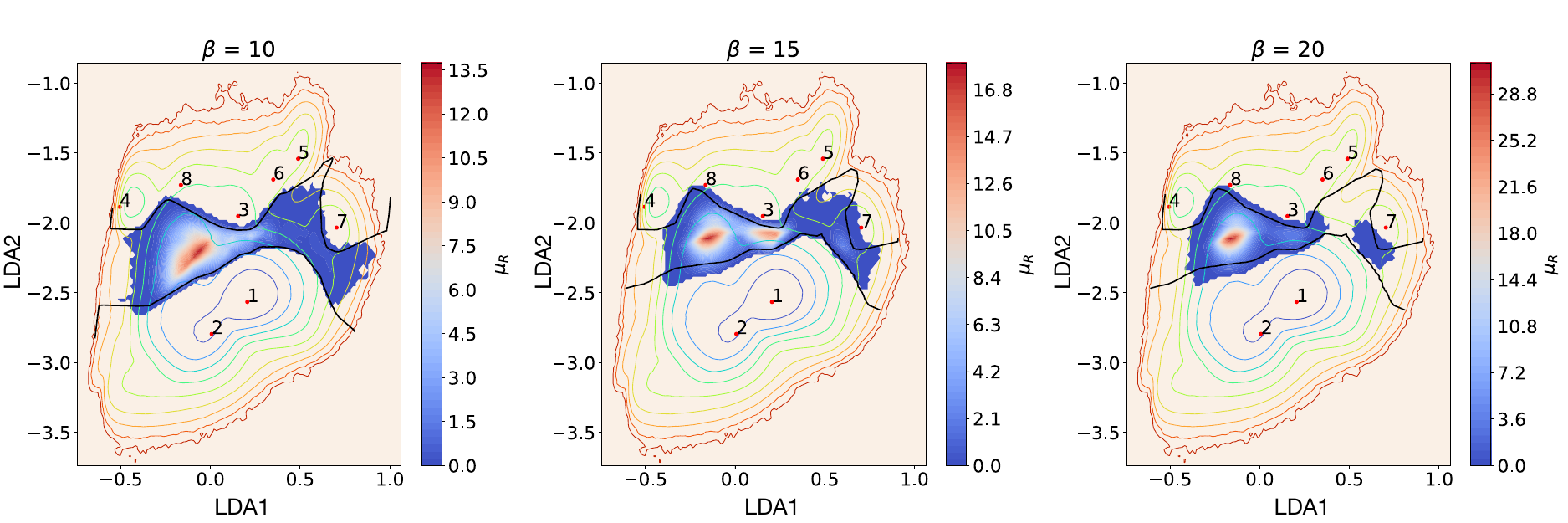}
    \caption{LJ8 in 3D. The estimated probability density of transition trajectories projected onto the CV space (LDA1,LDA2)  at $\beta = 10$ (left), $\beta = 15$ (middle), and $\beta = 20$ (right).}
    \label{fig:LJ8_LDA12_ProbDensityRTraj}
\end{figure}

The estimated probability density of transition trajectories for all cases considered is shown in Figs. \ref{fig:LJ7_ProbDensityRTraj}--\ref{fig:LJ8_LDA12_ProbDensityRTraj}.

\bibliographystyle{abbrv}

\end{document}